\documentstyle[11pt]{article}

\makeatletter
\def\section{\@startsection {section}{1}{\z@}{-3.5ex plus -1ex minus
 -.2ex}{2.3ex plus .2ex}{\large\bf}}
\def\subsection{\@startsection{subsection}{2}{\z@}{-3.25ex plus -1ex minus
 -.2ex}{1.5ex plus .2ex}{\normalsize\bf}}
\makeatother

\makeatletter
\@addtoreset{equation}{section}

\makeatother

\newcommand{\nc}{\newcommand}
\newcommand{\rnc}{\renewcommand}

\nc{\be}{\begin{equation}}
\nc{\ee}{\end{equation}}
\nc{\bea}{\begin{eqnarray}}
\nc{\eea}{\end{eqnarray}}

\rnc{\a}{\alpha}
\rnc{\b}{\beta}
\rnc{\gg}{\gamma}
\rnc{\d}{\delta}
\nc{\e}{\eta}
\nc{\eb}{\bar{\eta}}
\nc{\ep}{\epsilon}
\nc{\f}{\phi}
\nc{\fb}{\bar{\phi}}
\nc{\vf}{\varphi}
\nc{\p}{\psi}
\rnc{\pb}{\bar{\psi}}
\rnc{\c}{\chi}
\nc{\cb}{\bar{\c}}
\nc{\la}{\lambda}
\nc{\m}{\mu}
\nc{\n}{\nu}
\rnc{\o}{\omega}
\nc{\Om}{\Omega}
\rnc{\t}{\theta}
\nc{\eps}{\epsilon}

\rnc{\S}{\Sigma}
\nc{\Sg}{\Sigma_{g}}
\nc{\Sa}{\S\times\{0\}}
\nc{\Sb}{\S\times\{1\}}
\nc{\SI}{\S\times I}
\nc{\SS}{\S\times S^{1}}
\nc{\M}{{\cal M}}
\nc{\G}{{\cal G}}

\nc{\trac}[2]{{\textstyle\frac{#1}{#2}}}

\nc{\ex}[1]{{\rm e}^{\,\textstyle#1}}

\nc{\mat}[4]{\left(\begin{array}{cc}#1&#2\\#3&#4\end{array}\right)}

\nc{\som}[9]{\left(\begin{array}{ccc}#1&#2&#3\\#4&#5&#6\\#7&#8&#9%
\end{array}\right)}

\def\tr{\mathop{\rm tr}\nolimits}
\def\Tr{\mathop{\rm Tr}\nolimits}
\nc{\ra}{\rightarrow}
\nc{\Ra}{\Rightarrow}
\nc{\Lra}{\Leftrightarrow}
\nc{\ot}{\otimes}
\rnc{\ss}{\subset}
\nc{\nul}{\noindent\underline}
\nc{\ad}{\mbox{ad}}
\nc{\non}{\nonumber\\}

\nc{\subs}[1]{{\vspace*{0.5cm}}%
{\noindent\underline{#1}}{\addcontentsline{toc}{subsubsection}{#1}}%
{\vspace*{0.3cm}}}
\rnc{\lg}{{\bf g}}
\nc{\lt}{{\bf t}}
\nc{\lk}{{\bf k}}
\nc{\bft}{{\bf t}}
\nc{\bfk}{{\bf k}}
\nc{\bfg}{{\bf g}}
\nc{\del}{\partial}
\nc{\dbar}{\bar{\del}}
\nc{\dz}{\del_{z}}
\nc{\dzb}{\del_{\bar{z}}}
\nc{\az}{A_{z}}
\nc{\azb}{A_{\bar{z}}}
\nc{\bz}{B_{z}}
\nc{\bzb}{B_{\bar{z}}}
\nc{\ba}{{\bf A}}
\nc{\bb}{{\bf B}}
\nc{\g}{g^{-1}}
\nc{\h}{h^{-1}}
\nc{\dw}{\Delta_{W}}
\nc{\Det}{{\rm Det}\,}
\nc{\Ad}{{\rm Ad}}
\nc{\bG}{{\bf G}}
\nc{\bT}{{\bf T}}
\nc{\bK}{{\bf H}}
\nc{\bH}{{\bf H}}
\nc{\bP}{{\bf P}}
\nc{\gt}{G/T}
\nc{\map}[2]{{\rm Map}(#1,#2)}
\nc{\mg}{\map{M}{G}}
\nc{\mgr}{\map{M}{G_{r}}}
\nc{\mt}{\map{M}{T}}
\nc{\mtr}{\map{M}{T_{r}}}
\nc{\sg}{\map{\Sigma}{G}}
\nc{\sgr}{\map{\Sigma}{G_{r}}}
\nc{\st}{\map{\Sigma}{T}}
\nc{\str}{\map{\Sigma}{T_{r}}}
\nc{\mgt}{\map{M}{\gt}}
\nc{\sgt}{\map{\Sigma}{\gt}}
\nc{\ug}{\map{U}{G}}
\nc{\ugr}{\map{U}{G_{r}}}
\nc{\ut}{\map{U}{T}}
\nc{\utr}{\map{U}{T_{r}}}
\nc{\uag}{\map{\ua}{G}}
\nc{\uagr}{\map{\ua}{G_{r}}}
\nc{\uat}{\map{\ua}{T}}
\nc{\uatr}{\map{\ua}{T_{r}}}
\nc{\uabg}{\map{\uab}{G}}
\nc{\uabgr}{\map{\uab}{G_{r}}}
\nc{\uabt}{\map{\uab}{T}}
\nc{\uabtr}{\map{\uab}{T_{r}}}
\nc{\unA}{\underline{A}}
\nc{\C}{{\cal A}/{\cal G}}
\nc{\A}[1]{{\cal A}^{#1}/{\cal G}^{#1}}
\nc{\dx}{\dot{x}}
\rnc{\O}[2]{\Omega^{#1}({#2},\lg)}
\nc{\wif}{Weyl integral formula}
\nc{\CS}{Chern-Simons theory}
\nc{\tg}{\tilde{g}}
\nc{\tti}{\tilde{t}}
\nc{\th}{\tilde{h}}

\nc{\ga}{g_{\a}}
\nc{\gb}{g_{\b}}
\nc{\gc}{g_{\gg}}
\nc{\ha}{h_{\a}}
\nc{\hb}{h_{\b}}
\nc{\hc}{h_{\gg}}
\nc{\ka}{k_{\a}}
\nc{\kb}{k_{\b}}
\nc{\kc}{k_{\gg}}
\nc{\ta}{t_{\a}}
\nc{\tb}{t_{\b}}
\nc{\tc}{t_{\gg}}
\nc{\gab}{g_{\a\b}}
\nc{\gac}{g_{\a\gg}}
\nc{\gbc}{g_{\b\gg}}
\nc{\hab}{h_{\a\b}}
\nc{\hac}{h_{\a\gg}}
\nc{\hbc}{h_{\b\gg}}
\nc{\kab}{k_{\a\b}}
\nc{\kac}{k_{\a\gg}}
\nc{\kbc}{k_{\b\gg}}
\nc{\tab}{t_{\a\b}}
\nc{\tac}{t_{\a\gg}}
\nc{\tbc}{t_{\b\gg}}
\nc{\ua}{U_{\a}}
\nc{\ub}{U_{\b}}
\nc{\uc}{U_{\gg}}
\nc{\uab}{U_{\a}\cap U_{\b}}
\nc{\uac}{U_{\a}\cap U_{\gg}}
\nc{\ubc}{U_{\b}\cap U_{\gg}}

\nc{\RC}{{\cal R}_{\C}}                     %CURVATURES
\nc{\RM}{{\cal R}_{\M}}                     %----------
\nc{\RX}{{\cal R}_{X}}
\nc{\RY}{{\cal R}_{Y}}
\nc{\cs}{\c_{s}}                            %REGULARIZED EULER NUMBER
\nc{\F}{\Phi}
\nc{\Fn}{\Phi_{\nabla}}
\nc{\On}{\Omega_{\nabla}}
\nc{\tint}[1]{\int_{0}^{#1}\!dt\;}
\nc{\nt}{\nabla_{t}}
\def\s1{S^{1}}
\nc\en{e_{\nabla}}
\nc\ens{e_{s,\nabla}}
\nc\env{e_{V,\nabla}}
\def\tft{top\-o\-log\-i\-cal field the\-o\-ry}
\def\tgt{top\-o\-log\-i\-cal gauge the\-o\-ry}
\def\sqm{su\-per\-sym\-met\-ric quan\-tum mech\-an\-ics}
\def\mq{Mathai-Quillen}

\def\dh{Dui\-ster\-maat\--\-Heck\-man}
\def\dht{\dh\ theorem}

\makeatletter

\def\diagram{\leftwidth=\z@ \rightwidth=\z@ \topheight=\z@
\botheight=\z@ \setbox\@picbox\hbox\bgroup}

\def\enddiagram{\egroup\wd\@picbox\rightwidth\unitlength
\ht\@picbox\topheight\unitlength \dp\@picbox\botheight\unitlength
\hskip\leftwidth\unitlength\box\@picbox}

\def\bfig{\begin{diagram}}
\def\efig{\end{diagram}}
\newcount\wideness \newcount\leftwidth \newcount\rightwidth
\newcount\highness \newcount\topheight \newcount\botheight

\def\ratchet#1#2{\ifnum#1<#2 \global #1=#2 \fi}

\def\putbox(#1,#2)#3{%
\horsize{\wideness}{#3} \divide\wideness by 2
{\advance\wideness by #1 \ratchet{\rightwidth}{\wideness}}
{\advance\wideness by -#1 \ratchet{\leftwidth}{\wideness}}
\vertsize{\highness}{#3} \divide\highness by 2
{\advance\highness by #2 \ratchet{\topheight}{\highness}}
{\advance\highness by -#2 \ratchet{\botheight}{\highness}}
\put(#1,#2){\makebox(0,0){$#3$}}}

\def\putlbox(#1,#2)#3{%
\horsize{\wideness}{#3}
{\advance\wideness by #1 \ratchet{\rightwidth}{\wideness}}
{\ratchet{\leftwidth}{-#1}}
\vertsize{\highness}{#3} \divide\highness by 2
{\advance\highness by #2 \ratchet{\topheight}{\highness}}
{\advance\highness by -#2 \ratchet{\botheight}{\highness}}
\put(#1,#2){\makebox(0,0)[l]{$#3$}}}

\def\putrbox(#1,#2)#3{%
\horsize{\wideness}{#3}
{\ratchet{\rightwidth}{#1}}
{\advance\wideness by -#1 \ratchet{\leftwidth}{\wideness}}
\vertsize{\highness}{#3} \divide\highness by 2
{\advance\highness by #2 \ratchet{\topheight}{\highness}}
{\advance\highness by -#2 \ratchet{\botheight}{\highness}}
\put(#1,#2){\makebox(0,0)[r]{$#3$}}}

\def\adjust[#1]{} % For compatibility

\newcount \coefa
\newcount \coefb
\newcount \coefc
\newcount\tempcounta
\newcount\tempcountb
\newcount\tempcountc
\newcount\tempcountd
\newcount\xext
\newcount\yext
\newcount\xoff
\newcount\yoff
\newcount\gap%
\newcount\arrowtypea
\newcount\arrowtypeb
\newcount\arrowtypec
\newcount\arrowtyped
\newcount\arrowtypee
\newcount\height
\newcount\width
\newcount\xpos
\newcount\ypos
\newcount\run
\newcount\rise
\newcount\arrowlength
\newcount\halflength
\newcount\arrowtype
\newdimen\tempdimen
\newdimen\xlen
\newdimen\ylen
\newsavebox{\tempboxa}%
\newsavebox{\tempboxb}%
\newsavebox{\tempboxc}%

\newdimen\w@dth

\def\setw@dth#1#2{\setbox\z@\hbox{$#1$}\w@dth=\wd\z@
\setbox\@ne\hbox{$#2$}\ifnum\w@dth<\wd\@ne \w@dth=\wd\@ne \fi
\advance\w@dth by 1.2em}

\def\t@^#1_#2{\def\n@one{#1}\def\n@two{#2}\mathrel{\setw@dth{#1}{#2}
\mathop{\hbox to \w@dth{\rightarrowfill}}\limits
\ifx\n@one\empty\else ^{\box\z@}\fi
\ifx\n@two\empty\else _{\box\@ne}\fi}}
\def\t@@^#1{\@ifnextchar_ {\t@^{#1}}{\t@^{#1}_{}}}
\def\to{\@ifnextchar^ {\t@@}{\t@@^{}}}

\def\t@left^#1_#2{\def\n@one{#1}\def\n@two{#2}\mathrel{\setw@dth{#1}{#2}
\mathop{\hbox to \w@dth{\leftarrowfill}}\limits
\ifx\n@one\empty\else ^{\box\z@}\fi
\ifx\n@two\empty\else _{\box\@ne}\fi}}
\def\t@@left^#1{\@ifnextchar_ {\t@left^{#1}}{\t@left^{#1}_{}}}
\def\toleft{\@ifnextchar^ {\t@@left}{\t@@left^{}}}

\def\two@^#1_#2{\def\n@one{#1}\def\n@two{#2}\mathrel{\setw@dth{#1}{#2}
\mathop{\vcenter{\hbox to \w@dth{\rightarrowfill}\kern-1.7ex
                 \hbox to \w@dth{\rightarrowfill}}%
       }\limits
\ifx\n@one\empty\else ^{\box\z@}\fi
\ifx\n@two\empty\else _{\box\@ne}\fi}}
\def\tw@@^#1{\@ifnextchar_ {\two@^{#1}}{\two@^{#1}_{}}}
\def\two{\@ifnextchar^ {\tw@@}{\tw@@^{}}}

\def\tofr@^#1_#2{\def\n@one{#1}\def\n@two{#2}\mathrel{\setw@dth{#1}{#2}
\mathop{\vcenter{\hbox to \w@dth{\rightarrowfill}\kern-1.7ex
                 \hbox to \w@dth{\leftarrowfill}}%
       }\limits
\ifx\n@one\empty\else ^{\box\z@}\fi
\ifx\n@two\empty\else _{\box\@ne}\fi}}
\def\t@fr@^#1{\@ifnextchar_ {\tofr@^{#1}}{\tofr@^{#1}_{}}}
\def\tofro{\@ifnextchar^ {\t@fr@}{\t@fr@^{}}}

\def\mon{\mathop{\m@th\hbox to
      14.6\P@{\lasyb\char'51\hskip-2.1\P@$\arrext$\hss
$\mathord\rightarrow$}}\limits} % width of \epi
\def\leftmono{\mathrel{\m@th\hbox to
14.6\P@{$\mathord\leftarrow$\hss$\arrext$\hskip-2.1\P@\lasyb\char'50%
}}\limits} % width of \epi
\mathchardef\arrext="0200       % amr minus for arrow extension (see \into)

\def\settypes(#1,#2,#3){\arrowtypea#1 \arrowtypeb#2 \arrowtypec#3}
\def\settoheight#1#2{\setbox\@tempboxa\hbox{#2}#1\ht\@tempboxa\relax}%
\def\settodepth#1#2{\setbox\@tempboxa\hbox{#2}#1\dp\@tempboxa\relax}%
\def\settokens[#1`#2`#3`#4]{%
     \def\tokena{#1}\def\tokenb{#2}\def\tokenc{#3}\def\tokend{#4}}
\def\setsqparms[#1`#2`#3`#4;#5`#6]{%
\arrowtypea #1
\arrowtypeb #2
\arrowtypec #3
\arrowtyped #4
\width #5
\height #6
}
\def\setpos(#1,#2){\xpos=#1 \ypos#2}

\def\settriparms[#1`#2`#3;#4]{\settripairparms[#1`#2`#3`1`1;#4]}%

\def\settripairparms[#1`#2`#3`#4`#5;#6]{%
\arrowtypea #1
\arrowtypeb #2
\arrowtypec #3
\arrowtyped #4
\arrowtypee #5
\width #6
\height #6
}

\def\resetparms{\settripairparms[1`1`1`1`1;500]\width 500}%default values%

\resetparms

\def\mvector(#1,#2)#3{%%
\put(0,0){\vector(#1,#2){#3}}%
\put(0,0){\vector(#1,#2){26}}%
}
\def\evector(#1,#2)#3{{%%
\arrowlength #3
\put(0,0){\vector(#1,#2){\arrowlength}}%
\advance \arrowlength by-30
\put(0,0){\vector(#1,#2){\arrowlength}}%
}}

\def\horsize#1#2{%
\settowidth{\tempdimen}{$#2$}%
#1=\tempdimen
\divide #1 by\unitlength
}

\def\vertsize#1#2{%
\settoheight{\tempdimen}{$#2$}%
#1=\tempdimen
\settodepth{\tempdimen}{$#2$}%
\advance #1 by\tempdimen
\divide #1 by\unitlength
}

\def\putvector(#1,#2)(#3,#4)#5#6{{%
\ifnum3<\arrowtype
\putdashvector(#1,#2)(#3,#4)#5\arrowtype
\else
\ifnum\arrowtype<-3
\putdashvector(#1,#2)(#3,#4)#5\arrowtype
\else
\xpos=#1
\ypos=#2
\run=#3
\rise=#4
\arrowlength=#5
\ifnum \arrowtype<0
    \ifnum \run=0
        \advance \ypos by-\arrowlength
    \else
        \tempcounta \arrowlength
        \multiply \tempcounta by\rise
        \divide \tempcounta by\run
        \ifnum\run>0
            \advance \xpos by\arrowlength
            \advance \ypos by\tempcounta
        \else
            \advance \xpos by-\arrowlength
            \advance \ypos by-\tempcounta
        \fi
    \fi
    \multiply \arrowtype by-1
    \multiply \rise by-1
    \multiply \run by-1
\fi
\ifcase \arrowtype
\or \put(\xpos,\ypos){\vector(\run,\rise){\arrowlength}}%
\or \put(\xpos,\ypos){\mvector(\run,\rise)\arrowlength}%
\or \put(\xpos,\ypos){\evector(\run,\rise){\arrowlength}}%
\fi\fi\fi
}}

\def\putsplitvector(#1,#2)#3#4{%%
\xpos #1
\ypos #2
\arrowtype #4
\halflength #3
\arrowlength #3
\gap 140
\advance \halflength by-\gap
\divide \halflength by2
\ifnum\arrowtype>0
   \ifcase \arrowtype
   \or \put(\xpos,\ypos){\line(0,-1){\halflength}}%
       \advance\ypos by-\halflength
       \advance\ypos by-\gap
       \put(\xpos,\ypos){\vector(0,-1){\halflength}}%
   \or \put(\xpos,\ypos){\line(0,-1)\halflength}%
       \put(\xpos,\ypos){\vector(0,-1)3}%
       \advance\ypos by-\halflength
       \advance\ypos by-\gap
       \put(\xpos,\ypos){\vector(0,-1){\halflength}}%
   \or \put(\xpos,\ypos){\line(0,-1)\halflength}%
       \advance\ypos by-\halflength
       \advance\ypos by-\gap
       \put(\xpos,\ypos){\evector(0,-1){\halflength}}%
   \fi
\else \arrowtype=-\arrowtype
   \ifcase\arrowtype
   \or \advance \ypos by-\arrowlength
       \put(\xpos,\ypos){\line(0,1){\halflength}}%
       \advance\ypos by\halflength
       \advance\ypos by\gap
       \put(\xpos,\ypos){\vector(0,1){\halflength}}%
   \or \advance \ypos by-\arrowlength
       \put(\xpos,\ypos){\line(0,1)\halflength}%
       \put(\xpos,\ypos){\vector(0,1)3}%
       \advance\ypos by\halflength
       \advance\ypos by\gap
       \put(\xpos,\ypos){\vector(0,1){\halflength}}%
   \or \advance \ypos by-\arrowlength
       \put(\xpos,\ypos){\line(0,1)\halflength}%
       \advance\ypos by\halflength
       \advance\ypos by\gap
       \put(\xpos,\ypos){\evector(0,1){\halflength}}%
   \fi
\fi
}

\def\putmorphism(#1)(#2,#3)[#4`#5`#6]#7#8#9{{%
\run #2
\rise #3
\ifnum\rise=0
  \puthmorphism(#1)[#4`#5`#6]{#7}{#8}#9%
\else\ifnum\run=0
  \putvmorphism(#1)[#4`#5`#6]{#7}{#8}#9%
\else
\setpos(#1)%
\arrowlength #7
\arrowtype #8
\ifnum\run=0
\else\ifnum\rise=0
\else
\ifnum\run>0
    \coefa=1
\else
   \coefa=-1
\fi
\ifnum\arrowtype>0
   \coefb=0
   \coefc=-1
\else
   \coefb=\coefa
   \coefc=1
   \arrowtype=-\arrowtype
\fi
\width=2
\multiply \width by\run
\divide \width by\rise
\ifnum \width<0  \width=-\width\fi
\advance\width by60
\if l#9 \width=-\width\fi
\putbox(\xpos,\ypos){#4}%            %node 1
{\multiply \coefa by\arrowlength%      %node 2
\advance\xpos by\coefa
\multiply \coefa by\rise
\divide \coefa by\run
\advance \ypos by\coefa
\putbox(\xpos,\ypos){#5} }%
{\multiply \coefa by\arrowlength%      %label
\divide \coefa by2
\advance \xpos by\coefa
\advance \xpos by\width
\multiply \coefa by\rise
\divide \coefa by\run
\advance \ypos by\coefa
\if l#9%
   \putrbox(\xpos,\ypos){#6}%
\else\if r#9%
   \putlbox(\xpos,\ypos){#6}%
\fi\fi }%
{\multiply \rise by-\coefc%             %arrow
\multiply \run by-\coefc
\multiply \coefb by\arrowlength
\advance \xpos by\coefb
\multiply \coefb by\rise
\divide \coefb by\run
\advance \ypos by\coefb
\multiply \coefc by70
\advance \ypos by\coefc
\multiply \coefc by\run
\divide \coefc by\rise
\advance \xpos by\coefc
\multiply \coefa by140
\multiply \coefa by\run
\divide \coefa by\rise
\advance \arrowlength by\coefa
\ifcase\arrowtype
\or \put(\xpos,\ypos){\vector(\run,\rise){\arrowlength}}%
\or \put(\xpos,\ypos){\mvector(\run,\rise){\arrowlength}}%
\or \put(\xpos,\ypos){\evector(\run,\rise){\arrowlength}}%
\fi}\fi\fi\fi\fi}}

\newcount\numbdashes \newcount\lengthdash \newcount\increment

\def\howmanydashes{% Actually returns both number and length
\numbdashes=\arrowlength \lengthdash=40
\divide\numbdashes by \lengthdash
\lengthdash=\arrowlength
\divide\lengthdash by \numbdashes
\increment=\lengthdash
\multiply\lengthdash by 3
\divide\lengthdash by 5
}

\def\putdashvector(#1)(#2,#3)#4#5{%
\ifnum#3=0 \putdashhvector(#1){#4}#5
\else
\ifnum#2=0
\putdashvvector(#1){#4}#5\fi\fi}

\def\putdashhvector(#1,#2)#3#4{{%
\arrowlength=#3 \howmanydashes
\multiput(#1,#2)(\increment,0){\numbdashes}%
{\vrule height .4pt width \lengthdash\unitlength}
\arrowtype=#4 \xpos=#1
\ifnum\arrowtype<0 \advance\arrowtype by 7 \fi
\ifcase\arrowtype
\or \advance\xpos by 10
    \put(\xpos,#2){\vector(-1,0){\lengthdash}}
    \advance\xpos by 40
    \put(\xpos,#2){\vector(-1,0){\lengthdash}}
\or \advance \xpos by 10
    \put(\xpos,#2){\vector(-1,0){\lengthdash}}
    \advance\xpos by  \arrowlength
    \advance\xpos by  -50
    \put(\xpos,#2){\vector(-1,0){\lengthdash}}
\or \advance\xpos by 10
    \put(\xpos,#2){\vector(-1,0){\lengthdash}}
\or \advance\xpos by \arrowlength
    \advance\xpos by -\lengthdash
    \put(\xpos,#2){\vector(1,0){\lengthdash}}
\or {\advance\xpos by 10
    \put(\xpos,#2){\vector(1,0){\lengthdash}}}
    \advance\xpos by \arrowlength
    \advance\xpos by -\lengthdash
    \put(\xpos,#2){\vector(1,0){\lengthdash}}
\or \advance\xpos by \arrowlength
    \advance\xpos by -\lengthdash
    \put(\xpos,#2){\vector(1,0){\lengthdash}}
    \advance\xpos by -40
    \put(\xpos,#2){\vector(1,0){\lengthdash}}
   \fi
}}

\def\putdashvvector(#1,#2)#3#4{{%
\arrowlength=#3 \howmanydashes
\ypos=#2 \advance\ypos by -\arrowlength
\multiput(#1,#2)(0,\increment){\numbdashes}%
    {\vrule width .4pt height \lengthdash\unitlength}
\arrowtype=#4 \ypos=#2
\ifnum\arrowtype<0 \advance\arrowtype by 7 \fi
\ifcase\arrowtype
\or \advance\ypos by \arrowlength \advance\ypos by -40
    \put(#1,\ypos){\vector(0,1){\lengthdash}}
    \advance\ypos by -40
    \put(#1,\ypos){\vector(0,1){\lengthdash}}
\or \advance\ypos by 10
    \put(#1,\ypos){\vector(0,1){\lengthdash}}
    \advance\ypos by \arrowlength \advance\ypos by -40
    \put(#1,\ypos){\vector(0,1){\lengthdash}}
\or \advance\ypos by \arrowlength \advance\ypos by -40
    \put(#1,\ypos){\vector(0,1){\lengthdash}}
\or \advance\ypos by 10
    \put(#1,\ypos){\vector(0,-1){\lengthdash}}
\or \advance\ypos by 10
    \put(#1,\ypos){\vector(0,-1){\lengthdash}}
    \advance\ypos by \arrowlength \advance\ypos by -40
    \put(#1,\ypos){\vector(0,-1){\lengthdash}}
\or \advance\ypos by 10
    \put(#1,\ypos){\vector(0,-1){\lengthdash}}
    \advance\ypos by 40
    \put(#1,\ypos){\vector(0,-1){\lengthdash}}
\fi
}}

\def\puthmorphism(#1,#2)[#3`#4`#5]#6#7#8{{%
\xpos #1
\ypos #2
\width #6
\arrowlength #6
\arrowtype=#7
\putbox(\xpos,\ypos){#3\vphantom{#4}}%
{\advance \xpos by\arrowlength
\putbox(\xpos,\ypos){\vphantom{#3}#4}}%
\horsize{\tempcounta}{#3}%
\horsize{\tempcountb}{#4}%
\divide \tempcounta by2
\divide \tempcountb by2
\advance \tempcounta by30
\advance \tempcountb by30
\advance \xpos by\tempcounta
\advance \arrowlength by-\tempcounta
\advance \arrowlength by-\tempcountb
\putvector(\xpos,\ypos)(1,0)\arrowlength\arrowtype
\divide \arrowlength by2
\advance \xpos by\arrowlength
\vertsize{\tempcounta}{#5}%
\divide\tempcounta by2
\advance \tempcounta by20
\if a#8 %
   \advance \ypos by\tempcounta
   \putbox(\xpos,\ypos){#5}%
\else
   \advance \ypos by-\tempcounta
   \putbox(\xpos,\ypos){#5}%
\fi}}

\def\putvmorphism(#1,#2)[#3`#4`#5]#6#7#8{{%
\xpos #1
\ypos #2
\arrowlength #6
\arrowtype #7
\settowidth{\xlen}{$#5$}%
\putbox(\xpos,\ypos){#3}%
{\advance \ypos by-\arrowlength
\putbox(\xpos,\ypos){#4}}%
{\advance\arrowlength by-140
\advance \ypos by-70
\ifdim\xlen>0pt
   \if m#8%
      \putsplitvector(\xpos,\ypos)\arrowlength\arrowtype
   \else
   \putvector(\xpos,\ypos)(0,-1)\arrowlength\arrowtype
   \fi
\else
   \putvector(\xpos,\ypos)(0,-1)\arrowlength\arrowtype
\fi}%
\ifdim\xlen>0pt
   \divide \arrowlength by2
   \advance\ypos by-\arrowlength
   \if l#8%
      \advance \xpos by-40
      \putrbox(\xpos,\ypos){#5}%
   \else\if r#8%
      \advance \xpos by40
      \putlbox(\xpos,\ypos){#5}%
   \else
      \putbox(\xpos,\ypos){#5}%
   \fi\fi
\fi
}}

\def\putsquarep<#1>(#2)[#3;#4`#5`#6`#7]{{%
\setsqparms[#1]%
\setpos(#2)%
\settokens[#3]%
\puthmorphism(\xpos,\ypos)[\tokenc`\tokend`{#7}]{\width}{\arrowtyped}b%
\advance\ypos by \height
\puthmorphism(\xpos,\ypos)[\tokena`\tokenb`{#4}]{\width}{\arrowtypea}a%
\putvmorphism(\xpos,\ypos)[``{#5}]{\height}{\arrowtypeb}l%
\advance\xpos by \width
\putvmorphism(\xpos,\ypos)[``{#6}]{\height}{\arrowtypec}r%
}}

\def\putsquare{\@ifnextchar <{\putsquarep}{\putsquarep%
   <\arrowtypea`\arrowtypeb`\arrowtypec`\arrowtyped;\width`\height>}}
\def\square{\@ifnextchar< {\squarep}{\squarep
   <\arrowtypea`\arrowtypeb`\arrowtypec`\arrowtyped;\width`\height>}}
\def\squarep<#1>[#2`#3`#4`#5;#6`#7`#8`#9]{{%       %     #2------>#3
\setsqparms[#1]%                                   %      |       |
\diagram%                                          %      |       |
\putsquarep<\arrowtypea`\arrowtypeb`\arrowtypec`%  %    #7|       |#8
\arrowtyped;\width`\height>%                       %      |       |
(0,0)[#2`#3`#4`{#5};#6`#7`#8`{#9}]%                %      |       |
\enddiagram%                                       %      v       v
}}                                                 %     #4------>#5
\def\putptrianglep<#1>(#2,#3)[#4`#5`#6;#7`#8`#9]{{%
\settriparms[#1]%
\xpos=#2 \ypos=#3
\advance\ypos by \height
\puthmorphism(\xpos,\ypos)[#4`#5`{#7}]{\height}{\arrowtypea}a%
\putvmorphism(\xpos,\ypos)[`#6`{#8}]{\height}{\arrowtypeb}l%
\advance\xpos by\height
\putmorphism(\xpos,\ypos)(-1,-1)[``{#9}]{\height}{\arrowtypec}r%
}}

\def\putptriangle{\@ifnextchar <{\putptrianglep}{\putptrianglep
   <\arrowtypea`\arrowtypeb`\arrowtypec;\height>}}
\def\ptriangle{\@ifnextchar <{\ptrianglep}{\ptrianglep
   <\arrowtypea`\arrowtypeb`\arrowtypec;\height>}}
\def\ptrianglep<#1>[#2`#3`#4;#5`#6`#7]{{%%    %      #2----->#3
\settriparms[#1]%                             %      |      /
\diagram%                                     %      |     /
\putptrianglep<\arrowtypea`\arrowtypeb`%      %    #6|    /#7
\arrowtypec;\height>%                         %      |   /
(0,0)[#2`#3`#4;#5`#6`{#7}]%                   %      |  /
\enddiagram%%                                 %      v v
}}                                            %      #4

\def\putqtrianglep<#1>(#2,#3)[#4`#5`#6;#7`#8`#9]{{%
\settriparms[#1]%
\xpos=#2 \ypos=#3
\advance\ypos by\height
\puthmorphism(\xpos,\ypos)[#4`#5`{#7}]{\height}{\arrowtypea}a%
\putmorphism(\xpos,\ypos)(1,-1)[``{#8}]{\height}{\arrowtypeb}l%
\advance\xpos by\height
\putvmorphism(\xpos,\ypos)[`#6`{#9}]{\height}{\arrowtypec}r%
}}

\def\putqtriangle{\@ifnextchar <{\putqtrianglep}{\putqtrianglep
   <\arrowtypea`\arrowtypeb`\arrowtypec;\height>}}
\def\qtriangle{\@ifnextchar <{\qtrianglep}{\qtrianglep
   <\arrowtypea`\arrowtypeb`\arrowtypec;\height>}}
\def\qtrianglep<#1>[#2`#3`#4;#5`#6`#7]{{%%    %        #2----->#3
\settriparms[#1]%                             %         \      |
\width=\height                                %          \     |
\diagram%                                     %         #6\    |#7
\putqtrianglep<\arrowtypea`\arrowtypeb`%      %            \   |
\arrowtypec;\height>%                         %             \  |
(0,0)[#2`#3`#4;#5`#6`{#7}]%                   %              v v
\enddiagram%%                                 %               #4
}}

\def\putdtrianglep<#1>(#2,#3)[#4`#5`#6;#7`#8`#9]{{%
\settriparms[#1]%
\xpos=#2 \ypos=#3
\puthmorphism(\xpos,\ypos)[#5`#6`{#9}]{\height}{\arrowtypec}b%
\advance\xpos by \height \advance\ypos by\height
\putmorphism(\xpos,\ypos)(-1,-1)[``{#7}]{\height}{\arrowtypea}l%
\putvmorphism(\xpos,\ypos)[#4``{#8}]{\height}{\arrowtypeb}r%
}}

\def\putdtriangle{\@ifnextchar <{\putdtrianglep}{\putdtrianglep
   <\arrowtypea`\arrowtypeb`\arrowtypec;\height>}}
\def\dtriangle{\@ifnextchar <{\dtrianglep}{\dtrianglep
   <\arrowtypea`\arrowtypeb`\arrowtypec;\height>}}
\def\dtrianglep<#1>[#2`#3`#4;#5`#6`#7]{{%%    %                  / |
\settriparms[#1]%                             %                 /  |
\width=\height                                %              #5/   |#6
\diagram%                                     %               /    |
\putdtrianglep<\arrowtypea`\arrowtypeb`%      %              /     |
\arrowtypec;\height>%                         %             v      v
(0,0)[#2`#3`#4;#5`#6`{#7}]%                   %            #3----->#4
\enddiagram%%                                 %                #7
}}

\def\putbtrianglep<#1>(#2,#3)[#4`#5`#6;#7`#8`#9]{{%
\settriparms[#1]%
\xpos=#2 \ypos=#3
\puthmorphism(\xpos,\ypos)[#5`#6`{#9}]{\height}{\arrowtypec}b%
\advance\ypos by\height
\putmorphism(\xpos,\ypos)(1,-1)[``{#8}]{\height}{\arrowtypeb}r%
\putvmorphism(\xpos,\ypos)[#4``{#7}]{\height}{\arrowtypea}l%
}}

\def\putbtriangle{\@ifnextchar <{\putbtrianglep}{\putbtrianglep
   <\arrowtypea`\arrowtypeb`\arrowtypec;\height>}}
\def\btriangle{\@ifnextchar <{\btrianglep}{\btrianglep
   <\arrowtypea`\arrowtypeb`\arrowtypec;\height>}}
\def\btrianglep<#1>[#2`#3`#4;#5`#6`#7]{{%%   %              | \
\settriparms[#1]%                            %              |  \
\width=\height                               %            #5|   \#6
\diagram%                                    %              |    \
\putbtrianglep<\arrowtypea`\arrowtypeb`%     %              |     \
\arrowtypec;\height>%                        %              v      v
(0,0)[#2`#3`#4;#5`#6`{#7}]%                  %              #3----->#4
\enddiagram%%                                %                 #7
}}

\def\putAtrianglep<#1>(#2,#3)[#4`#5`#6;#7`#8`#9]{{%
\settriparms[#1]%
\xpos=#2 \ypos=#3
{\multiply \height by2
\puthmorphism(\xpos,\ypos)[#5`#6`{#9}]{\height}{\arrowtypec}b}%
\advance\xpos by\height \advance\ypos by\height
\putmorphism(\xpos,\ypos)(-1,-1)[#4``{#7}]{\height}{\arrowtypea}l%
\putmorphism(\xpos,\ypos)(1,-1)[``{#8}]{\height}{\arrowtypeb}r%
}}

\def\putAtriangle{\@ifnextchar <{\putAtrianglep}{\putAtrianglep
   <\arrowtypea`\arrowtypeb`\arrowtypec;\height>}}
\def\Atriangle{\@ifnextchar <{\Atrianglep}{\Atrianglep
   <\arrowtypea`\arrowtypeb`\arrowtypec;\height>}}
\def\Atrianglep<#1>[#2`#3`#4;#5`#6`#7]{{%%         %         /   \
\settriparms[#1]%                                  %        /     \
\width=\height                                     %     #5/       \#6
\diagram%                                          %      /         \
\putAtrianglep<\arrowtypea`\arrowtypeb`%           %     /           \
\arrowtypec;\height>%                              %    v             v
(0,0)[#2`#3`#4;#5`#6`{#7}]%                        %   #3------------>#4
\enddiagram%%                                      %          #7
}}

\def\putAtrianglepairp<#1>(#2)[#3;#4`#5`#6`#7`#8]{{%
\settripairparms[#1]%
\setpos(#2)%
\settokens[#3]%
\puthmorphism(\xpos,\ypos)[\tokenb`\tokenc`{#7}]{\height}{\arrowtyped}b%
\advance\xpos by\height
\puthmorphism(\xpos,\ypos)[\phantom{\tokenc}`\tokend`{#8}]%
{\height}{\arrowtypee}b%
\advance\ypos by\height
\putmorphism(\xpos,\ypos)(-1,-1)[\tokena``{#4}]{\height}{\arrowtypea}l%
\putvmorphism(\xpos,\ypos)[``{#5}]{\height}{\arrowtypeb}m%
\putmorphism(\xpos,\ypos)(1,-1)[``{#6}]{\height}{\arrowtypec}r%
}}

\def\putAtrianglepair{\@ifnextchar <{\putAtrianglepairp}{\putAtrianglepairp%
   <\arrowtypea`\arrowtypeb`\arrowtypec`\arrowtyped`\arrowtypee;\height>}}
\def\Atrianglepair{\@ifnextchar <{\Atrianglepairp}{\Atrianglepairp%
   <\arrowtypea`\arrowtypeb`\arrowtypec`\arrowtyped`\arrowtypee;\height>}}

\def\Atrianglepairp<#1>[#2;#3`#4`#5`#6`#7]{{%           %  #2a
\settripairparms[#1]%                         %           / | \
\settokens[#2]%                               %          /  |  \
\width=\height                                %       #3/  #4   \#5
\diagram%                                     %        /    |    \
\putAtrianglepairp                            %       /     |     \
<\arrowtypea`\arrowtypeb`\arrowtypec`%        %      v      v      v
\arrowtyped`\arrowtypee;\height>%             %     #2b---->#2c---->#2d
(0,0)[{#2};#3`#4`#5`#6`{#7}]%                 %         #6     #7
\enddiagram%%
}}

\def\putVtrianglep<#1>(#2,#3)[#4`#5`#6;#7`#8`#9]{{%
\settriparms[#1]%
\xpos=#2 \ypos=#3
\advance\ypos by\height
{\multiply\height by2
\puthmorphism(\xpos,\ypos)[#4`#5`{#7}]{\height}{\arrowtypea}a}%
\putmorphism(\xpos,\ypos)(1,-1)[`#6`{#8}]{\height}{\arrowtypeb}l%
\advance\xpos by\height
\advance\xpos by\height
\putmorphism(\xpos,\ypos)(-1,-1)[``{#9}]{\height}{\arrowtypec}r%
}}

\def\putVtriangle{\@ifnextchar <{\putVtrianglep}{\putVtrianglep
   <\arrowtypea`\arrowtypeb`\arrowtypec;\height>}}
\def\Vtriangle{\@ifnextchar <{\Vtrianglep}{\Vtrianglep
   <\arrowtypea`\arrowtypeb`\arrowtypec;\height>}}
\def\Vtrianglep<#1>[#2`#3`#4;#5`#6`#7]{{%%     %        #2------------->#3
\settriparms[#1]%                              %         \             /
\width=\height                                 %          \           /
\diagram%                                      %         #6\         /#7
\putVtrianglep<\arrowtypea`\arrowtypeb`%       %            \       /
\arrowtypec;\height>%                          %             \     /
(0,0)[#2`#3`#4;#5`#6`{#7}]%                    %              v   v
\enddiagram%%                                  %               #4
}}

\def\putVtrianglepairp<#1>(#2)[#3;#4`#5`#6`#7`#8]{{
\settripairparms[#1]%
\setpos(#2)%
\settokens[#3]%
\advance\ypos by\height
\putmorphism(\xpos,\ypos)(1,-1)[`\tokend`{#6}]{\height}{\arrowtypec}l%
\puthmorphism(\xpos,\ypos)[\tokena`\tokenb`{#4}]{\height}{\arrowtypea}a%
\advance\xpos by\height
\puthmorphism(\xpos,\ypos)[\phantom{\tokenb}`\tokenc`{#5}]%
{\height}{\arrowtypeb}a%
\putvmorphism(\xpos,\ypos)[``{#7}]{\height}{\arrowtyped}m%
\advance\xpos by\height
\putmorphism(\xpos,\ypos)(-1,-1)[``{#8}]{\height}{\arrowtypee}r%
}}

\def\putVtrianglepair{\@ifnextchar <{\putVtrianglepairp}{\putVtrianglepairp%
    <\arrowtypea`\arrowtypeb`\arrowtypec`\arrowtyped`\arrowtypee;\height>}}
\def\Vtrianglepair{\@ifnextchar <{\Vtrianglepairp}{\Vtrianglepairp%
    <\arrowtypea`\arrowtypeb`\arrowtypec`\arrowtyped`\arrowtypee;\height>}}
\def\Vtrianglepairp<#1>[#2;#3`#4`#5`#6`#7]{{%  %  #2a---->#2b---->#2c
\settripairparms[#1]%                          %   \      |      /
\settokens[#2]%                                %    \     |     /
\diagram%                                      %   #5\   #6    /#7
\putVtrianglepairp                             %      \   |   /
<\arrowtypea`\arrowtypeb`\arrowtypec`%         %       \  |  /
\arrowtyped`\arrowtypee;\height>%              %        v v v
(0,0)[{#2};#3`#4`#5`#6`{#7}]%                  %         #2d
\enddiagram%%
}}

\def\putCtrianglep<#1>(#2,#3)[#4`#5`#6;#7`#8`#9]{{%
\settriparms[#1]%
\xpos=#2 \ypos=#3
\advance\ypos by\height
\putmorphism(\xpos,\ypos)(1,-1)[``{#9}]{\height}{\arrowtypec}l%
\advance\xpos by\height
\advance\ypos by\height
\putmorphism(\xpos,\ypos)(-1,-1)[#4`#5`{#7}]{\height}{\arrowtypea}l%
{\multiply\height by 2
\putvmorphism(\xpos,\ypos)[`#6`{#8}]{\height}{\arrowtypeb}r}%
}}

\def\putCtriangle{\@ifnextchar <{\putCtrianglep}{\putCtrianglep
    <\arrowtypea`\arrowtypeb`\arrowtypec;\height>}}
\def\Ctriangle{\@ifnextchar <{\Ctrianglep}{\Ctrianglep
    <\arrowtypea`\arrowtypeb`\arrowtypec;\height>}}
\def\Ctrianglep<#1>[#2`#3`#4;#5`#6`#7]{{%%   %                / |
\settriparms[#1]%                            %             #5/  |
\width=\height                               %              /   |
\diagram%                                    %             v    |
\putCtrianglep<\arrowtypea`\arrowtypeb`%     %           #3     |#6
\arrowtypec;\height>%                        %             \    |
(0,0)[#2`#3`#4;#5`#6`{#7}]%                  %            #7\   |
\enddiagram%%                                %               \  |
}}                                           %                v v
\def\putDtrianglep<#1>(#2,#3)[#4`#5`#6;#7`#8`#9]{{%
\settriparms[#1]%
\xpos=#2 \ypos=#3
\advance\xpos by\height \advance\ypos by\height
\putmorphism(\xpos,\ypos)(-1,-1)[``{#9}]{\height}{\arrowtypec}r%
\advance\xpos by-\height \advance\ypos by\height
\putmorphism(\xpos,\ypos)(1,-1)[`#5`{#8}]{\height}{\arrowtypeb}r%
{\multiply\height by 2
\putvmorphism(\xpos,\ypos)[#4`#6`{#7}]{\height}{\arrowtypea}l}%
}}

\def\putDtriangle{\@ifnextchar <{\putDtrianglep}{\putDtrianglep
    <\arrowtypea`\arrowtypeb`\arrowtypec;\height>}}
\def\Dtriangle{\@ifnextchar <{\Dtrianglep}{\Dtrianglep
   <\arrowtypea`\arrowtypeb`\arrowtypec;\height>}}
\def\Dtrianglep<#1>[#2`#3`#4;#5`#6`#7]{{%%  %          | \
\settriparms[#1]%                           %          |  \#6
\width=\height                              %          |   \
\diagram%                                   %          |    v
\putDtrianglep<\arrowtypea`\arrowtypeb`%    %        #5|    #3
\arrowtypec;\height>%                       %          |    /
(0,0)[#2`#3`#4;#5`#6`{#7}]%                 %          |   /#7
\enddiagram%%                               %          |  /
}}                                          %          v v
\def\setrecparms[#1`#2]{\width=#1 \height=#2}%
\def\recursep<#1`#2>[#3;#4`#5`#6`#7`#8]{{%
\width=#1 \height=#2
\settokens[#3]
\settowidth{\tempdimen}{$\tokena$}
\ifdim\tempdimen=0pt
  \savebox{\tempboxa}{\hbox{$\tokenb$}}%
  \savebox{\tempboxb}{\hbox{$\tokend$}}%
  \savebox{\tempboxc}{\hbox{$#6$}}%
\else
  \savebox{\tempboxa}{\hbox{$\hbox{$\tokena$}\times\hbox{$\tokenb$}$}}%
  \savebox{\tempboxb}{\hbox{$\hbox{$\tokena$}\times\hbox{$\tokend$}$}}%
  \savebox{\tempboxc}{\hbox{$\hbox{$\tokena$}\times\hbox{$#6$}$}}%
\fi
\ypos=\height
\divide\ypos by 2
\xpos=\ypos
\advance\xpos by \width
\bfig
\putCtrianglep<-1`1`1;\ypos>(0,0)[`\tokenc`;#5`#6`{#7}]%
\puthmorphism(\ypos,0)[\tokend`\usebox{\tempboxb}`{#8}]{\width}{-1}b%
\puthmorphism(\ypos,\height)[\tokenb`\usebox{\tempboxa}`{#4}]{\width}{-1}a%
\advance\ypos by \width
\putvmorphism(\ypos,\height)[``\usebox{\tempboxc}]{\height}1r%
\efig
}}

\def\recurse{\@ifnextchar <{\recursep}{\recursep<\width`\height>}}

\def\puttwohmorphisms(#1,#2)[#3`#4;#5`#6]#7#8#9{{%
\puthmorphism(#1,#2)[#3`#4`]{#7}0a
\ypos=#2
\advance\ypos by 20
\puthmorphism(#1,\ypos)[\phantom{#3}`\phantom{#4}`#5]{#7}{#8}a
\advance\ypos by -40
\puthmorphism(#1,\ypos)[\phantom{#3}`\phantom{#4}`#6]{#7}{#9}b
}}

\def\puttwovmorphisms(#1,#2)[#3`#4;#5`#6]#7#8#9{{%
\putvmorphism(#1,#2)[#3`#4`]{#7}0a
\xpos=#1
\advance\xpos by -20
\putvmorphism(\xpos,#2)[\phantom{#3}`\phantom{#4}`#5]{#7}{#8}l
\advance\xpos by 40
\putvmorphism(\xpos,#2)[\phantom{#3}`\phantom{#4}`#6]{#7}{#9}r
}}

\def\puthcoequalizer(#1)[#2`#3`#4;#5`#6`#7]#8#9{{%
\setpos(#1)%
\puttwohmorphisms(\xpos,\ypos)[#2`#3;#5`#6]{#8}11%
\advance\xpos by #8
\puthmorphism(\xpos,\ypos)[\phantom{#3}`#4`#7]{#8}1{#9}
}}

\def\putvcoequalizer(#1)[#2`#3`#4;#5`#6`#7]#8#9{{%
\setpos(#1)%
\puttwovmorphisms(\xpos,\ypos)[#2`#3;#5`#6]{#8}11%
\advance\ypos by -#8
\putvmorphism(\xpos,\ypos)[\phantom{#3}`#4`#7]{#8}1{#9}
}}

\def\putthreehmorphisms(#1)[#2`#3;#4`#5`#6]#7(#8)#9{{%
\setpos(#1) \settypes(#8)
\if a#9 %
     \vertsize{\tempcounta}{#5}%
     \vertsize{\tempcountb}{#6}%
     \ifnum \tempcounta<\tempcountb \tempcounta=\tempcountb \fi
\else
     \vertsize{\tempcounta}{#4}%
     \vertsize{\tempcountb}{#5}%
     \ifnum \tempcounta<\tempcountb \tempcounta=\tempcountb \fi
\fi
\advance \tempcounta by 60
\puthmorphism(\xpos,\ypos)[#2`#3`#5]{#7}{\arrowtypeb}{#9}
\advance\ypos by \tempcounta
\puthmorphism(\xpos,\ypos)[\phantom{#2}`\phantom{#3}`#4]{#7}{\arrowtypea}{#9}
\advance\ypos by -\tempcounta \advance\ypos by -\tempcounta
\puthmorphism(\xpos,\ypos)[\phantom{#2}`\phantom{#3}`#6]{#7}{\arrowtypec}{#9}
}}

\def\setarrowtoks[#1`#2`#3`#4`#5`#6]{%
\def\toka{#1}
\def\tokb{#2}
\def\tokc{#3}
\def\tokd{#4}
\def\toke{#5}
\def\tokf{#6}
}
\def\hex{\@ifnextchar <{\hexp}{\hexp<1000`400>}}
\def\hexp<#1`#2>[#3`#4`#5`#6`#7`#8;#9]{%
\setarrowtoks[#9]
\yext=#2 \advance \yext by #2
\xext=#1 \advance\xext by \yext
\bfig
\putCtriangle<-1`0`1;#2>(0,0)[`#5`;\tokb``\tokd]
\xext=#1 \yext=#2 \advance \yext by #2
\putsquare<1`0`0`1;\xext`\yext>(#2,0)[#3`#4`#7`#8;\toka```\tokf]
\advance \xext by #2
\putDtriangle<0`1`-1;#2>(\xext,0)[`#6`;`\tokc`\toke]
\efig
}
\def\sAA{{\rm A\kern-0.85em A}} % simple version
\def\tAA{{\mathchoice
  {\sAA}
  {\sAA}
  {\rm A\kern-0.60em A}
  {\rm A\kern-0.50em A} }}
\def\sBB{{\rm I\kern-.17em{}B}}
\def\BB{{\mathchoice
  {\sBB}
  {\sBB}
  {\rm I\kern-.13em{}B}
  {\rm I\kern-.13em{}B} }}
\def\sCC{{\kern 0.27em\vrule height1.45ex width0.03em depth0em
          \kern-0.30em\rm C}}
\def\CC{{\mathchoice
  {\sCC}
  {\sCC}
  {\kern 0.225em \vrule height1.05ex width0.025em depth0em \kern-0.25em \rm C}
  {\kern 0.180em \vrule height0.78ex width0.02em depth0em \kern-0.2em \rm C}
        }}
\def\tCC{{\ooalign{C\crcr\kern0.27em\vrule height1.45ex width0.03em
depth0em\crcr}}}
\def\sDD{{\rm I\kern-.16em{}D}}
\def\DD{{\mathchoice
  {\sDD}
  {\sDD}
  {\rm I\kern-.13em{}D}
  {\rm I\kern-.13em{}D} }}
\def\sEE{{\rm I\kern-.17em{}E}}
\def\EE{{\mathchoice
  {\sEE}
  {\sEE}
  {\rm I\kern-.13em{}E}
  {\rm I\kern-.13em{}E} }}
\def\sFF{{\rm I\kern-.16em{}F}}
\def\FF{{\mathchoice
  {\sFF}
  {\sFF}
  {\rm I\kern-.13em{}F}
  {\rm I\kern-.13em{}F} }}
\def\sGG{{\kern 0.27em \vrule height1.45ex width0.03em depth0em
          \kern-0.30em \rm G}}
\def\GG{{\mathchoice
  {\sGG}
  {\sGG}
  {\kern 0.225em \vrule height1.05ex width0.025em depth0em \kern-0.25em \rm G}
  {\kern 0.180em \vrule height0.78ex width0.020em depth0em \kern-0.20em \rm G}
        }}
\def\sHH{{\rm I\kern-.16em{}H}}
\def\HH{{\mathchoice
  {\sHH}
  {\sHH}
  {\rm I\kern-.13em{}H}
  {\rm I\kern-.13em{}H} }}
\def\sII{{\rm I\kern-.16em{}I}}
\def\II{{\mathchoice
  {\sII}
  {\sII}
  {\rm I\kern-.12em{}I}
  {\rm I\kern-.10em{}I} }}
\def\sJJ{{\kern0.17em\vrule height1.5ex width 0.03em depth0em
          \kern-.20em\rm J}}
\def\JJ{{\mathchoice
  {\sJJ}
  {\sJJ}
  {\kern0.150em\vrule height1.05ex width 0.025em depth0em\kern-.175em\rm J}
  {\kern0.135em\vrule height0.78ex width 0.020em depth0em\kern-.155em\rm J} }}
\def\sKK{{\rm I\kern-.16em{}K}}
\def\KK{{\mathchoice
  {\sKK}
  {\sKK}
  {\rm I\kern-.12em{}K}
  {\rm I\kern-.10em{}K} }}
\def\sLL{{\rm I\kern-.16em{}L}}
\def\LL{{\mathchoice
  {\sLL}
  {\sLL}
  {\rm I\kern-.12em{}L}
  {\rm I\kern-.10em{}L} }}
\def\sMM{{\rm I\kern-.16em{}M}}
\def\MM{{\mathchoice
  {\sMM}
  {\sMM}
  {\rm I\kern-.12em{}M}
  {\rm I\kern-.10em{}M} }}
\def\sNN{{\rm I\kern-.16em{}N}}
\def\NN{{\mathchoice
  {\sNN}
  {\sNN}
  {\rm I\kern-.12em{}N}
  {\rm I\kern-.10em{}N} }}
\def\sOO{{\kern 0.27em \vrule height1.50ex width0.03em depth0em
					\kern-0.30em \rm O}}
\def\OO{{\mathchoice
  {\sOO}
  {\sOO}
  {\kern 0.225em \vrule height1.05ex width0.025em depth0em \kern-0.25em \rm O}
  {\kern 0.180em \vrule height0.78ex width0.020em depth0em \kern-0.20em \rm O}
        }}
\def\sPP{{\rm I\kern-.16em{}P}}
\def\PP{{\mathchoice
  {\sPP}
  {\sPP}
  {\rm I\kern-.12em{}P}
  {\rm I\kern-.10em{}P} }}
\def\sQQ{{\kern 0.27em \vrule height1.45ex width0.03em depth0em
          \kern-0.30em \rm Q}}
\def\QQ{{\mathchoice
	{\sQQ}
	{\sQQ}
  {\kern 0.225em \vrule height1.05ex width0.025em depth0em \kern-0.25em \rm Q}
  {\kern 0.180em \vrule height0.78ex width0.020em depth0em \kern-0.20em \rm Q}
        }}
\def\sRR{{\rm I\kern-0.16em{}R}}
\def\RR{{\mathchoice
  {\sRR}
  {\sRR}
  {\rm I\kern-0.12em{}R}
  {\rm I\kern-0.10em{}R} }}
\def\sSS{{\rm S\kern-.45em{}S}}
\def\sTT{{\rm T\kern-.60em{}T}}
\def\TT{{\mathchoice
  {\sTT}
  {\sTT}
  {\rm T\kern-.45em{}T}
  {\rm T\kern-.38em{}T} }}
\def\sUU{{\rm U\kern-.60em{}U}}
\def\UU{{\mathchoice
  {\sUU}
  {\sUU}
  {\rm U\kern-.46em{}U}
  {\rm U\kern-.38em{}U} }}
\def\sVV{{\rm V\kern-.62em{}V}}
\def\VV{{\mathchoice
  {\sVV}
  {\sVV}
  {\rm V\kern-.46em{}V}
  {\rm V\kern-.38em{}V} }}
\def\sWW{{\rm W\kern-.92em{}W}}
\def\WW{{\mathchoice
  {\sWW}
  {\sWW}
  {\rm W\kern-.80em{}W}
  {\rm W\kern-.67em{}W} }}
\def\sXX{{\rm X\kern-.58em{}X}}
\def\XX{{\mathchoice
  {\sXX}
  {\sXX}
  {\rm X\kern-.45em{}X}
  {\rm X\kern-.38em{}X} }}
\def\sYY{{\rm Y\kern-.58em{}Y}}
\def\YY{{\mathchoice
  {\sYY}
  {\sYY}
  {\rm Y\kern-.45em{}Y}
  {\rm Y\kern-.40em{}Y} }}
\def\sZZ{{\rm Z\kern-0.32em{}Z}}
\def\ZZ{{\mathchoice
  {\sZZ}
  {\sZZ}
  {\rm Z\kern-0.30em{}Z}
  {\rm Z\kern-0.25em{}Z} }}

\begin{document}
\global\parskip=4pt

%%%%%%%%% title page %%%%%%%%%%%%%%%%%%%%%%%%%%%%%%%%%%%%%%%%

\begin{titlepage}
%\newlength{\titlehead}
%\settowidth{\titlehead}{NIKHEF-H/91xxxxxx}
%\begin{flushright}
%\parbox{\titlehead}{
%\begin{flushleft}
%{\tt hep-th/9501075}\\
%IC/95/5\\
%\end{flushleft}
%}
%\end{flushright}
IC/95/5 \hfill {\tt hep-th/9501075}
\begin{center}
{\LARGE\bf Localization and Diagonalization}\footnote{To appear in the
{\em Journal of Mathematical Physics} Special Issue on Functional Integration
(May 1995)} \\
\vskip .40in
{\Large\bf A Review of Functional Integral Techniques}
\vskip .20in
{\Large\bf for Low-Dimensional Gauge Theories}
\vskip .20in
{\Large\bf and Topological Field Theories}
\vskip .40in
{\bf Matthias Blau}\footnote{e-mail: blau@ictp.trieste.it,
mblau@enslapp.ens-lyon.fr}'~\footnote{Present Address:
Lab.\ Physique Th\'eorique, \'Ecole Normale Sup\'erieure de Lyon, France}
and {\bf George Thompson}\footnote{e-mail: thompson@ictp.trieste.it}\\
\vskip .10in
ICTP\\
P.O. Box 586\\
I-34014 Trieste\\
Italy
\end{center}
\vskip .15in
\begin{abstract}
We review localization techniques for functional integrals which
have recently been used to perform calculations in and gain insight into the
structure of certain topological field theories and low-dimensional
gauge theories. These are the functional integral counterparts of the
Mathai-Quillen formalism, the Duistermaat-Heckman theorem, and the
Weyl integral formula respectively. In each case, we first introduce the
necessary mathematical background (Euler classes of vector bundles,
equivariant cohomology, topology of Lie groups),
and describe the finite dimensional integration formulae.
We then discuss some applications to path integrals and give an overview of
the relevant literature. The applications we deal with
include supersymmetric quantum mechanics, cohomological field theories,
phase space path integrals, and two-dimensional Yang-Mills theory.
\end{abstract}
\end{titlepage}

%%%%%%%%% end of title page %%%%%%%%%%%%%%%%%%%%%%%%%%%%%%

\setcounter{footnote}{0}

\begin{small}
\tableofcontents
\end{small}

\section{Introduction}

In recent years, the functional integral has become a very popular tool
in a branch of physics lying on the interface between string theory,
conformal field theory and topological field theory on the one hand
and topology and algebraic geometry on the other. Not only has it
become popular but it has also, because of the consistent reliability
of the results the functional integral can produce when handled with
due care, acquired a certain degree of respectability among
mathematicians.

Here, however, our focus will not be primarily on the results or
predictions obtained
by these methods, because an appreciation of these results would
require a rather detailed understanding of the mathematics and physics
involved. Rather, we want to explain some of the general features and
properties the functional integrals appearing in this context have
in common. Foremost among these is the fact that, due to a large number of
(super-)symmetries, these functional integrals essentially represent
finite-dimensional integrals.

The transition between the functional and finite dimensional integrals
can then naturally be regarded as a (rather drastic) localization
of the original infinite dimensional integral. The
purpose of this article is to give an introduction to and an
overview of some functional integral tricks and techniques which have turned
out to be useful in understanding these properties and which thus provide
insight into the structure of topological field theories
and some of their close relatives in general.

More specifically, we focus on three techniques which are
extensions to functional integrals of finite dimensional integration
and localization formulae which are quite
interesting in their own right, namely
\begin{enumerate}
\item the \mq\ formalism \cite{mq}, dealing with integral
      representations of Euler classes of vector bundles;
\item the \dh\ theorem \cite{dh} on the exactness of the stationary
      phase approximation for certain phase space path integrals,
      and its generalizations \cite{bv,bgv};
\item the classical Weyl integral formula, relating integrals over
      a compact Lie group or Lie algebra to integrals over a maximal
      torus or a Cartan subalgebra.
\end{enumerate}

We will deal with these three techniques in sections 2-4 of this
article respectively. In each case, we will first try to provide the
necessary mathematical background (Euler classes of vector bundles,
equivariant cohomology, topology of Lie groups). We then explain the
integration formulae in their finite dimensional setting before we
move on boldly to apply these techniques to some specific infinite
dimensional examples like supersymmetric quantum mechanics or
two-dimensional Yang-Mills theory.

Of course, the functional integrals we deal with in this article are very
special, corresponding to theories with no field theoretic degrees
of freedom. Moreover, our treatment of functional integrals is completely
formal as regards its functional analytic aspects.
So lest we lose the reader interested primarily in honest quantum
mechanics or field theory functional integrals at this point, we should
perhaps explain why we believe that it is, nevertheless, worthwile to
look at these examples and techniques.

First of all, precisely because the integrals we deal with are essentially
finite dimensional integrals, and the path integral manipulations
are frequently known to produce the correct (meaning e.g.\ topologically
correct) results, any definition of, or approach to, the functional
integral worth its salt should be able to reproduce these results
and to incorporate these techniques in some way.
This applies in particular to the infinite
dimensional analogue of the \mq\ formalism and to the \wif.

Secondly, the kinds of theories we deal with here allow one to study
kinematical (i.e.\ geometrical and topological) aspects of the path
integral in isolation from their dynamical aspects. In this sense,
these theories are complementary to, say, simple interacting theories.
While the latter are typically kinematically linear but dynamically
non-linear,
the former are usually dynamically linear (free field theories) but
kinematically highly non-linear (and the entire non-triviality of the
theories resides in this kinematic non-linearity). This is a feature
shared by all the three techniques we discuss.

Thirdly, in principle the techniques we review here are also applicable
to theories with field theoretic degrees of freedom, at least in the sense
that they provide alternative approximation techniques to the usual
perturbative expansion. Here we have in mind primarily the \wif\ and
the generalized WKB approximation techniques based on the \dh\ formula.

Finally, we want to draw attention to the fact that many topological field
theories can be interpreted as `twisted' versions of ordinary supersymmetric
field theories, and that this relation has already led to a dramatic
improvement of our understanding of both types of theories, see e.g.\
\cite{bcov}-\cite{cvew} for some recent developments.

Hoping to have persuaded the reader to stay with us, we will now sketch
briefly the organization of this paper. As mentioned above, we will deal
with the three different techniques in three separate sections, each one
of them beginning with an introduction to the mathematical background.
The article is written in such a way that these three sections can be
read independently of each other. We have also tried to set things up
in such a way that the reader primarily interested in the path integral
applications should be able to skip the mathematical introduction
upon first reading and to move directly to the relevant section,
going back to the more formal considerations with a solid example
at hand. Furthermore, ample references are given so that the interested
reader should be able to track down most of the applications of the
techniques that we describe.

Originally, we had intended to include a fifth section explaining
the interrelations among the three seemingly rather different techniques
we discuss in this article. These relations exist. A nice example
to keep in mind is two-dimensional Yang-Mills theory which can be
solved either via a version of the \dh\ formula or using the \wif,
but which is also equivalent to a certain two-dimensional cohomological
field theory which provides a field theoretic realization of the \mq\
formalism. Unfortunately, we haven't been able to come up with
some satisfactory general criteria for when this can occur and
we have therefore just added some isolated remarks on this in
appropriate places in the body of the text.

\section{Localization via the Mathai-Quillen Formalism}

The sort of localization we are interested in in this section is
familiar from classical differential geometry and topology.
Namely, let us recall that classically there exist two quite different
prescriptions for calculating the Euler number $\c(X)$
(number of vertices minus number of edges plus \ldots) of some manifold
$X$. The first is topological in
nature and instructs one to choose a vector field $V$ on $X$ with isolated
zeros and to count these zeros with signs (this is the Hopf theorem).
The second is differential
geometric and represents $\c(X)$ as the integral over $X$ of a density
(top form) $\en$ constructed from the curvature of some connection
$\nabla$ on $X$ (the Gauss-Bonnet theorem).

The integral over $X$ of $\en$ can hence be localized to
(in the sense of: evaluated in terms of contribution from)
the zero locus of $V$. This
localization can be made quite explicit by using a more general
formula for $\c(X)$, obtained by Mathai and Quillen \cite{mq}, which
interpolates between these two classical prescriptions. It relies on
the construction of a form $\env(X)$ which depends on both a vector field
$V$ and a connection $\nabla$. Its integral over $X$
represents $\c(X)$ for all $V$ and $\nabla$ and can be shown to
reduce to an integral over the zero locus $X_{V}$ of $V$ in the form
\be
\c(X)= \int_{X}\env(X) = \int_{X_{V}}e_{\nabla'}(X_{V})=\c(X_{V})\;\;.
\label{mqin1}
\ee
Here $\nabla'$ is the induced connection on $X_{V}$.

What makes this construction potentially interesting for functional
integrals is the following observation of Atiyah and Jeffrey \cite{aj}.
What they pointed out
was that, although $\en$ and $\int_{X}\en$ do not make sense for
infinite dimensional $E$ and $X$, the \mq\ form $\env$ can be
used to formally define {\em regularized Euler numbers} $\c_{V}(X)$
of such bundles by
\be
\c_{V}(X):=\int_{X}\env(X)  \label{mqin2}
\ee
for certain choices of $V$ (e.g.\ such that their zero locus is finite
dimensional so that the integral on the
right hand side of (\ref{mqin1}) makes sense).
Although not independent of $V$, these
numbers $\c_{V}(X)$ are naturally associated with $X$ for natural
choices of $V$ and are therefore likely to be of topological interest.
Furthermore, this construction (and its generalization to Euler classes
$e(E)$ of arbitrary real vector bundles over $X$) provides one with
an interesting class of functional integrals which are expected to
localize to, and hence be equivalent to, ordinary finite dimensional
integrals, precisely the phenomenon we want to study in these notes.

We should add that it is precisely such a representation of characteristic
classes or numbers  by functional integrals which is the characteristic
property of topological field theories. This suggests, that certain topological
field theories can be interpreted or obtained in this way which is indeed
the case. We will come back to this in the examples at the end of
this section.

A familiar theory in which this scenario is realized, and which preceded
and inspired both the \mq\ formalism and topological field theory is
\sqm, an example we will discuss in some detail in section 2.2.

\subsection{The Mathai-Quillen Formalism}

\subs{The Euler class of a finite dimensional vector bundle}

Consider a real vector bundle $\pi:E\ra X$ over a manifold $X$.
We will assume that $E$ and $X$ are orientable, $X$ is
compact without boundary, and that the rank (fibre dimension) of $E$ is
even and satisfies $rk(E)=2m\leq \dim(X)=n$.

The {\em Euler class} of $E$ is an integral cohomology class
$e(E)\in H^{2m}(X,\RR)\equiv H^{2m}(X)$ and there are two well-known and
useful ways of thinking about $e(E)$.

The first of these is in terms of sections of $E$. In general, a twisted
vector bundle will have no nowhere-vanishing non-singular sections and one
defines the Euler class to be the homology class of the zero locus of a generic
section of $E$. Its Poincar\'e dual is then a cohomology class $e(E)\in
H^{2m}(X)$.

The second makes use of the Chern-Weil theory of
curvatures and characteristic classes and produces an explicit
representative $\en(E)$ of $e(E)$ in terms of the curvature $\On$
of a connection $\nabla$ on $E$. Thinking of $\On$ as a matrix of
two-forms one has
\be
\en(E)=(2\pi)^{-m} {\rm Pf}(\On)\label{6}
\ee
where ${\rm Pf}(A)$ denotes the Pfaffian of the real antisymmetric matrix $A$.
Standard arguments
show that the cohomology class of $\en$ is independent of the choice
of $\nabla$.
For later use we note here that Pfaffians can be written as
fermionic (Berezin) integrals.  More precisely, if we
introduce real Grassmann odd variables $\c^{a}$, then
\be
\en(E)=(2\pi)^{-m}\int d\c \ex{\trac{1}{2}\c_{a}\On^{ab}\c_{b}}\label{25}\;\;.
\ee

If the rank of $E$ is equal to the dimension of $X$ (e.g.~if $E=TX$,
the tangent bundle of $X$)  then $H^{2m}(X)=H^{n}(X)=\RR$ and nothing
is lost by considering, instead of $e(E)$, its evaluation on
(the fundamental class $[X]$ of) $X$, the {\em Euler number\/}
$\c(E)=e(E)[X]$.
In terms of the two descriptions of $e(E)$ given above, this number can
be obtained either as the number of zeros of a generic section $s$ of $E$
(which are now isolated) counted with signs,
\be
\c(E)=\sum_{x_{k}:s(x_{k})=0}(\pm 1)\label{9}\;\;,
\ee
or as the integral
\be
\c(E)=\int_{X}\en(E)\label{10}\;\;.
\ee

If $2m < n $, then one cannot evaluate $e(E)$ on $[X]$ as above. One can,
however, evaluate it on homology $2m$-cycles or (equivalently) take the
product of $e(E)$ with elements of $H^{n-2m}(X)$ and evaluate this on $X$.
In this way one obtains {\em intersection numbers\/} of $X$ associated
with the vector bundle $E$ or, looked at differently, intersection numbers
of the zero locus of a generic section. The \mq\ form to be introduced
below also takes care of the situation when one chooses a non-generic
section $s$. In this case, the Euler number or class of $E$ can be expressed
as the Euler number or class of a vector bundle over the zero locus of $s$
(see \cite{ewnwzw} for an argument to that effect).

For $E=TX$, the Euler number can be defined as the
alternating sum of the Betti numbers of $X$,
\be
\c(X)=\sum_{k=0}^{n}(-1)^{k}b_{k}(X)\;\;,\;\;\;\;\;\;b_{k}(X)=\dim H^{k}(X,\RR)
\;\;.\label{betti}
\ee
In this context, equations (\ref{9}) and (\ref{10})  are the content of
the {\em Poincar\'e-Hopf theorem} and the {\em Gauss-Bonnet theorem}
respectively.
For $E=TX$ there is also an interesting generalization of (\ref{9})
involving  a possibly non-generic vector field $V$, i.e.\
with a zero locus $X_{V}$ which is not
necessarily zero-dimensional, namely
\be
\c(X)=\c(X_{V})\label{12}\;\;.
\ee
This reduces to (\ref{9}) when $X_{V}$ consists of isolated points
and is an identity when $V$ is the zero vector field.

\subs{The Mathai-Quillen representative of the Euler class}

One of the beauties of the \mq\ formalism is that
it provides a corresponding generalization of (\ref{10}), i.e.~an
explicit differential form representative $\ens$ of $e(E)$ depending
on both a section $s$ of $E$ and a connection $\nabla$ on $E$ such that
\be
e(E) = [\ens(E)]\;\;,
\ee
or (for $m=2n$)
\be
\c(E) = \int_{X}\ens(E)\label{13}
\ee
for any $s$ and $\nabla$.
The construction of $\ens$ involves what is known as the {\em Thom class}
$\F(E)$ of $E$ \cite{botu} or - more precisely - the explicit differential
form  representative $\Fn(E)$ of the Thom class found by Mathai and
Quillen \cite{mq}.  The construction of $\Fn(E)$ is best
understood within the framework of equivariant cohomology. What this means
is that one does not work directly on $E$, but that one realizes
$E$ as a vector bundle associated to some principal $G$-bundle
$P$ as $E=P\times_{G}V$ ($V$ the standard fibre of $E$) and works
$G$-equivariantly on $P\times V$. We will not go into the details here
and refer to \cite{aj,jk,mq2,mqlec} for discussions of various aspects of that
construction. Given $\Fn(E)$, $\ens$
is obtained by pulling back $\Fn(E)$ to $X$ via a section $s:X\ra E$ of $E$,
$\ens(E)=s^{*}\Fn(E)$. It is again
most conveniently represented as a Grassmann
integral, as in (\ref{25}), and is explicitly given by
\be
\ens(E)= (2\pi)^{-m}\int\! d\c\,
    \ex{-\trac{1}{2}|s|^{2}+\trac{1}{2}\c_{a}\On^{ab}\c_{b}+
    i\nabla s^{a}\c_{a}}\label{27}\;\;.
\ee
Here the norm $|s|^{2}$ refers to a fixed fibre metric on $E$ and
$\nabla$ is a compatible connection. Integrating out $\c$ one sees that
$\ens(E)$ is a $2m$-form on $X$. That $\ens(E)$ is closed
is reflected in the fact that the exponent in (\ref{27}) is invariant
under the transformation
\be
\d s = \nabla s \;\;,\;\;\;\;\;\;\d \c = is \;\;.
\label{trafo1}
\ee
To make contact with the notation
commonly used in the physics literature it will be useful to introduce
one more set of anti-commuting variables, $\p^{\mu}$, corresponding to
the one-forms $dx^{\mu}$ of a local coordinate basis. Given any form
$\omega$ on $X$, we denote by $\omega(\p)$ the object that one obtains
by replacing $dx^{\mu}$ by $\p^{\mu}$,
\be
\omega =
\trac{1}{p!}\omega_{\mu_{1}\cdots\mu_{p}}dx^{\mu_{1}}\ldots dx^{\mu_{p}}
\ra \omega(\p) =
\trac{1}{p!}\omega_{\mu_{1}\cdots\mu_{p}}\p^{\mu_{1}}\ldots \p^{\mu_{p}}
\;\;.\label{dxtop}
\ee
With the usual rules of Berezin integration, the integral of a
top-form $\omega^{(n)}$ can then be written as
\be
\int_{X}\omega^{(n)} = \int_{X}\!dx\,\int\!d\p\,\omega^{(n)}(\p)\;\;.
\label{psidx}
\ee
As the measures $dx$ and $d\p$ transform inversely to each other, the right
hand side is coordinate independent (and this is the physicist's way of saying
that integration of differential forms has that property). Anyway, with this
trick, the exponent of (\ref{27}) and the transformations (\ref{trafo1})
(supplemented by $\d x^{\mu}=\p^{\mu}$) then take the form of a supersymmetric
`action' and its supersymmetry.

We now take a brief look at $\ens(E)$ for various choices of $s$.
The first thing to note is that for $s$ the zero section of $E$ (\ref{27})
reduces to (\ref{25}), $e_{s=0,\nabla}(E)=\en(E)$.
As $\Fn(E)$ is closed, standard arguments now imply that $\ens(E)$ will be
cohomologous to $\en(E)$ for any choice of section $s$.
If $n=2m$ and $s$ is a generic section of $E$ transversal to the
zero section, we can calculate $\int_{X}\ens(E)$ by replacing
$s$ by $\gg s$ for $\gg\in{\bf R}$ and evaluating the integral in the limit
$\gg\ra\infty$. In that limit the curvature term in (\ref{27}) will
not contribute and one can use the stationary phase approximation
(see section 3)
to reduce the integral to a sum of contributions from the zeros of $s$,
reproducing equation (\ref{9}). The calculation is straightforward and
entirely analogous to similar calculations in \sqm\ (see e.g.~\cite{pr})
and we will not repeat it here. The important thing to remember, though,
is that one can use the $s$- (and hence $\gamma$-)independence
of $[\ens]$ to evaluate (\ref{13})
in terms of local data associated with the zero locus of some conveniently
chosen section $s$.

Finally, if $E=TX$ and $V$ is a non-generic vector field on $X$ with zero
locus $X_{V}$, the situation is a little bit more complicated. In
this case, $\int_{X}e_{\gg V,\nabla}$ for $\gg\ra\infty$ can be expressed in
terms of the Riemann curvature $\Omega_{\nabla'}$ of $X_{V}$ with
respect to the induced connection $\nabla'$.
Here $\Omega_{\nabla'}$ arises from the data $\On$ and $V$ entering
$e_{V,\nabla}$ via the classical {\em Gauss-Codazzi equations}.
Then equation (\ref{12}) is reproduced in the present setting in the form
(cf.\ (\ref{mqin2}))
\be
\c(X)=\int_{X}e_{V,\nabla}=(2\pi)^{-\dim(X_{V})/2}\int_{X_{V}}
{\rm Pf}(\Omega_{\nabla'})\label{34}\;\;.
\ee
Again the manipulations required to arrive at (\ref{34}) are exactly
as in \sqm, the Gauss-Codazzi version of which has been introduced and
discussed in detail in \cite{btqm}.

That the content of (\ref{27}) is quite non-trivial even in finite
dimensions where, as mentioned above, all the forms $\ens(E)$ are
cohomologous, can already be seen in the following simple example,
variants of which we will use throughout the
paper to illustrate the integration formulae (see e.g.\
(\ref{spins},\ref{spins2},\ref{spins3})).

Let us take $X=S^{2}$, $E=TX$, and equip $S^{2}$ with the standard constant
curvature metric $g= d\theta^{2}+ \sin^{2}\theta d\f^{2}$ and its
Levi-Civita connection $\nabla$. To obtain a
representative of the Euler class of $S^{2}$ which is different from the
Gauss-Bonnet representative, we pick some vector field $V$ on $S^{2}$.
Let us e.g.\ choose $V$ to be the vector field $\del_{\f}$
generating rotations about an axis of $S^{2}$.
Then the data entering (\ref{27}) can be readily
computed. In terms of an orthonormal frame $e^{a}=(d\theta, \sin\theta d\f)$
one finds
\bea
&&|V|^{2} = \sin^{2}\theta \;\;,\;\;\;\;\;\;
\nabla V^{a}=(\sin\theta\cos\theta d\f,-\cos\theta d\theta)\;\;,\non
&&\On^{12} = \sin\theta d\theta d\f\;\;,\;\;\;\;\;\;
\trac{1}{2}\c_{a}\On^{ab}\c_{b} = \c_{1}\c_{2}\sin\theta d\theta d\f\;\;.
\eea
To make life more interesting let us also replace $V$ by $\gg V$. The
$\c$-integral in (\ref{27}) is easily done and one obtains
\be
\env(TS^{2}) =(2\pi)^{-1} \ex{-\frac{\gg^{2}}{2}\sin^{2}\theta}
(1+ \gg^{2}\cos^{2}\theta)\sin\theta d\theta d\f\;\;. \label{spins0}
\ee
The quintessence of the above discussion, as applied to this example, is now
that the integral of this form over $S^{2}$ is the Euler number of $S^{2}$
and hence in particular independent of $\gg$. We
integrate over $\f$ (as nothing depends on it) and change variables from
$\theta$ to $x=\sin\theta$. Then the integral becomes
\be
\c(S^{2}) = \int_{-1}^{1}\ex{\frac{\gg^{2}}{2}(x^{2}-1)}(1+\gg^{2}x^{2})dx
\;\;.\label{exa2}
\ee
For $\gg=0$, the integrand is simply $dx$ and one obtains $\c(S^{2})=2$
which is (reassuringly) the correct result. A priori, however, it is far from
obvious that this integral is really independent of $\gg$. One way of making
this manifest is to note that the entire integrand is a total derivative,
\be
d(\ex{\frac{\gg^{2}}{2}(x^{2}-1)}x)
= \ex{\frac{\gg^{2}}{2}(x^{2}-1)}(1+\gg^{2}x^{2})dx\;\;.
\ee
This is essentially the statement that (\ref{spins0}) differs from the
standard representative $(1/2\pi)\sin\theta d\theta d\f$ at $\gg=0$
by a globally defined total derivative or, equivalently, that the
derivative of (\ref{spins0}) with respect to $\gg$ is exact. This makes the
$\gg$-independence of (\ref{exa2}) somewhat less mysterious. Nevertheless,
already this simple example shows that the content of the \mq\ formula
(\ref{27}) is quite non-trivial.

\subs{The Mathai-Quillen formalism for infinite dimensional vector bundles}

Let us recapitulate briefly what we have achieved so far.
We have constructed a family of
differential forms $\ens(E)$ parametrized by a section $s$ and a connection
$\nabla$, all representing the Euler class $e(E)\in H^{2m}(X)$. In
particular, for $E=TX$, the equation $\c(X)=\int_{X}\env(X)$ interpolates
between the classical Poincar\'e-Hopf and Gauss-Bonnet theorems.
It should be borne in mind, however, that all the forms $\ens$
are cohomologous so that this construction, as nice as it is, is not
very interesting from the cohomological point of view (although, as
we have seen, it has its charm also in finite dimensions).

To be in a situation where the forms $\ens$ are not necessarily all
cohomologous to $\en$, and where the \mq\ formalism thus `comes into its
own' \cite{aj}, we now consider infinite dimensional vector bundles
where $\en$ (an `infinite-form') is not defined at all. In that case
the added flexibility in the choice of $s$ becomes crucial and
opens up the possibility of obtaining well-defined, but $s$-dependent,
`Euler classes' of $E$. To
motivate the concept of a regularized Euler number of such a bundle,
to be introduced below, recall equation (\ref{12}) for the Euler number
$\c(X)$ of a manifold $X$.
When $X$ is finite dimensional this is an identity, while its left hand
side is not defined when $X$ is infinite dimensional. Assume, however,
that we can find a vector field $V$ on $X$ whose zero locus is a finite
dimensional submanifold of $X$. Then the right hand side of (\ref{12})
{\em is} well defined and we can use it to tentatively define a
{\em regularized Euler number} $\c_{V}(X)$ as
\be
\c_{V}(X):=\c(X_{V})\label{36}\;\;.
\ee
By (\ref{13}) and
the same localization arguments as used to arrive at (\ref{34}),
we expect this number to be given by the (functional) integral
\be
\c_{V}(X)=\int_{X}\env(X)\label{37}\;\;.
\ee
This equation can (formally) be confirmed by explicit calculation.
A rigorous proof can presumably be obtained in some
cases by probabilistic methods as used e.g.\ by Bismut \cite{bismut1,bismut2}
in related contexts. We will, however, content ourselves with verifying
(\ref{37}) in some examples below.

More generally, we are now led to define the regularized Euler number
$\cs(E)$ of an infinite dimensional vector bundle $E$ as
\be
\cs(E):=\int_{X}\ens(E)\;\;.\label{38}
\ee
Again, this expression turns out to make sense (for a physicist)
when the zero locus of
$s$ is a finite dimensional manifold $X_{s}$, in which case $\cs(E)$ is
the Euler number of some finite dimensional vector bundle over $X_{s}$
(a quotient bundle of the restriction $E|_{X_{s}}$, cf.~\cite{ewnwzw,aspin}).

Of course, there is no reason to expect $\cs(E)$ to be independent of $s$,
even if one restricts one's attention to those sections $s$ for which the
integral (\ref{38}) exists. However, if $s$ is a section of $E$ naturally
associated with $E$ (we will see examples of this below), then $\cs(E)$
is also naturally associated with $E$ and can be expected to
carry interesting topological information. This is indeed the case.

\subsection{The Regularized Euler Number of Loop Space,\protect\newline
             or: Supersymmetric Quantum Mechanics}

Our first application and illustration of the \mq\ formalism for infinite
dimensional vector bundles will be to the {\em loop space\/} $X=LM$ of a
finite dimensional manifold $M$ and its tangent bundle $E=T(LM)$. We will
see that this is completely equivalent to \sqm\ \cite{ewqm,ewqm1} whose
numerous attractive and interesting properties now find a neat explanation
within the present framework. As the authors of \cite{mq} were certainly
in part inspired by \sqm, deriving the latter that way may appear to be
somewhat circuitous. It is, nevertheless, instructive to do this because
it illustrates all the essential features of the \mq\ formalism in this
non-trivial, but manageable, quantum-mechanical context.

\subs{Geometry of loop space}

We denote by $M$ a smooth orientable Riemannian manifold with metric
$g$ and by $LM=\map{S^{1}}{M}$ the loop space of $M$.
Elements of $LM$ are denoted by
$x=\{x(t)\}$ or simply $x^{\m}(t)$,
where $t\in [0,1]$, $x^{\m}$ are (local) coordinates on $M$ and
$x^{\m}(0)=x^{\m}(1)$. In \sqm\ it is sometimes
convenient to scale $t$ such that $t\in [0,T]$ and
$x^{\m}(0)=x^{\m}(T)$ for some $T\in\RR$, and to regard
$T$ as an additional parameter of the theory.

A tangent vector $V(x)\in T_{x}(LM)$ at a loop $x\in LM$
can be regarded as a deformation of the loop, i.e.\ as a section of
the tangent bundle $TM$ restricted to the loop $x(t)$
such that $V(x)(t)\in T_{x(t)}(M)$. In more fancy terms this means
that
\be
T_{x}(LM) = \Gamma(x^{*}TM)\label{tlm}
\ee
is the infinite-dimensional space of sections of the pull-back  of the tangent
bundle $TM$ to $S^{1}$ via the map $x:S^{1}\ra M$.

There is a canonical vector field on $LM$ generating rigid rotations
$x(t)\ra x(t+\epsilon)$ of the loop, given by
$V(x)(t)=\dx(t)$ (or $V=\dx$ for short). A
metric $g$ on $M$ induces a metric $\hat{g}$ on $LM$ through
\be
\hat{g}_{x}(V_{1},V_{2}) = \tint{1} g_{\m\n}(x(t))
            V_{1}^{\m}(x)(t)V_{2}^{\n}(x)(t)
\label{46}\;\;.
\ee
Likewise, any differential form $\alpha$ on $M$ induces a differential
form $\hat{\alpha}$ on $LM$ via
\be
\hat{\alpha}_{x}(V_{1},\ldots,V_{p}) = \tint{1}
\alpha(x(t))(V_{1}(x(t)),\ldots,V_{p}(x(t)))\;\;.\label{hatf}
\ee
Because of the reparametrization invariance of these integrals, $\hat{g}$
and $\hat{\alpha}$ are invariant under the flow generated by $V=\dx$,
$L(V)\hat{g}=L(V)\hat{\alpha}=0$. We will make use of this observation
in section 3.2.

\subs{Supersymmetric quantum mechanics}

We will now apply the formalism developed in the previous section
to the data $X=LM$, $E=T(LM)$, and $V=\dx$ (and later on some variant
thereof). The zero locus $(LM)_{V}$ of $V$ is just the space of
constant loops, i.e.\ $M$ itself. If we therefore define the
regularized Euler number of $LM$ via (\ref{36})  by
\be
\c_{V}(LM):= \c((LM)_{V})=\c(M)\;\;,
\ee
we expect the functional integral $\int_{LM}\env(LM)$ to calculate the
Euler number of $LM$. Let us see that this is indeed the case.

The anticommuting variables $\c_{a}$ parametrize the fibres of
$T(LM)$ and we
write them as $\c_{a}=e_{a}^{\;\m}\pb_{\m}$ where $e_{a}^{\;\m}$ is the
inverse vielbein corresponding to $g_{\m\n}$. This change of variables
produces a Jacobian $\det[e]$ we will come back to below.
Remembering to substitute $dx^{\mu}(t)$ by $\p^{\mu}(t)$,
the exponent of the \mq\ form $\env(LM)$ (\ref{27}) becomes
\be
S_{M}=\tint{T}[-\trac{1}{2}g_{\m\n}\dx^{\m}\dx^{\n}
   +\trac{1}{4}R^{\m\n}_{\;\;\rho\sigma}\pb_{\m}\p^{\rho}\pb_{\n}\p^{\sigma}
            -i\pb_{\m}\nt\p^{\m}]\label{51}\;\;,
\ee
where $\nt$ is the covariant derivative along the loop $x(t)$ induced by
$\nabla$. This is precisely the standard action of de Rham (or $N=1$) \sqm\
\cite{ewqm,ewqm1,ag1,ag2,fw} (with the conventions as in \cite{pr}), the
supersymmetry given by (\ref{trafo1}). In order to reduce the measure
to the natural form $[dx][d\p][d\pb]$, one can introduce a multiplier field
$B^{\mu}$ to write the bosonic part of the action as
\be
\tint{T} g_{\m\n}(B^{\m}\dx^{\n} + B^{\m}B^{\n})\approx
\tint{T} [-\trac{1}{2}g_{\mu\nu}\dx^{\m}\dx^{\n}] \;\;,
\ee
because integration over $B$ will give rise to a determinant $\det[g]^{-1/2}$
that cancels the $\det[e]$ from above. With this understood,
the integral we are interested in is the partition function of this theory,
\be
\int_{LM}\env(LM) = Z(S_{M})  \label{39}
\ee
Now it is well known that the latter indeed calculates $\c(M)$
and this is our first non-trivial confirmation that (\ref{37}) makes sense
and calculates (\ref{36}). The explicit calculation of $Z(S_{M})$ is not
difficult, and we will sketch it below.

It may be instructive, however, to first
recall the standard {\em a priori\/} argument establishing
$Z(S_{M})=\c(M)$. One starts with the definition (\ref{betti}) of $\c(M)$.
As there is a one-to-one
correspondence between cohomology classes and harmonic forms on $M$,
one can write $\c(M)$ as a trace over the kernel ${\rm Ker}\;\Delta$
of the Laplacian $\Delta = d d^{*}+d^{*}d$ on differential forms,
\be
\c(M)={\rm tr}_{{\rm Ker}\;\Delta}(-1)^{F}\label{56}
\ee
(here $(-1)^{F}$ is $+1$ ($-1$) on even (odd) forms). As the operator
$d+d^{*}$ commutes with $\Delta$ and maps even to odd forms and vice-versa,
there is an exact pairing between `bosonic' and `fermionic' eigenvectors
of $\Delta$ with non-zero eigenvalue. It is thus possible to extend the
trace in (\ref{56}) to a trace over the space of all differential forms,
\be
\c(M)={\rm tr}_{\Omega^{*}(M)}(-1)^{F}\ex{-T\Delta}\label{57}\;\;.
\ee
As only the zero modes of $\Delta$ will contribute to the trace, it
is evidently independent of the value of $T$. Once one has put
$\c(M)$ into this form of a statistical mechanics partition function,
one can use the Feynman-Kac formula to represent it as a supersymmetric
path integral \cite{ag2} with the action (\ref{51}).

This Hamiltonian way of arriving at the action of \sqm\ should be
contrasted with the \mq\ approach. In the former
one starts with the operator whose index one
wishes to calculate (e.g.~$d+d^{*}$), constructs a corresponding Hamiltonian,
and then deduces the action. On the other hand,
in the latter one begins with a finite dimensional
topological invariant (e.g.~$\c(M)$) and represents that directly
as an infinite
dimensional integral, the partition function of a supersymmetric action.

What makes such a path integral representation of $\c(M)$ interesting is
that one can now go ahead and try to somehow evaluate it
directly, thus possibly obtaining alternative expressions for $\c(M)$.
Indeed, one can obtain path integral `proofs' of the Gauss-Bonnet and
Poincar\'e-Hopf theorems in this way. This is just the infinite dimensional
analogue of the fact discussed above that different choices of $s$
in $\int_{X}\ens(E)$ can lead to different expressions for $\c(E)$.

Let us, for example, replace the section $V=\dx$ in (\ref{51}) by
$\gg\dx$ for some $\gg\in\RR$. This has the effect of multiplying the
first term in (\ref{51}) by $\gg^{2}$ and the third by $\gg$.
In order to be able to take the limit
$\gg\ra\infty$, which would localize the functional integral to
$\dx=0$, i.e.\ to $M$, we proceed as follows. First of all, one
expands all fields in Fourier modes. Then one scales all non-constant
modes of $x(t)$ by $\gg$ and all the non-constant modes of $\p(t)$ and
$\pb(t)$ by $\gg^{1/2}$. This leaves the measure invariant.
Then in the limit $\gg\ra\infty$ only the
constant modes of the curvature term survive in the action, while the
contributions from the non-constant modes cancel identically between
the bosons and fermions because of supersymmetry. The net effect of this
is that one is left with a finite dimensional integral of the form
(\ref{25}) leading to
\be
Z(S_{M})=(2\pi)^{-\dim(M)/2}\int_{M}{\rm Pf}(\On(M))=
\c(M)\label{41}\;\;.
\ee
This provides a path integral derivation of the Gauss-Bonnet
theorem (\ref{10}). The usual \sqm\ argument
to this effect \cite{ag1,fw} makes use of the $T$-independence of the
partition function to suppress the non-constant modes as $T\ra 0$.
But the \mq\ formalism
now provides an understanding and explanation of the mechanism by which
the path integral (\ref{39}) over $LM$ localizes to the integral (\ref{41})
over $M$.

Other choices of sections are also possible, e.g.\ a vector field of the
form $V=\dx^{\mu} + \gg g^{\mu\nu}\del_{\nu}W$, where $W$ is some function
on $M$ and $\gg\in\RR$ a parameter. This
introduces a potential into the \sqm\ action (\ref{51}).
It is easy to see that the
zero locus of this vector field is the zero locus of the gradient
vector field $\del_{\mu}W$ on $M$ whose Euler number is the same as that
of $M$ by (\ref{12}).
Again this agrees with the explicit evaluation of the path integral
of the corresponding \sqm\ action which is, not unexpectedly,
most conveniently performed by considering the limit $\gg\ra\infty$.
In that limit, because of the term
$\gg^{2}g^{\mu\nu}\del_{\mu}W\del_{\nu}W$ in the action, the path integral
localizes around the critical points of $W$. Let us assume that these are
isolated. By supersymmetry, the fluctuations around the critical
points will, up to a sign, cancel between the bosonic and fermionic
contributions and one finds that the partition function is
\be
\c(M)=Z(S_{M})=\sum_{x_{k}:dW(x_{k}=0}{\rm sign}(\det H_{x_{k}}(W))
\label{61}\;\;,
\ee
where
\be
H_{x_{k}}(W)=(\nabla_{\m}\del_{\n}W)(x_{k})\label{62}
\ee
is the {\em Hessian\/} of $W$ at $x_{k}$. This is the Poincar\'e-Hopf theorem
(\ref{9}) (for a gradient vector field).

If the critical points of $W$ are not isolated then, by a combination of the
above arguments, one recovers the generalization
$\c(M)=\c(M_{W'})$ (\ref{12}) of the
Poincar\'e-Hopf theorem (for gradient vector fields) in the form (\ref{34}).

Finally, by considering a section of the form $V=\dx + v$, where $v$ is an
arbitrary vector field on $M$, one can also, to complete the circle,
rederive the general finite dimensional \mq\ form $e_{v,\nabla}(M)$
(\ref{27}) from \sqm\ by proceeding exactly as in the derivation
of (\ref{41}) \cite{btmq,mqlec}.

This treatment of \sqm\ has admittedly been somewhat sketchy and we should
perhaps, summarizing this section, state clearly what are the
important points to keep in mind as regards the \mq\ formalism and
localization of functional integrals:
\begin{enumerate}
\item Explicit evaluation of the \sqm\ path integrals obtained by formally
applying the \mq\ construction to the loop space $LM$
confirms that we can indeed represent the regularized Euler number
$\c_{V}(LM)$, as defined by (\ref{36}), by the functional integral
(\ref{37}).
\item In particular, it confirms that functional integrals
arising from or related to the \mq\ formalism (extended to infinite
dimensional bundles) localize to finite dimensional integrals.
\item This suggests that a large class of other quantum mechanics and field
theory functional integrals can be constructed which also have this
property. These integrals should be easier to understand
in a mathematically rigorous fashion than generic functional integrals
arising in field theory.
\end{enumerate}

\subsection{Other Examples and Applications - an Overview}

The purpose of this section is to provide an overview of other
applications and appearances of the \mq\ formalism in the physics
literature. None too surprisingly, all of these are related to
topological field theories or simple (non-topological) perturbations
thereof.

It was first shown by Atiyah and Jeffrey \cite{aj} that the
rather complicated looking action of
Witten's four-dimensional topological Yang-Mills theory \cite{ewdon}
(Donaldson theory) had a neat explanation in this framework.
Subsequently, this interpretation was also shown to be valid and useful
in topological sigma models coupled to topological gravity \cite{ewnwzw}.
In \cite{btmq,mqlec} this strategy was turned around to construct
topological field theories from scratch via the \mq\ formalism.
A nice explanation of the formalism from a slightly different perspective
can be found in the recent work of Vafa and Witten \cite{cvew}, and other
recent applications include \cite{gm,wu}.

As these applications thus range from topological gauge theory
(intersection numbers of moduli spaces of connections) over
topological sigma models (counting holomorphic curves) to topological
gravity (intersection theory on some moduli space of metrics) and the
string theory interpretation of 2d Yang-Mills theory, this section will
not and cannot be self-contained. What we will do is to try to provide
a basic understanding of why so-called cohomological topological field
theories can generally  be understood within and constructed from the
\mq\ formalism.

\subs{The basic strategy - illustrated by Donaldson theory}

On the basis of what has been done in the previous section, one possible
approach would be to ask if topological field theories can in some sense
be understood as \sqm\ models with perhaps infinite-dimensional target
spaces. For \tft\ models calculating the Euler number of some moduli space
this has indeed been shown to be the case in \cite{btqm}. In
principle, it should also be possible to extend the \sqm\ formalism of the
previous section to general vector bundles $E$ and their \mq\ forms
$\ens$. This generalization would then also cover other topological field
theories.

This is, however, not the point of view we are going to adopt in the
following. Rather, we will explain how the \mq\ formalism can be used
to construct a field theory describing intersection theory on some given
moduli space (of connections, metrics, maps, \ldots) of interest.

Before embarking on this, we should perhaps point out that, from a purely
pragmatic point of view, an approach to the construction of topological
field theories based on, say, BRST quantization may occasionally
be more efficient and straightforward. What one is gaining
by the \mq\ approach is geometrical insight.

Asume then, that the finite dimensional moduli space on which one would
like to base a topological field theory is given by the following data:
\begin{enumerate}
\item A space $\F$ of fields $\f$ and (locally) a defining equation
      or set of equations $\FF(\f)=0$ ;
\item A group $\G$ of transformations acting on $\F$ leaving invariant
      the set $\FF(\f)=0$, such that the moduli space is given by
\be
\M_{\FF}= \{\f\in\F: \FF(\f)=0\}/\G \ss \F/\G\equiv X\;\;.\label{mff}
\ee
\end{enumerate}
This is not the most general set-up one could consider but will
be sufficient for what follows.

The prototypical example to keep in mind is the moduli space of
anti-self-dual connections (instantons), where $\F$ is the space
$\cal A$ of
connections $A$ on a principal bundle $P$ on a Riemannian four-manifold
$(M,g)$,
$\FF(A)=(F_{A})_{+}$
is the self-dual part of the curvature of $A$, and $\G$ is the group
of gauge transformations  (vertical automorphisms of $P$).
In other cases either $\FF$ or $\G$ may be trivial - we will see examples
of that below.

Now, to implement the \mq\ construction, one needs to be able to
interpret the moduli space $\M_{\FF}$ as the zero locus of a section
$s_{\FF}$
of a vector bundle $E$ over $X$. In practice, it is often more convenient
to work `upstairs', i.e.\ equivariantly, and to exhibit the set
$\{\FF=0\}$ as the zero locus of a $\G$-equivariant vector bundle
over $\F$. This may seem to be rather abstract, but in practice
there are usually moderately obvious candidates for $E$ and $s$.
One also needs to choose some connection $\nabla$ on $E$.

Let us again see how this works for anti-self-dual connections.
In that case, $\FF$ can be regarded as a map from $\cal A$ to the
space $\Omega^{2}_{+}(M,{\rm ad}P)$, i.e.\  roughly speaking the
space  of Lie algebra valued self-dual two-forms
on $M$. The kernel of this map is the space of anti-self-dual
connections. To mod out by the gauge group we proceed as follows.
For appropriate choice of $\cal A$ or $\G$, $\cal A$
can be regarded as the total space of a principal $\G$-bundle over
$X=\C$. As $\G$ acts on $\Omega^{2}_{+}(M,{\rm ad}P)$, one
can form the associated infinite dimensional vector bundle
\be
{\cal E}_{+} = {\cal A}\times_{\G}\Omega^{2}_{+}(M,{\rm ad}P)\;\;.
\label{dobu}
\ee
Clearly, the map $\FF$ descends to give a well-defined section $s_{\FF}$
of ${\cal E}_{+}$ whose zero-locus is precisely the finite dimensional
moduli space of anti-self-dual connections and we can choose
$E={\cal E}_{+}$. This bundle also has a natural
connection coming from
declaring the horizontal subspaces in ${\cal A}\ra\C$ to be those
orthogonal to the $\G$-orbits in $\cal A$ with respect to the metric on
$\cal A$ induced by that on $M$.

Once one has assembled the data $X$, $E$ and $s=s_{\FF}$ one can - upon
choice of some connection $\nabla$ on $E$ - plug these into the
formula (\ref{27}) for the \mq\ form. In this way one obtains
a functional integral which localizes onto the moduli space $\M_{\FF}$.
A few remarks may help to clarify the status and role of the
`field theory action' one obtains in this way.
\begin{enumerate}
\item {\em A priori\/} this functional integral can be thought of
as representing a regularized Euler number $\cs(E)$ or Euler class
of the infinite dimensional vector bundle $E$. As mentioned in
section 2.1 (and as we have seen explicitly in section 2.2), this
can also be interpreted as the Euler number (class) of a vector bundle
over $\M_{\FF}$. If the rank of this vector bundle is strictly less
than the dimension of $\M_{\FF}$, than one needs to consider the
pairing of this Euler class with cohomology classes on $\M_{\FF}$.

{}From a mathematical point of view the necessity of this is clear while
in physics parlance this amounts to inserting operators (observables)
into the functional integral to `soak up the fermionic zero modes'.
\item Frequently, the connection and curvature of the bundle $E$
are expressed in terms of Green's functions on the underlying
`space-time' manifold. As such, the data entering the putative
field theory action (the exponent of (\ref{27})) are non-local
in space-time - an undesirable feature for a fundamental action.
The ham-handed way of eliminating this non-locality is to introduce
auxiliary fields. However, what this really amounts to is to working
`upstairs' and constructing an equivariant \mq\ form (cf.\ the
remarks before (\ref{27})). In practice, therefore, it is often
much simpler to start upstairs directly and to let Gaussian integrals
do the calculation of the curvature and connection terms of $E$
entering (\ref{27}).
\end{enumerate}

Let us come back once again to the moduli space of instantons.
Note that, among all the possible sections of
${\cal E}_{+}$ the one we have chosen is really the only natural one.
Different sections could of course be used in the \mq\ form. But in the
infinite dimensional case there is no reason to believe that the
result is independent of the choice of $s$ and for other choices of $s$
it would most likely not be of any mathematical interest.
With the choice $s_{\FF}$, however, the resulting action is interesting
and, not unexpectedly, that of Donaldson theory \cite{don,ewdon}.
We refer to \cite[pp.\ 198-247]{pr} for a partial collection of
things that can and should be said about this theory.

\subs{A topological gauge theory of flat connections}

With the \mq\ formalism at one's disposal, it is now relatively easy to
construct other examples. In the following we will sketch
two of them, one for a moduli space of flat connections and
the other related to holomorphic curves.

If one is, for instance, interested in the moduli space $\M^{3}$
of flat connections
$F_{A}=0$ on a principal $G$-bundle on a three-manifold $M$, one could
go about constructing an action whose partition function localizes onto
that moduli space as follows. The flatness condition expresses the
vanishing of the two-form $F_{A}$ or - by duality - the one-form
$*F_{A}\in \Omega^{1}(M,{\rm ad}P)$. Now, as the space $\cal A$ is an
affine space modelled on $\Omega^{1}(M,{\rm ad}P)$, we can think of
$*F_{A}$ as a section of the tangent bundle of $\cal A$, i.e.\ as
a vector field on $\cal A$. In fact, it is the gradient vector field
of the {\em Chern-Simons functional\/}
\be
S_{CS}(A)=\int_{M}\!\Tr\, A\,dA + \trac{2}{3}A^{3} \;\;.
\ee
This vector field passes down to a vector
field on $\C$ whose zero locus is precisely the moduli
space $\M^{3}$ of flat connections and our data are therefore
$X=\C$, $E=TX$, and $V(A)=*F_{A}$ (in accordance with section 2.1 we denote
sections of the tangent bundle by $V$). Following the steps outlined
above to construct the action form the \mq\ form (\ref{27}), one
finds that it coincides with that constructed e.g.\ in
\cite{ewtop,bbt,ms,btmq}. Again, as in \sqm, one finds full agreement of
\be
\c_{V}(\C)\equiv \c((\C)_{V}) = \c(\M^{3})\label{43}
\ee
with the partition function of the action which gives $\c(\M^{3})$ in the
form (\ref{34}), i.e.\ via the Gauss-Codazzi equations for the embedding
$\M^{3}\ss\C$ \cite{btmq}. We conclude this example with some remarks.
\begin{enumerate}

\item In \cite{aj} the partition function of this theory was first
identified with a regularized Euler number of $\C$. We can now
identify it more specifically with the Euler number of $\M^{3}$.

\item In \cite{tau} it was shown that for certain
three-manifolds (homology spheres) $\c_{V}(\A{3})$ is the Casson invariant.
This has led us to propose the Euler number of $\M^{3}$ as a
generalization of the Casson invariant
for more general three-manifolds. See \cite{btmq} for
a preliminary investigation of this idea.

\item Note that the chosen moduli space $\M^{3}$ itself, via the
defining equation $F_{A}=0$, determined the bundle $E$ to be used
in the \mq\ construction. Hence it also determined the fact that we ended up
with a \tft\ calculating the Euler number of $\M^{3}$ rather than
allowing us to model intersection theory on $\M^{3}$. While this
may appear to be a shortcoming of this procedure, there are also
other reasons to believe that intersection theory on $\M^{3}$  is
not a particularly natural and meaningful thing to study.

\item  Nevertheless, in two
dimensions one can construct both a \tgt\ corresponding
to intersection theory on the moduli space $\M^{2}$ of flat connections
(this is just the 2d analogue of Donaldson theory) and a \tgt\
calculating the Euler number of $\M^{2}$. For the former, the relevant
bundle is ${\cal E}={\cal A}\times_{\G}\Omega^{0}(M,{\rm ad}P)$ (cf.\
(\ref{dobu})) with
section $*F_{A}$. In the latter case, life turns out not be so simple.
The base space $X$ cannot possibly be $\C$
as $*F_{A}$ does not define a section of its tangent bundle. Rather,
the base space turns out to be ${\cal E}$ with $E=T{\cal E}$ and an
appropriate section. Because
the fibre directions of ${\cal E}$ are topologically trivial, this indeed
then calculates the Euler number of $\M^{2}$ - see the final remarks in
\cite{mqlec}.
\end{enumerate}

\subs{Topological sigma models}

Finally, we will consider an example of a \tft\ which is not a gauge theory
but rather, in a sense, the most obvious field theoretic generalization of
\sqm, namely the topological sigma model \cite{ewsig,ewmirr}. In its simplest
form this is a theory of maps from a Riemann surface $\S$ to a K\"ahler
manifold $M$ localizing either onto holomorphic maps (in the so-called
$A$-model) or onto constant maps (in the $B$-model). As an obvious
generalization of \sqm, the interpretation of the $A$-model in terms of the
\mq\ formalism is completely straightforward. It is already implicit
in the Langevin equation approach to the construction of the model
\cite{brt} and has recently been spelled out in detail in \cite{gm,wu}.
We will review this construction below. The \mq\ interpretation
of the $B$-model is slightly more subtle and we will discuss that
elsewhere.

In terms of the notation introduced at the beginning of this section
(cf.\ (\ref{mff})), the space of fields is the space $\F=\map{\S}{M}$
of maps from a Riemann surface $\S$ (with complex structure $j$) to
a K\"ahler manifold $M$ (with complex structure $J$ and hermitian metric $g$).
We will denote
by $T^{(1,0)}\S$ etc.\ the corresponding decomposition
of the complexified tangent and cotangent bundles into their holomorphic and
anti-holomorphic parts.

The defining
equation is the condition of holomorphicity of $\f\in\F$
with respect to both $j$ and $J$, $\FF(\f)=\dbar_{J}\f=0$.
%\be
%\FF(\f) =\dbar_{J}\f\equiv \trac{1}{2}(d\f + J \circ d\f \circ j) = 0
%\;\;. \label{hol1}
%\ee
In terms of local complex coordinates on $\S$ and $M$, $\f$ can be
represented as a map $(\f^{k}(z,\bar{z}),\f^{\bar{k}}(z,\bar{z}))$ and
the holomorphicity condition can be written as
\be
\dzb\f^{k} = \dz \f^{\bar{k}}=0 \;\;.\label{hol}
\ee
Solutions to (\ref{hol}) are known as {\em holomorphic curves} in $M$.
The moduli space of interest is just the space $\M_{\dbar_{J}}$ of
holomorphic curves,
\be
\M_{\dbar_{J}}=\{\f\in\map{\S}{M}: \dbar_{J}\f = 0\}\;\;,\label{modhol}
\ee
so that this is an example where the symmetry group $\G$ is trivial.

Let us now determine the bundle $E$ over $X=\map{\S}{M}$.
In the case of \sqm, the map $x(t)\ra\dx(t)$ could be regarded as a
section of the tangent bundle (\ref{tlm}) of $LM=\map{S^{1}}{M}$.
Here the situation is, primarily notationally,
a little bit more involved. First of all, as
in \sqm, the tangent space at a map $\f$ is
\be
T_{\f}(\map{\S}{M})=\Gamma(\f^{*}TM)\;\;.
\ee
The map $\f\ra d\f$ can hence be regarded as a section of a vector bundle
$\cal E$ whose fibre at $\f$ is $\Gamma(\f^{*}TM\otimes T^{*}\Sigma)$.
The map
\be
s_{\FF}:\;\;\f\ra(\dzb\f^{k},\dz\f^{\bar{k}})
\ee
we are interested in can then be regarded as a
section of a subbundle ${\cal E}^{(0,1)}$ of that bundle spanned by
$T^{(1,0)}M$-valued $(0,1)$-forms $\pb^{k}_{\bar{z}}$ and $T^{(0,1)}M$-valued
$(1,0)$-forms $\pb^{\bar{k}}_{z}$ on $\S$. Evidently, the moduli space
$\M_{\dbar_{J}}$ is the zero locus of the section
$s_{\FF}(\f)=\dbar_{J}\f$ of $E={\cal E}^{(0,1)}$.

Since the bundle ${\cal E}^{(0,1)}$ is contructed from the tangent and
cotangent bundles of $M$ and $\S$, it inherits a connection from the
Levi-Civit\`a connections on $M$ and $\Sigma$. With the convention of
replacing one-forms on $X$ by their fermionic counterparts (cf.\
(\ref{dxtop},\ref{51})), the exterior covariant derivative of $s_{\FF}$ can
then be written as
\be
\nabla s_{\FF}(\p)= (D_{\bar{z}}\p^{k},D_{z}\p^{\bar{k}})\;\;.
\ee
One can now obtain the action $S_{TSM}$ for the topological sigma model from
the \mq\ form (\ref{27}),
\be
S_{TSM}(\F)= \int_{\S}\!d^{2}z\,
g_{k\bar{l}}(\trac{1}{2}\dzb\f^{k}\dz\f^{\bar{l}}
+i\pb_{z}^{\bar{l}}D_{\bar{z}}\p^{k} + i\pb_{\bar{z}}^{k}D_{z}\p^{\bar{l}})
-R_{k\bar{k}l\bar{l}}\pb_{\bar{z}}^{k}\pb_{{z}}^{\bar{k}}\p^{l}\p^{\bar{l}}
\label{tsm}\;\;.
\ee
For $M$ a Ricci-flat K\"ahler manifold (i.e. a Calabi-Yau manifold), this
action and its $B$-model partner have recently been studied intensely in
relation with {\em mirror symmetry}, see e.g.\ \cite{ewmirr,aspin} and the
articles in \cite{mirror}. Usually, one adds
a topological term $S_{TOP}$ to the action which is the pullback
of the K\"ahler form $\omega$ of $M$,
\be
S_{TOP}=\int_{\S}\f^{*}\omega = \int_{\S}\!d^{2}z\, g_{k\bar{l}}
        (\dz\f^{k}\dzb\f^{\bar{l}}-\dzb\f^{k}\dz\f^{\bar{l}})\;\;.
\ee
This term keeps track of the instanton winding number. Its inclusion
is also natural from the conformal field theory point of view as the
`topological twisting' of the underlying supersymmetric sigma model
automatically gives rise to this term. We refer to \cite{ewsig} and
\cite{pr} to background information on the topological sigma model in
general and to \cite{ewmirr,aspin} for an introduction to the subject
of mirror manifolds and holomorphic curves close in spirit to the point
of view presented here.

The topological sigma model becomes even more interesting when it is
coupled to topological gravity \cite{ewint}, as such a coupled model
can be regarded as a topological string theory. A similar model,
with both $\Sigma$ and $M$ Riemann surfaces, and with localization onto
a rather complicated (Hurwitz) space of branched covers,
has recently been studied as a candidate for a string theory realization of
2d Yang-Mills theory \cite{gm}.

\subs{BRST fixed points and localization}

This has in part been an extremely brief survey of some applications
of the \mq\ formalism
to \tft. The main purpose of this section, however, was to draw attention to
the special properties the functional integrals of these theories enjoy.
We have seen that it is the interpretation of the
geometrical and topological features of the functional integral in terms of the
\mq\ formalism which allows one to conclude that it is {\em a
priori\/} designed to represent a finite dimensional integral.

In order to develop the subject further, and  directly from a functional
integral
point of view, it may also be useful to know how physicists read off this
property as a sort of fixed point theorem from the action or the `measure'
of the functional integral and its symmetries. The key point is that all
these models have a Grassmann odd scalar symmetry $\d$
of the kind familiar from BRST symmetry. This is essentially the field theory
counterpart of (\ref{trafo1}), which we repeat here in the form
\be
\d \c = i s_{\FF} (x)\;\;,\;\;\;\;\;\;
\d s_{\FF}(x) = \nabla s_{\FF} (\p)\;\;.
\label{trafo2}
\ee
The `essentially' above refers to the fact that this may be the $\d$
symmetry of the complete action only modulo gauge transformations
(equivariance) and equations of motion.

Now, roughly one can argue as follows (see \cite{ewnwzw,ewmirr}).
If the $\d$ action were free, then one could introduce an anti-commuting
collective coordinate $\theta$ for it. The path integral would then contain
an integration $\int\!d\theta\ldots$. But $\d$-invariance implies
$\theta$-independence of the action, and hence the integral would be zero by
the rules of Berezin integration. This implies that the path integral only
receives contributions from some arbitrarily small $\d$-invariant
tubular neighbourhood of the fixed point set
of $\d$. While the integral over the fixed point set itself has to be
calculated exactly, the integral over the `tranverse' directions can then
(for generic fixed points)
be calculated in a stationary phase (or one-loop) approximation.
If one wants to make something obvious look difficult, one can also apply this
reasoning to the ordinary BRST symmetry of gauge fixing. In that case the
above prescription says that one can set all the ghost and multiplier fields
to zero provided that one takes into account the gauge fixing condition and
the Faddeev-Popov determinant (arising form the quadratic ghost-fluctuation
term) - a truism.

{}From (\ref{trafo2}) one can read off that, in the case at hand,
this fixed point set is precisely the desired
moduli space described by the zero locus of $s_{\FF}$ and its tangents $\p$
satisfying the linearized equation $\nabla s_{\FF} (\p)=0$. In this way the
functional integral reduces to an integral of differential
forms on $\M_{\FF}$.

This point of view has the virtue of being based on easily verifiable
properties of the BRST like symmetry. It also brings out the analogy
with the functional integral generalization of the \dh\ theorem, the
common theme being a supersymmetry modelling an underlying geometric
or topological structure which is ultimately responsible for
localization.

\section{Equivariant Localization and the Stationary Phase Approximation,
\protect\newline or: The Duistermaat-Heckman Theorem}

In this section we will discuss a localization formula which, roughly speaking,
gives a criterion for the stationary phase approximation to an oscillatory
integral to be exact.

To set the stage, we will first briefly recall the ordinary stationary
phase approximation. Thus, let
$X$ be a smooth compact manifold of dimension $n=2l$, $f$ and $dx$
a smooth function and a smooth density on $X$. Let $t\in\RR$ and consider
the function (integral)
\be
F(t)=\int_{X}\ex{itf}dx\;\;.                              \label{sp1}
\ee
The stationary phase approximation expresses the fact that for large $t$
the main contributions to the integral come from neighbourhoods of the
critical points of $f$. In particular, if the critical points of $f$ are
isolated and non-degenerate (so that the determinant of the Hessian
$H_{f}=\nabla df$
of $f$ is non-zero there), then for $t\ra\infty$ the integral (\ref{sp1})
can be approximated by
\be
F(t) =(\frac{2\pi}{t})^{l} \sum_{x_{k}:df(x_{k})=0}
\ex{\frac{i\pi}{4}\sigma(H_{f}(x_{k}))}|\det(H_{f}(x_{k})|^{-1/2}\ex{itf(x_{k})}
+ O(t^{-l-1})\;\;,\label{sp2}
\ee
where $\sigma(H_{f})$ denotes the signature of $H_{f}$ (the number of
positive minus the number of negative eigenvalues).

Duistermaat and Heckman \cite{dh} discovered a class of examples where this
approximation gives the exact result and the error term in the above
vanishes. To state these examples and the \dh\ formula in their
simplest form, we take $X$ to be a symplectic
manifold with symplectic form $\omega$ and Liouville measure $dx(\omega)
= \omega^{l}/l!$. We will also assume that the function $f$
generates a circle action on $X$ via its Hamiltonian vector field $V_{f}$
defined by $i(V_{f})\omega = df$, and denote by $L_{f}(x_{k})$
the infinitesimal action
induced by $V_{f}$ on the tangent space $T_{x_{k}}X$ (essentially, $L_{f}$
can be represented by the matrix $H_{f}$). Then the \dh\ formula reads
\be
\int_{X}\ex{itf}dx(\omega) = (\frac{2\pi i}{t})^{l}\sum_{x_{k}:df(x_{k})=0}
\det(L_{f}(x_{k}))^{-1/2}\ex{itf(x_{k})}\;\;. \label{sp3}
\ee
Being careful with signs and the definition of the square root of the
determinant, it can be seen that (\ref{sp3}) expresses nothing other than
the fact that the stationary phase approximation to $F(t)$ is exact for all
$t$.

As a simple example consider a two-sphere of unit radius centered at the
origin of $\RR^{3}$, with $f(x,y,z)=z$ (or, more generally, $z+a$ for
some constant $a$) the generator of rotations about
the $z$-axis and $\omega$ the volume form.
This integral can of course easily be done exactly, e.g.\
by converting to polar coordinates, and one finds
\be
\int_{S^{2}} dx(\omega) \ex{itf} =
\int \sin\theta d\theta d\f \ex{it(\cos\theta+a)}
= \frac{2\pi i}{t}\left(\ex{-it(1-a)}-\ex{it(1+a)}\right)\;\;.\label{spins}
\ee
We see that the result can be expressed as a sum of two terms, one from
$\theta = \pi$ and the other form $\theta=0$.
This is just what one expects form the \dh\ formula, as these are just the
north and south poles, i.e.\ the fixed points of the $U(1)$-action. The
relative sign between the two contributions is due to the fact that $f$ has
a maximum at the north pole and a minimum at the south pole. This example
can be considered as the classical partition function of a spin system.
Thinking of the two-sphere as $SU(2)/U(1)$, it has a nice generalization
to homogeneous spaces of the form $G/T$, where $G$ is a compact Lie group
and $T$ a maximal torus. The corresponding quantum theories are also all
given exactly by their stationary phase (WKB) approximation and evaluate
to the Weyl character formula for $G$. For an exposition of these
ideas see the beautiful paper \cite{stone} by Stone.

It is now clear that the \dh\ formula (\ref{sp3}) (and its generalization
to functions with non-isolated critical points) can be regarded as a
localization formula as it reduces the integral over $X$ to a sum (integral)
over the critical point set. As the stationary phase approximation is very
much like the semi-classical or WKB approximation investigated extensively
in the physics (and, in particular, the path integral) literature, it is
of obvious interest to inquire if or under
which circumstances a functional integral analogue of the formula (\ref{sp3})
can be expected to exist. In this context, (\ref{sp3}) would now express
something like the exactness of the one-loop approximation to the path
integral.

The first encouraging evidence for the existence
of such a generalization was pointed out by Atiyah and Witten \cite{atew}.
They showed that a formal
application of the \dht\ to the infinite-dimensional loop space $X=LM$ of
a manifold $M$ and, more specifically, to the partition
function of $N=\frac{1}{2}$ supersymmetric quantum mechanics
representing the index of the Dirac operator on $M$, reproduced the known
result correctly. This method of evaluating the quantum mechanics
path integral has been analyzed by Bismut \cite{bismut2} within a
mathematically rigorous framework. Subsequently, the work
of Stone \cite{stone} brought the \dht\ to the attention of a wider
physicists audience. Then, in \cite{csqm,bkn}, a general supersymmetric
framework for investigating \dh- (or WKB-)like localization theorems for
(non-supersymmetric) phase space path integrals was introduced,
leading to a fair amount of activity in the
field - see e.g.\ \cite{picken,an1,an2,anot,lykken,semenoff,rajeev}.
We will review some of these matters below.

While Duistermaat and Heckman originally discovered their formula within the
context of symplectic geometry, it turns out to have
its most natural explanation in the setting of equivariant
cohomology and equivariant characteristic classes \cite{ab,bv,bgv}.
This point of view also suggests some generalizations of  the \dht,
e.g.\ the Berline-Vergne formula \cite{bv}  valid for Killing vectors
on general compact Riemannian manifolds $X$.
Strictly speaking, the example considered by Atiyah and Witten falls into this
category as the free loop space of a manifold is not quite a symplectic
manifold
in general. We will discuss the interpretation of these formulae in terms
of equivariant cohomology below. An excellent survey of these and many other
interesting matters can be found in \cite{bgv}.

There is also a rather far-reaching generalization of the localization
formula which is due to Witten \cite{ew2d}. It applies to non-Abelian
group actions and to `actions' (the exponent on the lhs of (\ref{sp3}))
which are polynomials in the `moment map' $f$ rather than just linear
in the moment map as in (\ref{sp3}). This generalization is potentially much
more interesting  for field-theoretic applications, and we give a brief
sketch of the set-up in section 3.3, but as a detailed discussion
of it would necessarily go far beyond the elementary and introductory scope
of this paper we refer to \cite{ew2d,wu2,ljfk} for further information.

\subsection{Equivariant Cohomology and Localization}

As mentioned above, the localization theorems we are interested in,
in this section, are most conveniently understood in terms of equivariant
cohomology and we will first introduce the necessary concepts. Actually,
all we will need is a rather watered down version of this, valid for
Abelian group actions. The following discussion is intended as a
compromise between keeping things elementary in this simple setting
and retaining the flavour and elegance of the general theory.
We will see that the localization formulae of Berline-Vergne \cite{bv}
and \dh\ \cite{dh} are then fairly immediate consequences of the general
formalism. Everything that is said in this section can be found in
\cite{bgv}.

\subs{Equivariant cohomology}

Let $X$ be a compact smooth $(n=2l)$-dimensional
manifold, and let $G$ be a compact
group acting smoothly on $X$ (the restriction to even dimension is by no
means necessary - it will just allow us to shorten one of the arguments below).
We denote by $\lg$ the Lie algebra of $G$,
by $\CC[\lg]$ the algebra of complex valued polynomial functions on $\lg$,
and by $\Omega^{*}(X)$ the algebra of complex valued differential forms on
$X$. For any $V\in\lg$ there is an associated vector field on $X$ which we
denote by $V_{X}$. We note that any metric $h$ on $X$ can be made
$G$-invariant by averaging $h$ over $G$. Hence we can assume without loss
of generality that $G$ acts on the Riemannian manifold $(X,h)$ by
isometries and that all the vector fields $V_{X}$ are Killing vectors of
$h$.

The $G$-equivariant cohomology of $X$, a useful generalization of the
cohomology of $X/G$  when the latter is not a smooth manifold, can be
defined by the following construction (known as the Cartan model of
equivariant cohomology). We consider the tensor product $\CC[\lg]\otimes
\Omega^{*}(X)$. As there is a natural (adjoint)
action of $G$ on $\CC[\lg]$ and the
$G$-action on $X$ induces an action on $\Omega^{*}(X)$,
one can consider
the space $\Omega^{*}_{G}(X)$ of $G$-invariant elements of $\CC[\lg]
\otimes\Omega^{*}(X)$. Elements of $\Omega^{*}_{G}(X)$ will be called
equivariant differential forms. They can be regarded as $G$-equivariant
maps $\mu:V\ra\mu(V)$ from $\lg$ to $\Omega(X)$. The $\ZZ$-grading of
$\Omega^{*}_{G}(X)$ is defined by assigning degree two to the elements of
$\lg$. $G$-equivariance implies
that the operator
\bea
d_{\lg}:&& \CC[\lg]\otimes\Omega^{*}(X)\ra\CC[\lg]\otimes\Omega^{*}(X)
\nonumber\\
  && (d_{\lg}\mu)(V)= d(\mu(V))- i(V_{X})(\mu(V))\;\;,
\eea
which raises the degree by one, squares to zero on $\Omega^{*}_{G}(X)$.
The $G$-equivariant
cohomology $H_{G}^{*}(X)$ of $X$ is then defined to be the cohomology of
$d_{g}$ acting on $\Omega^{*}_{G}(X)$. In analogy with the ordinary notions
of cohomology an equivariant differential form $\mu$
is said to be equivariantly closed (respectively equivariantly exact)
if $d_{\lg}\mu=0$ ($\mu = d_{\lg}\lambda$ for some
$\lambda\in\Omega_{G}^{*}(X)$).

It follows from these definitions that $H_{G}^{*}(X)$ coincides with the
ordinary cohomology if $G$ is the trivial group, and that the $G$-equivariant
cohomology of a point is the algebra of $G$-invariant polynomials on $\lg$,
$H_{G}^{*}(pt)=\CC[\lg]^{G}$.

Integration of equivariant differential forms can be defined as a map
from $\Omega^{*}_{G}(X)$ to $\CC[\lg]^{G}$ by
\be
(\int_{X}\mu)(V) = \int_{X}\mu(V)\;\;, \label{intedf}
\ee
where it is understood that the integral on the right hand side picks out
the top-form component $\mu(V)^{(n)}$ of $\mu(V)$ and the integral is
defined to be zero if $\mu$ contains no term of form-degree $n$.

A final thing worth pointing out is that
one can also generalize the standard notions of characteristic classes,
Chern-Weil homomorphism etc.\ to the equivariant case. The equivariant
Euler form of a vector bundle $E$ over $X$ to which the
action of $G$ lifts, with $\nabla$ a $G$-invariant connection on $E$, will
make a brief appearance later. It is defined in the standard way
by (\ref{6}), with the curvature $\On$ replaced by the
equivariant curvature
\be
\On^{\lg}(V) = \On + (L^{E}(V) -\nabla_{V})  \label{eqcu}
\ee
of $E$. Here $L^{E}(V)$ denotes the infinitesimal $G$-action on $E$ and
the term in brackets is also known as the equivariant moment of the
action - in analogy with the moment map of symplectic geometry.
It can be checked directly that
the differential form one obtains in this way
is equivariantly closed and the usual arguments carry
over to this case to establish that its cohomology class does not
depend on the choice of $G$-invariant connection $\nabla$.

For $E$ the tangent bundle of $X$ and $\nabla$ the
Levi-Civit\'a connection one finds that, as a consequence of
$\nabla_{V}W-\nabla_{W}V=[V,W]$ (no torsion),  the Riemannian
moment map is simply
$-(\nabla V)_{\mu\nu}= -(\nabla_{\mu} h_{\nu\lambda}V^{\lambda})$.
Note that this matrix is anti-symmetric because $V$ is Killing. Hence
the Riemannian equivariant curvature is simply
\be
\On^{so(n)}(V) = \On - \nabla V \;\;.\label{eqrcu}
\ee
The equivariant $\hat{{\rm A}}$-genus $\hat{{\rm A}}(\On^{so(n)})$
of the tangent bundle appears in the
localization theorem of \cite{anot} which we will discuss below.

This is as far as we will follow the general story. In our applications
we will be interested in situations where we have a single Killing vector
field $V_{X}\equiv V$ on $X$, perhaps corresponding to an action of
$G=S^{1}$. This means that we will be considering equivariant differential
forms $\mu(V)$ evaluated on a fixed $V\in\lg$. In those cases, nothing
is gained by carrying around $\CC[\lg]$ and we will simply be considering
the operator $d(V)=d - i(V)$ on $\Omega^{*}(X)$. By the Cartan formula,
the square of this operator is (minus) the Lie derivative $L(V)$ along $V$,
\be
(d(V))^{2}= (d-i(V))^{2}= -(i(V)d + di(V)) = -L(V)\;\;,
\ee
so that it leaves invariant the space $\Omega_{V}^{*}(X)$ of $V$-invariant
forms on $X$ (the kernel of $L(V)$) and squares to zero there.
Thus it makes sense to consider the cohomology of $d(V)$
on $\Omega^{*}_{V}(X)$ and we will call forms on $X$ satisfying
$d(V)\alpha=0$  $d(V)$-closed or equivariantly closed etc.
The final observation we will
need is that if a differential form is $d(V)$-exact, $\alpha=d(V)T$,
then its top-form
component is actually exact in the ordinary sense. This is obvious
because the $i(V)$-part of $d(V)$ lowers the form-degree by one
so there is no way that one can produce a top-form by acting with $i(V)$.

\subs{The Berline-Vergne and \dh\ localization formulae}

We now come to the localization formulae themselves. We will see that
integrals over $X$ of $d(V)$-closed forms localize to the zero locus
of the vector
field $V$. There are two simple ways of establishing this localization
and as each one has its merits we will present both of them. It requires
a little bit more work to determine the precise contribution of each
component of $X_{V}$ to the integral, and here we will only sketch the
required calculations and quote the result.

The essence of the localization theorems is the fact that equivariant
cohomology is determined by the fixed point locus of the $G$-action.
The first argument for localization essentially boils down to an explicit
proof of this fact at the level of differential forms. Namely, we will show
that any equivariantly closed
form $\a$, $d(V)\a=0$, is equivariantly exact away from the zero locus
$X_{V}=\{x\in X: V(x)=0\}$ of $V$. To see this, we will construct explicitly
a differential form $\beta$ on $X\setminus X_{V}$ satisfying $d(V)\beta
=\alpha$. As this implies that the top-form component of $\alpha$ is
exact, it then follows from Stokes theorem that the integral $\int_{X}\alpha$
only receives contributions from an arbitrarily small neighbourhood of
$X_{V}$ in $X$.

It is in the construction of $\beta$ that the condition enters that $V$
be a Killing vector. Using the invariant metric $h$ we can construct the
metric dual one-form $h(V)$. It satisfies $d(V)h(V) = d(h(V))-|V|^{2}$ and
(since $V$ is a Killing vector) $L(V)h(V)=0$. Away from $X_{V}$, the
zero-form part of $d(V)h(V)$ is non-zero and hence $d(V)h(V)$ is
invertible. Here the inverse of an inhomogeneous differential form with
non-zero scalar term is defined by analogy with the formula $(1+x)^{-1}
=\sum_{k} (-x)^{k}$. We now define the (inhomogeneous) differential form
$\Theta$ by
\be
\Theta =h(V)(d(V)h(V))^{-1}       \label{theta} \;\;.
\ee
It follows immediately that
\be
d(V)\Theta = 1\;\;,\;\;\;\;\;\;L(V)\Theta = 0\;\;.
\ee
Hence, 1 is equivariantly exact away from $X_{V}$ and we can choose $\beta =
\Theta\alpha$,
\be
d(V)\alpha=0 \;\;\;\;\;\;\Ra \;\;\;\;\;\;
\alpha =  (d(V)\Theta)\alpha =  d(V)(\Theta\alpha)
  \;\;\mbox{on}\;\;X\setminus X_{V}\;\;,
\ee
so that any equivariantly closed form is equivariantly exact away from $X_{V}$.
In particular, the top-form component of an equivariantly closed form
is exact away from $X_{V}$.

Let us see what this amounts to to in the example (\ref{spins}) of the
introduction to this section (cf.\ also (\ref{spins0},\ref{spins3})).
Using the symplectic form $\omega = \sin\theta d\theta d\f$,
the Hamiltonian vector field corresponding to $\cos\theta$ is $V=\del_{\f}$.
This vector field generates an isometry of the standard metric $d\theta^{2}
+ \sin^{2}\theta d\f^{2}$ on $S^{2}$, and the corresponding $\Theta$ is
$\Theta = -d\f$. This form is, as expected, ill defined at the two poles of
$S^{2}$. Now the integral (\ref{spins}) can be written as
\be
\int_{S^{2}}\omega \ex{itf} = (it)^{-1}\int_{S^{2}}d(\ex{itf}d\f)\;\;,
\label{spins2}
\ee
thus receiving contributions only from the critical points $\theta=0,\pi$,
the endpoints of the integration range for $\cos\theta$, in agreement with
the explicit evaluation.

Alternatively, to establish localization, one can use the fact that the
integral $\int_{X}\alpha$ of a $d(V)$-closed form only depends on its
cohomology class $[\alpha]$ to evaluate the integral using a particularly
suitable representative of $[\alpha]$ making the localization manifest.
Again with the help of an invariant metric, such a representative can be
constructed. Consider the inhomogeneous form
\be
\alpha_{t}= \alpha\ex{td(V) h(V)}\;\;,\;\;\;\;\;\;t\in\RR\;\;.
\label{alphat}
\ee
Clearly, $\alpha_{t}$ is cohomologous to $\alpha$ for all $t$. Equally
clearly, as $d(V)h(V) = -|V|^{2} + d(h(V))$, the form $\alpha_{t}$ is
increasingly sharply Gaussian peaked around $X_{V}\ss X$ as $t\ra\infty$.
Hence, evaluating the integral as
\be
\int_{X}\alpha = \lim_{t\ra\infty}\int_{X}\alpha_{t}  \label{intalpha}
\ee
reestablishes the localization to $X_{V}$.

Perhaps it is good to point out here that there is nothing particularly
unique about the choice $\a_{t}$ to localize the integral of $\alpha$
over $X$. Indeed, instead of $h(V)$  one could choose some other
$L(V)$-invariant form $\beta$ in the exponent of (\ref{alphat}) and try
to evaluate $\int_{X}$ as some limit of
\be
\int_{X}\alpha = \int_{X}\alpha\ex{td(V)\beta}\;\;. \label{intbeta}
\ee
This can potentially localize the integral to something other than $X_{V}$
(we will see an example of this in the context of path integrals later on),
and so lead to a seemingly rather different expression for the integral.
In principle this argument for localization could also work
without the assumption
that $V$ is a Killing vector, but it appears to be difficult to make any
general statements in that case.
However, as everything in sight is clearly $d(V)$-closed, it is possible
to reduce the resulting expression further to $X_{V}$ by applying the above
localization arguments once more, now to the localized expression.

One
{\em caveat} to keep in mind is that the above statements require some
qualifications when the manifold $X$ is not compact. In that case,
$t$-independence of the rhs of (\ref{intbeta}) is only ensured if the
asymptotic behaviour of $\alpha$ is not changed by the replacement
$\alpha\ra\alpha\exp d(V)\beta$. For a version of the \dh\ formula
for non-compact manifolds see \cite{epwu}. The extension to
non-compact groups is also not immediate. For example \cite{bgv},
consider the nowhere vanishing vector field
\be
V = (1 + \trac{1}{2} \sin x)\del_{y}
\ee
on the two-torus ($x,y\in \RR/2\pi\ZZ$). Then one can easily check
that
\be
\a(V)=-\trac{1}{2}(7 \cos x + \sin 2x) + (1-4\sin x) dx dy
\ee
is equivariantly closed. Its integral over the torus, on the other
hand, is equal to $(2\pi)^{2}$, so that the two-form component of
$\a(V)$ cannot possibly be exact.

Another ambiguity can arise when $\alpha$ is equivariantly
closed with resepct to two vector fields $V$ and $W$, $d(V)\alpha =
d(W)\alpha = 0$. In this case one can get seemingly different expressions
by localizing to either $X_{V}$ or $X_{W}$ or $X_{V}\cap X_{W}$.
Of course, in whichever way one chooses to evaluate the
integral, the results are guaranteed to agree, even if not manifestly so.

Having established that the integral will indeed localize,
to obtain a localization formula one needs to determine the
contributions to the integral from the connected components of $X_{V}$.
This is most readily done when $X_{V}$ consists of isolated points
$x_{k}$, $V(x_{k})=0$. In that case the Lie derivative $L(V)$ induces
invertible linear transformations $L_{V}(x_{k})$
on the tangent spaces $T_{x_{k}}X$. One can introduce local coordinates
in a $G$-invariant neighbourhood of these points such that the localized
integrals become essentially Gaussian. Then one finds that the contribution
of $x_{k}$ to the integral is just $(-2\pi)^{l}\det(L_{V}(x_{k}))^{-1/2}$.
Restricting the equivariantly closed form $\alpha$ to $X_{V}$, only the
scalar part $\alpha^{(0)}(x_{k})$ survives and one finds the Berline-Vergne
localization formula
\be
d(V)\alpha = 0 \;\;\;\;\;\;\Ra\;\;\;\;\;\;
 \int_{X}\alpha = (-2\pi)^{l}\sum_{x_{k}\in X_{V}}
\det(L_{V}(x_{k}))^{-1/2} \alpha^{(0)}(x_{k}) \label{bv}\;\;.
\ee
If $X_{V}$ has components of non-zero dimension, things become a little bit
more complicated. In that case, in the above discussion the tangent space
$T_{x_{k}}X$ has to be replaced by the normal bundle $N$ to $X_{V}$ in $X$ and
$L_{V}(x_{k})$ by its equivariant curvature (\ref{eqcu}). As $V=0$ on $X_{V}$,
the scalar part of (\ref{eqcu}) reduces to $L^{N}(V)$, thought of as an
endomorphism of $N$. The localization formula one obtains in this case is then
\be
d(V)\alpha = 0 \;\;\;\;\;\;\Ra\;\;\;\;\;\; \int_{X}\alpha =
    \int_{X_{V}}\det(L^{N}(V)+\On)^{-1/2}\alpha|_{X_{V}}\label{bv2}
\ee
(a sum over the components of $X_{V}$ being understood). As pointed out before,
the form appearing in the denominator of this formula can be regarded as the
equivariant Euler form of the normal bundle $N$. We refer to \cite{bgv} and
\cite{audin}
for details and some of the beautiful applications of this formula.

We will now consider a particular case of the Berline-Vergne formula. Namely,
let $(X,\omega)$ be a symplectic manifold and assume that there is a
Hamiltonian action of $G$ on $X$. This means that the vector fields $V_{X}$
generating the action of $G$ on $X$ are Hamiltonian.
That is, there is a moment map $\mu:X\ra\lg^{*}$
such that $\langle \mu, V\rangle \equiv \mu(V)\in C^{\infty}(X)$
generates the action of $G$ on $X$
via the Hamiltonian vector fields $V_{X}$.
Since $\omega$ is closed, the defining equation
\be
i(V_{X})\omega = d(\mu(V))
\ee
implies that $\mu + \omega$ is an equivariantly closed form,
\be
d_{\lg}(\mu + \omega)(V) = d(\mu(V)) - i(V_{X})\omega = 0\;\;.
\ee
In fact, finding an equivariantly closed extension of $\omega$ is equivalent
to finding a moment map for the $G$-action \cite{ab}.

If $\omega$ is integral, it can be regarded as the curvature of a line bundle
$E$ on $(M,\omega)$, the prequantum line bundle of geometric quantization
\cite{wood}. In that case, the equivariant curvature form as defined in
(\ref{eqcu}) agrees with the equivariant extension of $\omega$ given above.
Indeed in geometric quantization it is well known that the `prequantum
operator' $L^{E}(V)$ can be realized as  $L^{E}(V) = \nabla_{V} + \mu(V)$.

To produce from this
an equivariantly closed form on $X$ which we can integrate (which has a
top-form
component), we consider the form $\exp it(\mu(V) + \omega)$ which is
annihilated by $d(V)$. Its integral is
\bea
\int_{X}\ex{it(\mu(V) + \omega)} &=&
        \int_{X}\frac{(it\omega)^{l}}{l!}\ex{it\mu(V)}\nonumber\\
      &=&(it)^{l}\int_{X}dx(\omega)\ex{it\mu(V)}\;\;.\label{sympint}
\eea
Assuming that the zeros of $V_{X}$ are isolated,
we can apply the Berline-Vergne formula (\ref{bv}) to obtain
\be
\int_{X} dx(\omega)\ex{it\mu(V)} = (it)^{-l}(-2\pi)^{l}\sum_{x_{k}\in X_{V}}
\det(L_{V}(x_{k}))^{-1/2}\ex{it\mu(V)(x_{k})}\;\;. \label{dhint}
\ee
As the critical points of $\mu(V)$ are precisely the zeros of $V_{X}$ ($\omega$
is non-degenerate) this is the \dh\ formula (\ref{sp3}) if $V$ is a
generator of a circle action for some $U(1)\ss G$. For an example
we refer back to the classical spin system discussed in the introduction
to this section. Of course there is  a corresponding
generalization of the \dh\ formula for the case that the zeros of $V_{X}$
are not isolated which follows from applying (\ref{bv2}) to
(\ref{sympint}).

\subsection{Localization Formulae for Phase Space Path Integrals}

In quantum mechanics there are not too many
path integrals that can be evaluated explicitly and exactly, while
the semi-classical approximation can usually be obtained quite readily.
It is therefore of obvious interest
to investigate if there is some path integral
analogue of the \dh\ and Berline-Vergne formulae.

One class of quantum mechanics models for which the \dh\ formula clearly
seems to make sense is $N=\frac{1}{2}$ \sqm. We will come back to this
in section 3.3.
What one would really like, however, is to have some version of the
equivariant localization formulae available which can be applied to
non-supersymmetric models and when the partition
functions cannot be calculated directly (or only with difficulty)
by some other means. One non-trivial field theoretic example in which
(a non-Abelian version of) the \dht\ has been applied successfully is
two-dimensional Yang-Mills theory \cite{ew2d}. This theory has a
natural symplectic interpretation because the space of gauge
fields in two dimensions is symplectic and the Yang-Mills action
is just the square of the moment map generating gauge transformations
of the gauge fields. Hence one is in principle in the right framework
to apply equivariant localization. We will sketch an Abelian version
of the localization formula for the square of the moment map in
section 3.3.

A large class of examples where one also has an
underlying equivariant cohomology which could be
responsible for localization is provided by
phase space path integrals, i.e.\ the direct loop space analogues
of the lhs of (\ref{sp3}).
Now phase space path integrals are of course
notoriously awkward objects, and it is for this reason that the localization
formulae we will obtain in this way should not be regarded as definite
predictions but rather as suggestions for what kind of results to expect.
Because of the lack of rigour that goes into the derivation of these
localization formulae it is perhaps surprising that nevertheless some of the
results that have been obtained are not only conceptually
interesting but also physically reasonable.

\subs{Equivariant cohomology for phase space path integrals}

In this section we choose our manifold $X$ to be the loop space
$X=LM$ (see section 2.2) of a finite-dimensional symplectic manifold
$(M,\omega)$, the phase space of a classical system.
$LM$ is also a symplectic manifold in the sense
that $\hat{\omega}$ (cf.\ (\ref{hatf})) defines a closed and non-degnerate
two-form on $LM$. We denote by $S= S[x]$ the phase space action and by $H$
its Hamiltonian. Thus, if we denote by $\theta$ a local symplectic potential
for $\omega$, $d\theta = \omega$, then the action is essentially of the form
\be
S[x]=\tint{T} (\theta_{\mu}(x(t))\dx^{\mu}(t) - H(x(t))\;\;.
\ee
Of course, when
$\omega$ is not globally exact this has to be suitably defined by regarding
the first term as a Wess-Zumino like term for $\omega$, but this is a
standard procedure which will be tacitly adopted in the following.

Consider now the phase space path integral
\be
Z(T) = \int_{LM} [dx(\hat{\omega})] \ex{\frac{i}{\hbar}S[x]}\;\;.\label{z1}
\ee
As in (\ref{sympint}), we will lift the symplectic form into the exponent
which will then take the form $S + \hat{\omega}$. Clearly, therefore, the
integrand is equivariantly closed with respect to the flow generated on
the loop space by the Hamiltonian vector field $V_{S}$ of the action $S$,
\be
d(V_{S})(S + \hat{\omega}) = 0 \;\;. \label{dvs}
\ee
The zeros of this vector field are precisely the critical points of the
action, i.e.\ the classical solutions. This is the reason why this set-up
has the potential to produce WKB-like localization formulae \cite{csqm,bkn}.

Let us make the above formulae a little bit more explicit. First of all,
if we denote by $H$ the Hamiltonian corresponding to $S$, then the
components of $V_{S}$ can be written as
\be
V_{S}^{\mu}(x(t)) = \dx^{\mu}(t) - \omega^{\mu\nu}(x(t))\del_{\nu}H(x(t))
\;\;, \label{vs}
\ee
so that clearly $V_{S}(x)=0$ when $x(t)$ satisfies the classical equations
of motion. The first term is just the canonical vector field $V=\dx$ on
$LM$, while the second is the Hamiltonian vector field $V_{H}$ of $H$ on $M$,
regarded as a vector field on $LM$.

Using also the trick of section 2 (\ref{psidx}) to replace the loop space
one-forms $dx(t)$ by anticommuting variables $\p(t)$, we can write the
partition function (\ref{z1}) as
\be
Z(T)= \int_{LM}[dx][d\p]\ex{\frac{i}{\hbar}(S + \hat{\omega}(\p))}\label{z2}
\;\;.
\ee
Then the fact (\ref{dvs}) that $S+\hat{\omega}$ is equivariantly closed
translates into the statement that the augmented action $S[x,\p]$
appearing in (\ref{z2}) is invariant under the supersymmetry
\be
\d x^{\mu} = \p^{\mu}\;\;,\;\;\;\;\;\;\d\p^{\mu} = -V_{S}^{\mu}\;\;\;\;\;\;
\Ra\;\;\;\;\;\;\d (S+\hat{\omega}(\p)) = 0\;\;.
\ee
We will mostly set $\hbar=1$ in the following.

\subs{Localization formulae}

The above is of course not yet sufficient to establish that (formally) the path
integral localizes. As we know from the previous section, at the very least
we need a metric on $LM$ with respect to which $V_{S}$ is Killing. To
investigate when this is the case, we choose a metric $\hat{g}$ on $LM$ of
the ultra-local form (\ref{46}). As any such metric is invariant under
$V=\dx$, $L(V)\hat{g}=0$, the condition $L(V_{S})\hat{g}=0$ reduces to
the condition that the Hamiltonian vector field $V_{H}$ of $H$ on $M$ be
a Killing vector of the metric $g$,
\be
L(V_{S})\hat{g}=0 \;\;\;\;\;\;\Leftrightarrow \;\;\;\;\;\;L(V_{H})g = 0\;\;.
\ee
This is exactly the same condition we encountered in the
finite-dimensional case. For these examples, then, for which there
are localization formulae for the classical  partition function,
can we hope to find analogous formulae for the quantum partition function
as well.

We now proceed as in the finite-dimensional case and localize the integral
by adding some appropriate $d(V_{S})$-exact term $d(V_{S})\beta$ with
$L(V_{S})\beta=0$ to the action. First of
all we need to show that this does not change the value of the path integral.
In the finite dimensional case this (or the $t$-independence of integrals
like (\ref{intalpha},\ref{intbeta})) relied, at least implicitly, on
Stokes' theorem which is not directly
available for integrals over $LM$. However,
instead of that one has a kind of Stokes' theorem in the form of a
Ward identity associated with the
supersymmetry $\d$ (or, in more elementary terms, a change of variables
argument \cite{bkn}) to reach the desired conclusion provided that the
supersymmetry is non-anomalous. We will assume this and proceed
with fingers crossed.

The obvious candidate for $\beta$ is, as in the finite dimensional case,
the metric dual one-form $\hat{g}(V_{S})$ of the Hamiltonian vector field
$V_{S}$ itself \cite{bkn} which, by our assumption $L(V_{H})g=0$, satisfies
$L(V_{S})\hat{g}(V_{S})=0$. This choice will formally lead to a localization
onto the critical points of $S$ and hence onto classical trajectories
$x_{c}(t)$.
If these are isolated and non-degenerate this will lead to the exact analogue
of the \dh\ formula (\ref{sp3}), namely
\be
Z(T) \sim \sum_{x_{c}(t)}
\det[L_{S}(x_{c})]^{-1/2}\ex{\frac{i}{\hbar}S[x_{c}]}\;\;. \label{pidh}
\ee
Here the determinant is essentially the functional determinant of the Jacobi
operator (the Hessian of the action).
%\be
%L_{S}(x_{c}) = \left( \d^{\mu}_{\nu}\trac{d}{dt} -
%\del_{\nu}(\omega^{\mu\lambda}(x_{c}(t))\del_{\lambda}H(x_{c}(t)))\right)
%\;\;.
%\ee
This formula can be interpreted as saying that the WKB
approximation to the path integral is exact, the only difference to the
usual WKB approximation being that here one is to sum over all critical
points of the action and not just the local minima.

The validity of
(\ref{pidh}) has been well investigated in the path integral literature
(although more for configuration space path integrals),
and we have little to add to that discussion. We only want to point out
that in those examples we know of where the semi-classical approximation
is known to be correct, the assumptions that went into the derivation of
(\ref{pidh}) here, like the existence of an invariant phase space metric,
are indeed satisfied (see e.g.\ \cite{schulman,dowker,picken} for the
propagator of a particle moving on a group manifold).

Nevertheless, (\ref{pidh}) should be taken with a grain of salt. First of
all, as (\ref{pidh}) is a sum over classical periodic trajectories with
period $T$, these will typically either occur only at the critical
points of the Hamiltonian (in which case (\ref{pidh}) resembles even more
closely the classical \dh\ formula) or on a non-zero dimensional
submanifold of $M$ (in which case one should really use the degenerate
version of the \dh\ formula). Moreover, more care has to be exercised
(already in the finite dimensional case) when one is dealing with non-compact
phase spaces, which is the rule in systems of physical interest.

However, even if this formula or its obvious modifications are not correct
as they stand, it
would be interesting to uncover the reason for that. At the very least this
could then provide one with a systematic geometric method for analyzing
corrections to the WKB approximation. We are not aware of any work that
has been done in this direction yet.

One of the strengths
of the present setting is the ability to produce other, different,
localization formulae, with perhaps different ranges of validity,
by exploiting the flexibility in the choice of $\beta$. For example,
one can easily convince oneself that the two terms of $\hat{g}(V_{S})$
(\ref{vs}), namely $\hat{g}(V=\dx)$ and $\widehat{g(V_{H})}$, are separately
invariant under $V_{S}$ so that one can choose $\beta$ to be a general
linear combination of them \cite{an1,an2},
\be
\beta = s \hat{g}(\dx) - r \widehat{g(V_{H})}\;\;.
\ee
For $s=r$ this reduces to what we did above. But, by choosing e.g.\
$r=0$ and $s\ra\infty$ one can localize the path integral (\ref{z2})
to an integral over the classical phase space $M$. Alternatively,
by choosing $s=0$ and $r\ra\infty$, one can obtain an expression for
$Z(T)$ in terms of the critical points of the Hamiltonian. We will only
consider the first of these possibilities here, as it leads to a nice
expression in terms of equivariant characteristic classes. The action one
obtains in this way is
\bea
S^{s}[x,\p] &=& S[x] + \hat{\omega}(\p) + s d(V_{S})\hat{g}(\dx)\label{ss}\\
&=& S[x] + \hat{\omega}(\p) + s d(\hat{g}(\dx))(\p) - s \hat{g}(\dx,\dx)
                     + s \hat{g}(\dx,V_{H})\nonumber\\
&=& S[x] + \hat{\omega}(\p) +
s \tint{T}  \p(t)\nabla_{t}\p(t) + g(\dx(t),\dx(t)) - g(\dx(t),V_{H}(x(t)))
\nonumber
\eea
As in (\ref{51}), $\nabla_{t}$ denotes the covariant derivative along the
loop $x(t)$ induced by the Riemannian connection $\nabla$ on $M$. This
action can be evaluated as in \sqm. One possibility is to expand all fields
in Fourier modes and to scale the non-constant modes by $s^{-1/2}$ so as
to eliminate positive powers of $s$ from the action.
Then the limit $s\ra\infty$
can be taken with impunity, the integration over the non-constant modes
gives rise to a ratio of determinants and in the exponent one is left with
the contribution from the constant modes, namely $T(\omega(\p)-H)$. The
result one obtains is
\be
Z(T)= \int_{M}\! dx\,d\p\,
\det\left[\frac{\frac{T}{2}(\On + \nabla V_{H})}%
{\sinh[\frac{T}{2}(\On + \nabla V_{H})]}\right]^{1/2}
\ex{iT(\omega(\p)-H)}\label{nt1}
\ee
(see (\ref{ahat1}) and (\ref{ahat2}) in section 3.3 for a brief explanation
of the appearance of this particular determinant in supersymmetric path
integrals).
This is the Niemi-Tirkkonen localization formula \cite{anot}.
In this expression we recognize the $d(-V_{H})$-equivariant Riemannian
curvature (\ref{eqrcu}) and its $\hat{{\rm A}}$-genus (the occurrence of minus
signs at awkward places is unavoidable in symplectic geometry \ldots).
The exponent can also
be interpreted in equivariant terms. Namely, as we have seen in the
discussion of the \dht, as the $(-V_{H})$-equivariant extension
of the symplectic form $\omega$. Roughly speaking
(modulo the fact that $\omega$ need not be integral) the exponential
can then be regarded as an equivariant Chern character and the result
can then be written more succinctly and elegantly in the notation of
(\ref{intedf}) as
\be
Z(T) =\left(\int_{M} \hat{{\rm A}}(T\On^{so(n)})
{\rm Ch}(T\omega^{so(n)})\right)(-V_{H})\;\;.\label{nt2}
\ee
This differs from the classical partition function for the dynamical
system described by $H$ by the $\hat{{\rm A}}$-term which can be thought
of as encoding the information due to the quantum fluctuations.

As the integrand is clearly $d(-V_{H})$-closed, this integral can be
localized further to an integral over the critical points of $H$
provided that the zero locus of $V_{H}$ is non-degenerate.
The possible
advantage of the present formula is that no such assumption appears
to have been nececssary in the derivation of (\ref{nt1}) which thus
potentially applies to examples where the ordinary WKB approximation
breaks down (e.g.\ when classical paths coalesce).

This possibility has been analyzed in \cite{lykken} for some
simple one-dimensional quantum mechanics examples,
the harmonic oscillator and the `hydrogen atom', i.e.\ a particle
moving in a $1/|x|$-potential. The latter example, in particular,
is interesting because there the classical paths are known to coalesce
so that the traditional WKB formula is not applicable.
The results obtained in \cite{lykken} as well as some further considerations
in \cite{semenoff} illustrate the potential usefulness of these
generalized localization formulae.  However, in order to establish to what
extent formulae like the above are trustworthy and can be turned into
reliable calculational tools, other (higher dimensional) examples will
need to be worked out.

\subsection{Other Examples and Applications - an Overview}

In this section we shall sketch some of the other applications of equivariant
localization and, in particular, the \dh\ formula.
Most prominent among these are applications to \sqm\ and index
theorems.

\subs{Supersymmetric quantum mechanics and index theorems}

As mentioned in the introduction to this section, the first infinite
dimensional application of the \dht\ is due to Atiyah and Witten \cite{atew}
(see also \cite{bismut2} and \cite{jones})
who applied it to the loop space $LM$ of a Riemannian manifold and
$N=\frac{1}{2}$
\sqm. This is the model which represents the index of the Dirac operator
\cite{ag1,ag2,fw} in
the same way that
the $N=1$ model we discussed in section 2.2  represents the index
of the de Rham complex. Just like the \mq\ formalism
provides the appropriate framework for understanding the topological and
localization properties of $N=1$ (de Rham) \sqm, those of $N=\frac{1}{2}$
(Dirac) \sqm\ find their natural explanation within the framework of
loop space equivariant localization. And, just as in the case of the \mq\
formalism, this way of looking at the $N=\frac{1}{2}$
models provides some additional
insights and flexibility in the evaluation of the path integrals
\cite{an2}.

Roughly speaking, the action of the $N=\frac{1}{2}$ model can be obtained
from that of the $N=1$ model (\ref{51}) by setting $\p=\pb$. Then the
four-fermi curvature term drops out by the Bianchi identity
$R_{\mu[\nu\rho\sigma]}=0$ and (with a suitable rescaling of the fermi
fields) one is left with
\be
S_{M}[x,\p]=
\tint{\beta}g_{\m\n}\left(\dx^{\m}(t)\dx^{\n}(t)-\p^{\m}(t)\nt\p^{\n}(t)\right)
  \label{dirac}\;\;.
\ee
This action has the supersymmetry
\be
\d x^{\mu}(t)=\p^{\mu}(t)\;\;,\;\;\;\;\;\;\d\p^{\mu}(t)=-\dx^{\mu}(t)
\;\;\;\;\;\;
\Ra\;\;\;\;\;\;\d S_{M}[x,\p]=0\;\;,
\ee
which we can now recognize immediately as the action of the equivariant
exterior derivative $d(V=\dx)= d - i(\dx)$ on $LM$. Hence the action
of $N=\frac{1}{2}$ \sqm\ defines an equivariant differential form on $LM$ and
on the basis of the general arguments we expect its integral to localize
to an integral over $M$, the zero locus of $\dx$. This is of course well
known to be the case \cite{ag1,ag2,fw}.

One new feature of this model is
that the action is actually equivariantly exact,
\be
S_{M}=d(V)\hat{g}(V)\;\;\;\;\;\;\Lra\;\;\;\;\;\;
S_{M}[x,\p] = \delta \tint{\beta} g_{\mu\nu}\dx^{\mu}(t)\p^{\nu}(t)\;\;,
\label{sdg}
\ee
so that the partition function will not depend on the coefficient of the
action (and thus manifestly localizes onto $\dx=0$).
As one can think of this coefficient as $\hbar$, this is another
way of seeing that the semi-classical approximation to this model is exact.
Nevertheless one can of course not simply set the coefficient to zero
as the integral is ill-defined in this case (infinity from the $x$-integral
times zero from the $\p$-integral).

There are now at least three different ways of calculating the partition
function
\be
Z[M]=\int\![dx][d\p]
\ex{i\tint{\beta}g_{\m\n}\left(\dx^{\m}(t)\dx^{\n}(t)-
\p^{\m}(t)\nt\p^{\n}(t)\right)}\;\;.
\ee
The traditional method is to make use of the presumed
$\beta$-independence of the theory to  evaluate $Z$ in the limit $\beta\ra 0$
using a normal coordinate expansion \cite{ag1,ag2,fw}. One can also apply
directly the Berline-Vergne formula (\ref{bv2}), valid when the zero locus
is not zero-dimensional. In that case one has to calculate the
equivariant Euler  form of the normal bundle $N$ to $M$ in $LM$ (this is
the bundle spanned by the non-constant modes) and evaluate the determinant
appearing in (\ref{bv2}) using e.g.\ a zeta-function prescription. This
was the approach adopted in \cite{atew}. Finally, one can scale the
action (\ref{sdg}) by some parameter $s$, scale the non-constant modes of
$x$ and $\p$ by $s^{-1/2}$ and take the limit $s\ra\infty$.
The remaining integral is then Gaussian. In whichever way one proceeds
one obtains the result
\be
Z[M] = \int_{M} \hat{{\rm A}}(M) \;\;,
\ee
which is the index of the Dirac operator on $M$ if $M$ is a spin-manifold.

The reason for the ubiquitous appearance of the $\hat{{\rm A}}$-character
in \sqm\ path integrals is, that it arises whenever one calculates the
determinant of a first-order differential operator on the circle.
Let $D_{a}=\del_{t} + a$ be such an operator. We can without loss
of generality assume that $a$ is constant as the non-constant modes
of $a$ could always be removed by conjugating the entire operator by
$\exp if(t)$ for some function $f$, an operation that does not change
the determinant of $D_{a}$. Acting on periodic functions on $S^{1}$,
the eigenvalues of $D_{a}$ are $a+ 2\pi in$ for $n\in\ZZ$ (and $n\neq 0$ if
one excludes the constant mode). The determinant of $D_{a}$ is now
formally defined as the product of the eigenvalues. Of course, this
requires some regularization and a suitable prescription is zeta-function
regularization. Then the determinant $\det{}'D_{a}$ over the non-constant
modes can be written as
\bea
\det{}'D_{a} &=& \prod_{n\neq 0} (2\pi in+a) =
\prod_{n>0} (a^{2}+(2\pi n)^{2})\nonumber\\
&=& (\prod_{n>0} (2\pi n)^{2}) \prod_{n>0} (1+a^{2}/(2\pi n)^{2})=
\prod_{n>0} (1+a^{2}/(2\pi n)^{2})\;\;,\label{ahat1}
\eea
as formally the infinite prefactor is equal to one by zeta-function
regularization.
The function defined by the infinite product in the last line
has zeros for $a\in 2\pi i\ZZ$,
$a\neq 0$, and is equal to one at $a=0$. It is nothing other than the
function $(\sinh  a/2)/(a/2)$, so that we can write
\be
\det{}'D_{a} = \frac{\sinh a/2}{a/2} = \hat{{\rm A}}(a)^{-1} \;\;,\label{ahat2}
\ee
where we define the function $\hat{{\rm A}}(x) = (x/2)/\sinh(x/2)$.

This can of course be generalized in various ways to include the coupling
of fermions to a gauge field on a vector bundle $E$, yielding the
result
\be
Z[M,E]=\int_{M}{\rm Ch}(E)\hat{{\rm A}}(M)\;\;.
\ee
The actual calculations involved in these three different approaches
are practically identical and it is really only in the interpretation of
what one is doing that they differ. However, there seems to be one instance
where method three appears to be superior to method one \cite{an3},
namely when one is dealing with odd-dimensional open-space index theorems
like that of Callias and Bott. In that case it is simply not true
that the partition function is independent of $\beta$ (the index is
obtained for $\beta\ra\infty$) so that one cannot simplify matters by
going to the $\beta\ra 0$ limit. In \cite{an3} it was shown that
the correct result can be obtained by introducing the parameter $s$ and
calculating the partition function as $s\ra\infty$.

\subs{A localization formula for the square of the moment map}

In \cite{ew2d}, Witten introduced a new non-Abelian localization formula
for finite dimensional integrals and applied it to path integrals. He
was able in this way to deduce the intersection
numbers of the moduli space of flat connections on a two-surface $\Sigma$
from the solution of Yang-Mills theory on $\Sigma$. Subsequently, certain cases
of this localization formula have been derived rigorously by Jeffrey and Kirwan
\cite{ljfk} and Wu \cite{wu2}.

In its simplest version, this formula applies to integrals
over symplectic manifolds of the form
\be
Z(\epsilon) =  (2\pi\epsilon)^{-m/2}\int_{X}\ex{-I/2\epsilon + \omega}\;\;,
\label{ze1}
\ee
where $I=(\mu,\mu)$ is the square with resepct to some invariant
scalar product on $\lg^{*}$ of the moment map $\mu:X\ra\lg^{*}$
of a Hamiltonian $G$-action on $M$ and $\dim G = m$.
By introducing an auxiliary integral over $\lg$, (\ref{ze1}) can be put
into a form very similar to the \dh\ integral (\ref{sp3},\ref{dhint}),
namely
\be
Z(\epsilon) = (2\pi)^{-m}\int_{\lg}\! d\f \ex{-\epsilon(\f,\f)/2}
\int_{M} \ex{\omega +i(\mu,\f)}\label{ze2}\;\;.
\ee
Here we recognize again the equivariant extension of the symplectic form
(the factor of $i$ is irrelevant).
Hence, as the integrand is equivariantly closed, one can again attempt
to localize it by adding a $d_{\lg}$-exact term $d_{\lg}\beta$ to the
exponent, where $\beta\in\Omega^{*}_{G}(X)$. A reasonably canonical choice
for $\beta$ is $\beta= J(dI)/2$, where $J$ is some positive
$G$-invariant almost complex structure on $X$. In that case it can be shown
that the integral localizes to the critical points of $I$. These are
either critical points of $\mu$, as in the \dh\ formula, or zeros of $\mu$.
The latter are a new feature of this localization theorem and are interesting
for a number of reasons. For one, as the absolute minima of $I$ they give
the dominant contribution to the integral in the `weak coupling' limit
$\epsilon\ra 0$, the contributions form the other points being roughly
of order $\exp(-1/\epsilon)$. Furthermore, the contribution from $\mu=0$
is, by $G$-equivariance, related to the Marsden-Weinstein reduced phase
space (or symplectic quotient) $X//G = \mu^{-1}(0)/G$. Hence,
integrals like (\ref{ze1}) can detect the cohomology of $X//G$.

This is
of interest in 2d Yang-Mills theory, where $X=\cal A$ is the space of gauge
potentials $A$, and $\mu(A)=F_{A}$ is the curvature of $A$. The Yang-Mills
action is therefore just the square of the moment map and the
symplectic quotient is the moduli space of flat connections.

In general, the contribution from the non-minimal critical points
of $I$ can be quite complicated, even for $G$ Abelian.
In that case the integral over $M$ in (\ref{ze2}) can also be evaluated
via the \dh\ formula and a comparison of the expressions
one obtains by following either route has been performed
in \cite{wu2}. To illustrate the complicated structure that arises,
the example of the spin system (\ref{spins},\ref{spins2})
will suffice. In that
case we have to study the integral
\bea
Z(\epsilon)& =& (2\pi)^{-1}\int_{-\infty}^{\infty}\!d\f
\ex{-\epsilon\f^{2}/2}
                \int_{X}\ex{\omega + i\f(\cos\theta + a)}\nonumber\\
&=& (2\pi/\epsilon)^{1/2}\int_{-1}^{1}\!dx\ex{-(x+a)^{2}/2\epsilon}\;\;.
\label{spins3}
\eea
For $|a|<1$ this can be written as
\be
Z(\epsilon) =2\pi( 1 - I_{+}-I_{-})\;\;,
\ee
where
\be
I_{\pm} = \pm(2\pi\epsilon)^{-1/2}\int_{\pm 1}^{\pm\infty}dx
    \ex{-(x+a)^{2}/2\epsilon}\;\;.
\ee
These terms correspond to the contributions from the three critical points
of $I=\mu^{2}$, the absolute minimum at $\cos\theta=-a$ contributing the
simple first term, and the other two contributions coming from the
critical points of $\mu$ at $\theta=0$ and $\theta=\pi$. The appearance
of the error function in this example is in marked contrast with the
elementary functions that appear as the contributions from the critical
points in the \dh\ formula. The above integral should be
compared with the (closely related) integral (\ref{spins0}) which we
discussed in the context of the \mq\ formalism.
There is much more that should be said about
these localization formulae, but for this we refer to \cite{ew2d}.

\subs{Character Formulae and other applications}

In this section we will just mention some other applications of the
\dh\ formula and other equivariant localization theorems in the
physics literature. We have already mentioned the path integral
derivation of the Weyl character formula by Stone
\cite{stone} and the application of the \dh\ theorem to
the particle on a group manifold by Picken \cite{picken}
(recovering old results by Schulman \cite{schulman} and
Dowker \cite{dowker} in this way).
In a similar
setting, the action invariant of Weinstein \cite{Weinstein}, an
invariant probing the first cohomology group of the symplectomorphism
group of a symplectic manifold, has been related to
what is known as Chern-Simons quantum mechanics using the \dh\ formula
in \cite{csqm}.
The derivation of Stone has been generalized by Perret \cite{perret}
to the Weyl-Kac character formula for Kac-Moody algebras.

Because
of their classical properties, coherent states are particularly well
suited for studying semi-classical properties of quantum systems,
and in \cite{rajeev} the predictions of the \dh\ theorem have been
verified for coherent state path integrals associated with $SU(2)$
and $SU(1,1)$. A very careful analysis of the WKB approximation for
these coherent state path integrals has recently been performed in
\cite{fks}.

It is a longstanding conjecture that the quantum theories of
classically integrable systems are given approximately by their
semi-classical approximation. An application of the finite-dimensional
\dh\ formula to certain integrable models can be found in \cite{franc}
and some suggestive formulae for the phase space path integrals of
integrable models have been obtained in \cite{anin}.

Finally, we have recently found \cite{btloc}
that a localization formula of Bismut
\cite{bismut3} for equivariant K\"ahler geometry has a field theoretic
realization in the $G/G$ gauged Wess-Zumino-Witten model and can be
used to shed some light on and
give an alternative derivation to \cite{btver} of the
Verlinde formula from the $G/G$ model.

\section{Gauge Invariance and Diagonalization - the Weyl Integral Formula}

In this section we discuss a technique for solving or simplifying path
integrals which is quite different in spirit to those we encoutered in
sections 2 and 3. Both the \mq\ formalism and the \dh\ and Berline-Vergne
localization theorems are fundamentally
cohomological in nature and can be understood in
terms of a supersymmetry allowing one to deform the integrand without
changing the integral. The technique we will discuss here, on the other
hand, requires not a supersymmetry but an ordinary non-Abelian symmetry of the
integrand (as in non-Abelian gauge theories), the idea being to reduce such
an integral to one with a (much more tractable) Abelian symmetry.

More
precisely, the classical integration formula, which we wish to generalize to
functional integrals,  is what is known in group theory and harmonic
analysis as the Weyl integral formula. To state this formula, we need
some notation (and refer to the next section for details). We denote by
$G$ a compact Lie group and by $T$ a maximal torus of $G$. As every element
of $G$ is conjugate to some element of $T$ (in other words, every $g\in G$
can be `diagonalized'), a conjugation invariant function $f$ on $G$, $f(g)
= f(\h g h)$ for all $h \in G$, is determined by its restriction to $T$.
In particular, therefore, the integral of $f$ over $G$ can be expressed
as an integral over $T$, and the Weyl integral formula gives an expression
for the integrand on $T$,
\be
\int_{G}\!dg\,f(g) = \int_{T}\!dt\, \det\dw(t) f(t)\;\;.\label{w01}
\ee
Here $\dw(t)$ is the Weyl determinant whose precise form we will give below.
One can read this formula as expressing the fact that an integral of a
function with a non-Abelian (conjugation) symmetry can be reduced to an
integral of a function with an Abelian symmetry. In physics parlance one
would say that one has integrated over the $G/T$ part of the gauge volume
or partially fixed the gauge using an abelianizing gauge condition.

It is this formula that we wish to generalize to functional integrals, i.e.\
to integrals over spaces of maps $\mg$ from some manifold $M$ into $G$.
What is interesting about this generalization is the fact that the naive
extension of (\ref{w01}) to this case is definitely wrong for basic topological
reasons. This is in marked contrast with the other path integral formulae
we have discussed above which appear to correctly capture the topological
aspects of the situation without any modifications. The correct formula
in this case (correct in the sense that the topology comes out right)
turns out to include a summation over isomorphism classes of
non-trivial $T$-bundles on $M$ on the right hand side of (\ref{w01}), so
that (very roughly) one has
\be
\mbox{``}
\;\;\int_{\mg}\![dg]\,F[g] = \sum_{T-{\rm bundles}}\int_{\mt}\![dt]\,
   \det\dw[t] F[t]\;\;\mbox{''}\label{w02}
\ee
(see (\ref{weyl4}) for a less outrageous rendition of this formula).
To see how this comes about, suffice it to note here that in order to
apply (\ref{w01}) to spaces of maps, one first needs to establish that
maps from $M$ to $G$ can be diagonalized. It turns out that there
are topological obstructions to doing this globally (again, in physics
parlance, to choose this abelianizing gauge globally), and the
summation over topological sectors reflects this fact.

The classical finite dimensional version of the Weyl integral formula
(\ref{w01}) and its various ramifications have played an important
role in physics in the context of matrix models for a long time
\cite{mehta}, and in particular recently in view of the
connections between matrix models and quantum gravity (see e.g.\
\cite{dijk} for a review). Some of these integration formulae \cite{daci}
can also be understood in terms of the \dh\ theorem applied to integrals
over coadjoint orbits \cite{stone}, which provides an intriguing link
between these two types of localization.

The path integral version (\ref{w02}) of the \wif\ was first used in
\cite{btver},  where we calculated the partition function and
correlation functions
of some low-dimensional gauge theories (Chern-Simons theory and the $G/G$
gauged Wess-Zumino-Witten model) from the path integral. Subsequently,
we also applied it to 2d Yang-Mills theory \cite{btlec}, rederiving the
results which had been previously obtained by other  methods
\cite{rusakov,ewym,fine,btym}. The method has been applied by Witten
\cite{ewver} to solve the Grassmannian $N=2$ sigma model, and the
topological obstructions to diagonalization have been analysed in
detail in \cite{btdia}.

In the following we will first recall the necessary background from the
theory of Lie groups and Lie algebras and
then discuss the topological aspects to the
extent that the more precise version of (\ref{w02}) acquires some degree
of plausibility. To keep the presentation as simple as possible, we will
here only deal with simply connected groups. We then discuss 2d Yang-Mills
theory as an example.
For more details, the reader is referred to the review \cite{btver} and to
\cite{btdia}.

\subsection{The Weyl Integral Formula}

\subs{Background from the theory of Lie groups}

Let $G$ be a compact connected and simply-connected Lie group. We denote
by $T$ a maximal torus of $G$, i.e.\ a maximal compact connected Abelian
subgroup of $G$, and by $r$ the rank of $G$, $r=\mbox{rk}(G)=\dim T$.
We also denote by $G_{r}$ the set of regular elements of $G$, i.e.\ those
lying  in one and only one maximal torus of $G$, and set $T_{r}=G_{r}\cap T$.
The set of non-regular elements of $G$ is of codimension 3 and
$\pi_{1}(G)=0$ implies $\pi_{1}(G_{r})=0$.

The crucial information we need is that any element $g$ of $G$ lies in
some maximal torus and that any two maximal tori are conjugate to each
other. We will henceforth choose one maximal torus $T$ arbitrarily and
fix it. It follows that any element $g$ of $G$ can be conjugated into $T$,
$\h g h \in T$ for some $h\in G$.
Such an $h$ is of course not unique. First of all,
$h$ can be multiplied on the right by any element of $T$, $h \ra ht, t \in T$
as $T$ is Abelian. To specify the residual ambiguity in $h$, we need to
introduce the Weyl group $W$. It can be defined as the quotient $W=N(T)/T$,
where $N(T)= \{g\in G: \g t g\in T \;\forall t \in T\}$ denotes the normalizer
of $T$ in $G$. If $\h gh = t \in T$, then $(hn)^{-1}g(hn) =
n^{-1}tn \in T$, $n\in N(T)$,
is one of the finite number of images $w(t)$ of $t$ under the
action of the Weyl group $W$. It follows that for regular elements $g\in G_{r}$
the complete ambiguity in $h$ is $h\ra hn$. For
non-regular elements this ambiguity is larger (e.g.\ for $g$ the
identity element $h$ is completely arbitrary).

A useful way of summarizing the above result for regular elements,
one which will allow us to pose the
question of diagonalizability of $G_{r}$-valued maps in a form amenable to
topological considerations, is to say that the conjugation map
\bea
q: G/T \times T_{r} &\ra& G_{r} \nonumber\\
([h],t) &\mapsto& h t \h \label{w1}
\eea
is a $|W|$-fold covering onto $G_{r}$ (see e.g.\  \cite{btd,hel}).
As $G_{r}$ is simply-connected, this
covering is trivial and we can think of $\gt\times T_{r}$ as the total
space of a trivial $W$-bundle over $G_{r}$. Thus any element of
$G_{r}$ can be lifted to $G/T\times T_{r}$ (with a $|W|$-fold ambiguity).
As coverings induce isomorphisms on the higher homotopy groups, it also
follows that $\pi_{2}(G_{r})=\pi_{2}(\gt)=\pi_{1}(T)=\ZZ^{r}$,
to be contrasted with $\pi_{2}(G)=0$.

All of the above results are also true for Lie algebras of compact
Lie groups.
In particular, if we denote by $\lg$ the Lie algebra of $G$ and by $\lt$ a
Cartan subalgebra of $\lg$, there is a trivial $W$-fibration
\bea
q: G/T \times \lt_{r} &\ra& \lg_{r} \nonumber\\
([h],\tau) &\mapsto& h \tau \h \label{wl1}\;\;.
\eea

In order to state the Weyl integral formula, which can be thought
of as a formula relating an integral over $G_{r}$ to an integral
over $G/T\times T_{r}$ via (\ref{w1}), we will need some more
information.
First of all, we choose an invariant metric $(.,.)$ on $\lg$ and will use it
to identify $\lg$ with its dual $\lg^{*}$ whenever convenient. This
metric also induces natural Haar measures $dg$ and $dt$ on $G$ and $T$,
normalized to $\int_{ G}dg= \int_{ T}dt = 1$.

For the purpose of
integration over $ G$ we may restrict ourselves to $ G_{r}$
and we can
thus use (\ref{w1}) to pull back the measure $dg$ to $ G/ T \times  T$.
To calculate the Jacobian of $q$, we need to know the infinitesimal
conjugation action of $ T$
on $ G/ T$. Corresponding to a choice of $ T$
we have an orthogonal direct sum decomposition of the Lie algebra
$\lg$ of $ G$, $\lg = \lt \oplus \lk$.
$G$ acts on $\lg$ via the adjoint representation $\rm Ad$. This induces an
action of $ T$ which acts trivially
on $\lt$ and leaves $\lk$ invariant (the isotropy representation
${\rm Ad}_{\lk}$ of $ T$ on $\lk$, the tangent space to $ G/ T$).
Therefore the Jacobian matrix is
\be
\dw(t) =  ({\bf 1}- {\rm Ad}_{\lk}(t)) \label{w3}
\ee
and one finds the Weyl integral formula
\be
\int_{ G}\!dg\;f(g) =\trac{1}{|W|}\int_{ T}\!dt\; \det\dw(t)
                    \int_{ G/ T}\!dh\;f(h t \h)\;\;.\label{w2}
\ee
In particular, if $f$ is conjugation invariant, this reduces to
\be
\int_{G}\!dg\;f(g) = \trac{1}{|W|}
           \int_{ T}\!dt\; \det\dw(t) f(t) \;\;, \label{w4}
\ee
which is the version of the \wif\ which we will make use of later on.
The infinitesimal version of (\ref{w3}) is the
determinant of $\dw(\tau) = {\rm ad}_{\lk}(\tau)$, where $\tau\in\lt$.
It appears in the
corresponding formulae for integration over Lie algebras which are
otherwise the exact analogues of (\ref{w2},\ref{w4}).
A more explicit version of (\ref{w4}) for $G=SU(2)$ is given in
(\ref{w12}).

The determinant can be calculated by using the Cartan decomposition of
$\lg$. The complexified Lie algebra $\lg_{\CC}$ splits
into $\lt_{\CC}$ and the one-dimensional eigenspaces $\lg_{\a}$ of the
isotropy representation, labelled by the roots $\a$. It
follows that the Jacobian (Weyl determinant) can be written as
\bea
\det\dw(t) &=&\prod_{\a} (1-e^{\a}(t))\;\;,\nonumber\\
\det\dw(\tau)&=& \prod_{\alpha}(\alpha,\tau)\label{w7}\;\;.
\eea
For later use we note that $\dw(t)$ vanishes precisely on the
non-regular elements of $T$, this being the mechanism by which
non-regular maps (i.e. maps taking values also in the non-regular
elements of $G$) should be suppressed in the path integral.

\subs{Topological obstructions to diagonalization}

We have seen above that the crucial ingredient in the \wif\ is the
fact that any element of $G$ is conjugate to some element of $T$.
In \cite{btdia}
we investigated to which extent this property continues to hold for spaces
of (smooth) maps $\mg$ from a manifold $M$ to a compact Lie group $G$.
Here we will summarize the relevant results for simply-connected $G$,
as they will be essential
for the generalization of the \wif\ to integrals over $\mg$ and
$\map{M}{\lg}$.

Thus the question we need  to address is if a smooth map $g\in\mg$
can be written as
\be
g(x) = h(x)t(x)h(x)^{-1} \label{conj1}
\ee
for $t\in\mt$ and $h\in\mg$. Of course there is no problem with doing this
pointwise, so the question is really if it can be done consistently
in such a way that
the resulting maps are smooth. Intuitively it is clear that problems can
potentially arise from the ambiguities in $h$ and $t$. And it is indeed
easy to see
(by examples) that (\ref{conj1}) cannot be achieved globally and smoothly
on $M$ in general. In fact, the situation turns out to be manageable only
for regular maps, i.e.\ for maps taking values in $G_{r}$. The reason
for this is that non-regular maps may not be smoothly diagonalizable
in any open neighbourhood of a preimage of a non-regular element even if
the map takes on non-regular values only at isolated points (we will see
an example of this below). Clearly then, methods of (differential) topology
do not suffice
to deal with the problems posed by non-regular maps. Henceforth we will
deal almost exclusively with maps taking values in the dense set $G_{r}$
of regular elements of $G$. For these regular maps, the
problem can be solved completely and for simply-connected groups the results
can be summarized as follows:
\begin{enumerate}
\item Conjugation into $T$ can always be achieved locally, i.e.\ in
     open contractible neighbourhoods of $M$.
\item The diagonalized map $t$ can always be chosen to be smooth globally.
\item Non-trivial $\bT$-bundles on $M$ are the obstructions to
      finding smooth functions $h$ which accomplish (\ref{conj1}) globally.
     In particular, when $M$ and $\bG$
     are such that there are no non-trivial $\bG$ bundles on $M$ (e.g.\ for
     $G$ simply-connected and $M$ two-dimensional), all isomorphism
     classes of torus bundles appear as obstructions.
\end{enumerate}
We actually want to go a little bit further than that. Namely, as we
want to use (\ref{conj1}) as a change of variables (gauge transformation)
in the path integral,
we need to inquire what the effect of the occurrence of these obstructions
will be on the other fields in the theory. In particular, when gauge fields
are present, a transformation $g\ra\h gh$ has to be accompanied by a gauge
transformation $A\ra A^{h}=\h Ah + \h dh$.
E.g.\ in \cite{btver} it was found that the Weyl integral
formula leads to the correct results for the gauge theories investigated there
only when the gauge field integral includes a sum over $T$-connections on
all isomorphism classes of $T$-bundles on $M$ (a two-manifold in
those examples), even though the original bundle was trivial. One would
therefore like to know if the above obstructions to diagonalization can
account for this.
It turns out that the results concerning the behaviour of gauge fields
are indeed in complete agreement with what
one expects on the basis of the results of \cite{btver,btlec}, namely:
\begin{enumerate}
\setcounter{enumi}{3}
\item  If $P_{T}$ is the (non-trivial) principal $T$-bundle which is the
       obstruction to the global smoothness of $h$, then the $\lt$-component
       of $A^{h}$ defines a connection on $P_{T}$.
\item Hence the Weyl integral formula, when applied to gauge theories,
      should contain a sum over all those isomorphism classes of $T$-bundles
      on $M$ which arise as obstructions to diagonalization.
\end{enumerate}
Rather than present proofs of all these statements (which can be found in
\cite{btdia}), we will illustrate the above in two different ways. On the
one hand we will present a very hands-on example which allows one to see
explicitly the obstruction and how it is related to connections on non-trivial
$T$-bundles. On the other hand we will set up the general problem in such a
way that the reader with a background in topology will be able to read off
the answers to at least the questions (1)-(3) almost immediately.

Our example will be a particular map from $S^{2}$ to $SU(2)$. We parametrize
elements of $SU(2)$ as
\be
x_{4}{\bf 1} + \sum_{k=1}^{3}x_{k}\sigma_{k} =
\mat{x_{4}+ix_{3}}{x_{1} + i x_{2}}{-x_{1} + i x_{2}}{x_{4}-i x_{3}}\;\;,
\ee
subject to $\sum_{k=1}^{4}(x_{k})^{2}=1$, and consider the identity map
$g(x) = \sum_{k=1}^{3}x_{k}\sigma_{k}$ from the two-sphere to the equator
$x_{4}=0$ of $SU(2)$. As the only non-regular elements of $SU(2)$ are
$\pm{\bf 1}$, $g$ is a smooth regular map.
To detect a possible obstruction to diagonalizing $g$ smoothly,
we proceed as follows.
To any map $f$ from $S^{2}$ to $S^{2}$ we can assign an integer,
its winding number $n(f)$. It can be realized as the integral of the
pull-back of the normalized volume-form $\omega$ on the target-$S^{2}$,
$n(f)=\int_{S^{2}}f^{*}\omega$. This winding number is invariant under smooth
and continuous homotopies of $f$. Clearly for the identity map $g$ we have
$n(g)=1$. Representing (as above) $f$ as $f = \sum_{k}f_{k}\sigma_{k}$
with $\sum_{k}(f_{k})^{2}=1$, a more explicit realisation of $n(f)$ is
\be
n(f) = -\trac{1}{32\pi}\int_{S^{2}} \tr f[df,df] \;\;. \label{wno}
\ee
Here and in the following expressions like
$[df,df] = df_{k}\,df_{l} [\sigma_{k},\sigma_{l}]$
denote the wedge product of forms combined with the
commutator in the Lie algebra.
Now suppose that one can smoothly conjugate the map $g$ into
$U(1)$ via some $h$, $\h g h =t$. As the space of maps from $S^{2}$ to
$SU(2)$ is connected ($\pi_{2}(G)=0$), $g$ is homotopic to $t$ and one has
$n(g) = n(t)$. But, since $g^{2}=-{\bf 1}$, $t$ is a constant map (in fact,
$t=\pm\sigma_{3}$, the two being related by a Weyl transformation) so that
$n(t)=0$, a contradiction. This shows that there can be no smooth or continuous
$h$ satisfying $\h g h =t$. As both $g$ and its diagonalization
$\pm\sigma_{3}$ may just as well be regarded as Lie algebra valued maps,
this example establishes that obstructions to diagonalization
will also arise in the (seemingly topologically trivial) case of Lie algebra
valued maps.

To see how connections on non-trivial bundles arise, we consider
a slight generalisation $n(f,A)$ of $n(f)$,
\be
n(f,A) = -\trac{1}{32\pi}\int_{S^{2}} \tr f [df,df] -
\trac{1}{2\pi}\int_{S^{2}}  \tr [d(fA)] \, , \label{gno}
\ee
depending on both $f$ and an $SU(2)$-connection $A$. Because the
second term is a total derivative, (\ref{gno})
obviously coincides with (\ref{wno}) when $A$ is globally defined.
The crucial property of $n(f,A)$ is that it is gauge invariant, i.e.\
invariant under simultaneous transformation of $f$ and $A$,
\be
n(h^{-1}fh, A^{h})=n(f,A) \label{ginv} \;\;,
\ee
even for discontinous $h$
(the key point being that no integration by parts is required in establishing
(\ref{ginv}).
The advantage of a formula with such an invariance is that it
allows one to relate maps which are not necessarily homotopic.
In particular, let us now choose some (possibly discontinuous) $h$ such
that it conjugates
$g$ into $U(1)$, say $ g = h\sigma_{3} \h$ (since this can be done pointwise,
some such $h$ will exist). Using (\ref{ginv}) we find
\be
n(g,A)=1= -\trac{1}{2\pi}\int_{S^{2}} \tr \sigma_{3} d(A^{h})\;\;.
\label{chno}
\ee
In particular, if we introduce the Abelian gauge field $a=-\tr\sigma_{3}
A^{h}$, we obtain
\be
n(g,A) = 1 = \trac{1}{2\pi}\int_{S^{2}} da\;\;.
\ee
We now see the price of conjugating into the torus. The first Chern class of
the $U(1)$ component of the gauge field $A^{h}$ is equal to the winding
number of the original map! We have picked up the sought for non-trivial
torus bundles. In this case it is just the pull-back of the $U(1)$-bundle
$SU(2)\ra SU(2)/U(1)\sim S^{2}$ via $g$ and this turns out to be more or
less what happens in general.

Finally, we will show that a certain non-regular extension of this map
provides us with an example of a map which cannot
be smoothly diagonalized in any open neighbourhood of a non-regular point.
As a preparation, we make the obvious observation that if two maps from
$M$ to $G_{r}$ are regularly homotopic, i.e.\ homotopic in $\mgr$,
then if there are no obstructions to diagonalizing one of them,
there are also none for the other. Incidentally, this also
shows why winding numbers play a role at all in this discussion although
the space of maps from $S^{2}$ to $SU(2)$ is connected. As we have seen
in our discussion of $G_{r}$, $\pi_{2}(G_{r}) =\ZZ^{r}$, so that the
space of regular maps is not connected but decomposes into disjoint sectors
labelled by an $r$-tuple of winding numbers.

Consider now
the extension $\tilde{g}$ of $g$ to the identity map from the three-sphere to
$SU(2)$, $\tilde{g}(x)=x_{4}{\bf 1} + \sum_{k}x_{k}\sigma_{k}$. This
map takes on non-regular values only at $x_{4}=\pm 1$. There is clearly
no smooth diagonalization of the restriction of this map to any open set
containing the north-pole $\{x_{4}=1\}$. If there were, this would
in particular imply the existence of a global smooth diagonalization of
a map from $S^{2}$  to $SU(2)_{r}$ which is regularly homotopic to
$g$, as any neighbourhood of the north pole contains a surrounding two-sphere
  - a contradiction.

We will now return to the general situation and briefly describe how to
set up the problem concerning obstructions to diagonalization in such
a way that it can be solved by standard topological arguments.

Being able to (locally) conjugate smoothly into the maximal
torus is the statement that one can (locally) find smooth maps $h$
and $t$ such that $g_{U} = h t \h$. In other words, one
is looking for a (local) lift of the map $g\in\mgr$ to a map
$(h,t)\in\mg\times\mtr$. It will be convenient to break this problem
up into parts and to establish the (local)
existence of this lift in a two-step procedure, as indicated in the diagram
below.
\be
\bfig
\putsquare<1`-1`1`1;1000`1000>(500,0)%
[G\times T_{r}`G/T\times T_{r}`M`G_{r};p\times 1`(h,t)`q`g]
\putmorphism(1500,1000)(-1,-1)[``(f,t)]{1000}{-1}{l}
\efig
\ee
Here, on the right hand side of this diagram we recognize the conjugation
map (\ref{w1}) and its trivial $W$-fibration. The top arrow, on the other
hand, is essentially the non-trivial fibration $G\ra\gt$.
In the first step one attempts to lift $g$ along the diagonal, i.e.\ to
construct a pair $(f,t)$, where $f\in\mgt$,
which projects down to $g$ via the projection $q$.
By a standard argument the obstruction to doing this globally
is the possible non-triviality of the $W$-bundle
$g^{*}(\gt\times T_{r})$ on $M$. Happily, when $G$ is simply connected,
(\ref{w1}) is trivial and this obstruction is absent. Thus we have
just established the existence of a globally smooth diagonalization $t$ of
$g$.

In the second step, dealing with the upper triangle, one needs to lift $f$
locally to $\mg$. The obstruction to doing this globally is
the possible non-triviality of the principal $T$-bundle $f^{*}(G)$ over $M$
(as in the above example).
If this bundle is trivial (e.g.\ when $H^{2}(M,\ZZ)=0$, so that there
are no non-trivial torus bundles at all on $M$), a smooth $h$
accomplishing the diagonalization will exist globally. Locally, a lift
$h$ can of course always be found.

The principal
$T$-bundles which appear as obstructions are precisely those that can be
obtained from the trivial principal $G$-bundle $P_{G}$ by restriction
of the structure group from $G$ to $T$
(or, more informally, that sit inside $P_{G}$).
Under such a restriction, the $\lt$-part of a connection on $P_{G}$
becomes a connection on $P_{T}$, while the $\lk$-part becomes
a one-form with values in the bundle associated to $P_{T}$ via the
isotropy representation of $T$ on $\lk$.

Because of the similarity of (\ref{w1}) and (\ref{wl1}), the situation
for Lie algebra valued maps is exactly the same as above (differences
only occur for non-simply connected groups).

\subs{The Weyl integral formula for path integrals}

We have now collected all the results we need to state the \wif\
for integrals over $\mg$ which correctly takes into account the
topological considerations of the previous section.

For concreteness, consider a local functional $S[g;A]$ (the `action')
of maps $g\in\mg$ and gauge fields $A$ on a trivial principal $G$-bundle
$P_{G}$ on $M$ (a dependence on other fields could of course be
included as well). Assume that $\exp iS[g,A]$ is gauge invariant,
\be
\ex{i S[g,A]} = \ex{ i S[h^{-1}gh, A^{h}]} \;\;\;\;\;\;\forall\;h\in\mg\;\;,
\label{weyl3}
\ee
at least for smooth $h$. If e.g.~a partial integration is
involved in establishing the gauge invariance (as in Chern-Simons theory),
this may fail for
non-smooth $h$'s and more care has to be exercised when such a gauge
transformation is performed.
Then the functional $F[g]$ obtained by integrating  $\exp iS[g;A]$ over $A$,
\be
F[g]:=\int\![dA]\,\ex{ iS[g,A]}\;\;,
\ee
is conjugation invariant,
\be
F[h^{-1}gh] = F[g]\;\;.
\ee
It is then tempting to use a formal analogue of (\ref{w4}) to reduce the
remaining integral over $g$ to an integral over maps taking values in the
Abelian group $T$. In field theory language this amounts to using the
gauge invariance (\ref{weyl3}) to impose the `gauge condition' $g(x)\in T$.
The first modification of (\ref{w4}) will then be the replacement of
the Weyl determinant $\det\dw(t)$ by a functional determinant $\det\dw[t]$
of the same
form  which needs to be regularized appropriately (see the Appendix of
\cite{btlec}).

However, the crucial point is, of course, that this is not the
whole story. We already know that this `gauge condition' cannot necessarily
be achieved smoothly and globally. Insisting on achieving this `gauge'
nevertheless, albeit via non-continuous field
transformations, turns the $\lt$-component $A^{\lt}$
of the transformed gauge field $A^{h}$ into a gauge field on a
possibly non-trivial $T$ bundle $P_{T}$ (and the $\lk$-component
turns out to transforms as a section of the associated bundle $P_{T}
\times_{T}\lk$).
Moreover, we know that
all those $T$ bundles will contribute which arise as restrictions of
the (trivial) bundle $P_{G}$. Let us denote the set of isomorphism classes
of these $T$ bundles by $[P_{T};P_{G}]$. Hence the `correct'
(meaning correct modulo the analytical difficulties inherent in making any
field theory functional integral rigorous) version of the Weyl integral
formula, capturing the topological aspects of the situation,
is one which includes a sum over the contributions from the
connections on all the isomorphism classes of bundles in $[P_{T};P_{G}]$.

Let us denote
the space of connections on $P_{G}$ and on a principal $T$ bundle $P_{T}^{l}$
representing an element $l\in [P_{T};P_{G}]$ by ${\cal A}$ and ${\cal A}[l]$
respectively and the space of one-forms with values in the sections of
$P_{T}^{l}\times_{T}\lk$ by ${\cal B}[l]$.
Then, with
\be
Z[P_{G}] = \int\! [dA]\int\! [dg]\ex{ i S[g,A]} \;\;,
\ee
the Weyl integral formula for functional integrals reads
\be
Z[P_{G}] = \sum_{l\in [P_{T};P_{G}]}\int_{{\cal A}[l]} [dA^{\lt}]
                         \int_{{\cal B}[l]} [dA^{\lk}]
                         \int\! [dt] \dw[t] \ex{ i S[t,A^{\lt},A^{\lk}]}
                         \label{weyl4}
\ee
(modulo a normalization constant on the right hand side).
The $t$-integrals carry no $l$-label as the spaces
of sections of ${\rm Ad}P_{T}^{l}$ are all isomorphic to the space of
maps into $\bT$. There is an exactly analogous formula
generalizing the Lie algebra version of the Weyl integral formula.
Namely, assuming that we have a gauge invariant action $S[\f,A]$,
where $\f\in\map{M}{\lg}$, we can use the \wif\ to reduce its partition
function to
\be
Z[P_{G}] = \sum_{l\in [P_{T};P_{G}]}\int_{{\cal A}[l]}\! [dA^{\lt}]
                         \int_{{\cal B}[l]}\! [dA^{\lk}]
                  \int\! [d\tau] \dw[\tau] \ex{ i S[\tau,A^{\lt},A^{\lk}]}
                         \label{weyl5}
\ee

In the examples considered in \cite{btver,btlec}, the fields $A^{\lk}$
entered purely quadratically in the reduced action $S[t,A^{\lt},A^{\lk}]$
and could be integrated out directly, leaving behind an effective Abelian
theory depending on the fields $t$ and $A^{\lt}$
(respectively $\tau$ and $A^{\lt}$).
The information on the
non-Abelian origin of this theory is contained entirely in the
measure determined by $\dw[t]$ and the (inverse) functional determinant
coming from the $A^{\lk}$-integration. In all the examples considered so far,
e.g.\ in \cite{btver,btlec,ewver}, this procedure simplified the
path integral to the extent that it could be calculated explicitly.
One does of course not expect to be able to get that far in general.
Nevertheless, the simplification brought about by replacing an
interacting non-Abelian theory with a theory with only an Abelian
gauge symmetry should allow one to obtain at least some new information
from the path integral. We will see the formula (\ref{weyl4}) at work
in the next section, where we will use it to solve 2d Yang-Mills theory.

\subsection{Solving 2d Yang-Mills Theory via Abelianization}

In this section we will apply the \wif\ (\ref{weyl4}) to Yang-Mills
theory on a two-dimensional closed surface $\S$ \cite{btlec}. As mentioned
before, this theory has recently been solved by a number of other
techniques as well \cite{ewym,rusakov,fine,btym}, but it appears
to us that none of them is as simple and elementary as the one based
on Abelianization.  As the partition function of Yang-Mills theory
for a gauge group $G$ can be expressed entirely in terms of the
representation theory (more precisely the dimensions and the quadratic
Casimirs of the representations) of $G$, we will first briefly recall the
relevant formulae below. Then we will calculate the partition function.
To keep the Lie algebra theory as simple as possible (as it is only
of tangential interest here) we will assume that $G$ is not only
simply-connected but also simply-laced.
For the relation and application of these results to topological Yang-Mills
theory  see \cite{ew2d,btlec}.

\subs{Some Lie algebra theory}

It turns out that the partition function of Yang-Mills theory for a group
$G$ can be expressed in terms of the dimensions and quadratic Casimirs of
the unitary irreducible representations of $G$. We will recall here the
relevant formulae, see e.g.\ \cite{btd}.

First of all, we choose a positive Weyl chamber $C^{+}$ and
introduce a corresponding set of simple roots
$\{\alpha_{a}, a =1,\ldots,r\}$, a dual set of fundamental weights
$\{\lambda^{a}, a=1,\ldots,r\}$ and denote the weight lattice
by $\Lambda=\ZZ[\lambda^{a}]$. Then the highest weights of the unitary
irreducible representations can be identified with the elements of
$\Lambda^{+}=\Lambda\cap\bar{C}^{+}$, $\bar{C}^{+}$ denoting the closure
of the Weyl chamber. For $\mu\in\Lambda^{+}$ we denote by $d(\mu)$ and
$c(\mu)$ the dimension and the quadratic Casimir of the corresponding
representation. The formulae for these are
\bea
&&d(\mu) = \prod_{\alpha >0}\frac{(\alpha,\mu+\rho)}{(\alpha,\rho)}
\label{weyldim}\\
&& c_{\mu} = (\mu + \rho,\mu + \rho) - (\rho,\rho)\;\;,\label{cas}
\eea
where $\rho = \frac{1}{2}\sum_{\alpha > 0}\alpha$ is the Weyl vector.

Let us illustrate this and the formulae obtained in our discussion of
the Weyl integral formula in the case $G=SU(2)$ and $T=U(1)$.
We use the trace to identify the Lie algebra $\lt$ of $T$ with its dual.
Then we can choose the positive root $\a$ and the fundamental weight $\la$
to be $\a = \mbox{diag}(1,-1)$ and $\la =\alpha/2$, so that $\rho = \lambda$.
We also parametrize elements of $T$ as
$t = \exp i\la\f=\mbox{diag}(\exp i\f/2,\exp -i\f/2)$.
It follows that the Weyl determinant $\det\dw(t)$  (\ref{w3}) is
$\det\dw(t) = 4 \sin^{2}\f/2$. Hence the \wif\ for class functions
(conjugation invariant functions) is
\be
\int_{\bG}\!dg\;f(g) = \frac{1}{2\pi}\int_{0}^{4\pi}\!d\f\,
\sin^{2}(\f/2) f(\f) \;\;. \label{w12}
\ee
For the spin $j$ representation of $SU(2)$ (with highest
weight $\mu(j)=2j\la\in\Lambda^{+}$), one finds $d(\mu(j))=2j+1$ and
$c(\mu(j))=2j(j+1)$.

\subs{2d Yang-Mills theory}

The action of Yang-Mills theory on a two-dimensional surface $\Sg$ of
genus $g$ is
\be
S[A] = \trac{1}{2\ep}\int_{\Sg} \tr F_{A}*F_{A}\;\;. \label{ymact}
\ee
Here $F_{A}=dA + \frac{1}{2}[A,A]$ is the curvature of a gauge field
$A$ on a trivial principal $G$-bundle $P_{G}$ on $\Sg$, $*$ denotes the
Hodge duality operator with respect to some metric on $\Sg$,
$\tr$ refers to the trace in the fundamental representation for $G=SU(n)$.
$\ep$ represents the coupling constant of the theory. A $\tr$ will henceforth
be understood in integrals of Lie algebra valued forms. It is readily seen
that, since $*F_{A}$ is a scalar, the action does not depend on the details
of the metric but only on the area $A(\Sg)=\int_{\Sg}*1$
of the surface. As a change in the metric
can thus be compensated by a change in the coupling constant, we can
without loss of generality choose a metric and a corresponding volume
two-form
$\omega$ with unit area. $\omega$ is a symplectic form on $\Sg$
representing the generator of $H^{2}(\Sg,\ZZ)\sim\ZZ$.

The action is invariant
under gauge transformations $A\ra A^{g}$ and the
path integral that we would like to compute is
\be
Z_{\Sg}( \ep) = \int_{{\cal A}}\! [dA]\,
\ex{\frac{1}{2\ep}\int_{\Sg} F_{A}*F_{A}} \;\;.
\ee
It will be convenient to rewrite this in first order form by introducing
an additional $\lg$-valued scalar field $\f\in\map{\Sg}{\lg}$
(cf.\  (\ref{ze2})), so that the partition function becomes
\be
Z_{\Sg}(\eps) = \int_{{\cal A}}\! [dA]\int_{\map{\Sg}{\lg}}\! [d\f]
\ex{\int_{\Sg} i\f F_{A}+ \frac{\ep}{2}\f*\f}\;\;. \label{pi}
\ee
The gauge invariance of the action $S[\f,A]$ appearing in (\ref{pi}) is
\be
S[\g\f g,A^{g}]=S[\f,A]\;\;,
\ee
and we are thus precisely in a situation where we can apply the
considerations of the previous section. Before embarking on that,
it will however be useful to recall some properties of the quantum field
theory defined by the classical action (\ref{ymact}).

Yang-Mills theory in two-dimensions is super-renormalisable and there
are no ultraviolet infinities associated with diagrams involving
external $A$ or $\f$ fields. Furthermore, on a compact
manifold $\Sg$  there are no infrared divergences associated with
these diagrams either. However, because of the coupling of $\f$ to the metric,
there are diagrams which do not involve external $\f$ or $A$ legs but do
have external background graviton legs and which require regularisation.
These terms arise in the determinants that are being calculated and they
depend only on the area and topology of $\Sg$, i.e.\ they have the
form  $\a_{1}(\ep)A(\Sg) + \a_{2}(\ep)\c(\Sg)$,
where $\c(\Sg)=2-2g$ is the Euler number of $\Sg$. These
may be termed area and topological standard renormalisations.
We wish to ensure that the scaling invariance that allowed us to move
all of the metric dependence into $\ep$ is respected. Hence only
those regularisation schemes which preserve this symmetry are to be
considered. Then the dependence of $\a_{1}$ on $\ep$ is
fixed to be $\a_{1}(\ep) = \ep \beta$ and $\a_{2}$ and $\beta$ are
independent of $\ep$.

With these remarks in mind,
we can now proceed by using the gauge freedom to
conjugate $\f$ into $\map{\Sg}{\lt}$ (i.e.\ to impose $\f^{\lk}=0$). This
simply amounts to plugging $S[\f,A]$ into (\ref{weyl5}). Alternatively,
one can of course implement this gauge condition by following the standard
Faddeev-Popov procedure. The ghost determinant one obtains in this way is
precisely the Weyl determinant $\det\dw[\tau]$ appearing in (\ref{weyl5}).

We should perhaps stress
once more that in choosing this gauge we are assuming that $\f$ is regular.
We have discussed this issue at some length in \cite{btver,btlec} and have
nothing to add to this here. We just want to point out that only constant
modes of $\f$ contribute to the path integral and that a non-regular constant
$\f$ would give a divergent contribution to the partition function from the
gauge fields (and a zero from the Weyl determinant). Thus any way of
regularizing this divergence will set the contribution from the non-regular
points to zero (by the ghost contribution) and is therefore
tantamount to discarding the non-regular maps which is what we will do.

Let us now determine the ingredients that go into (\ref{weyl5}).
The Weyl determinant $\det\dw[\tau]$
is the functional determinant of the adjoint action
by $\tau$ on the space of maps $\map{\Sg}{\lk} = \Omega^{0}(\Sg,\lk)$,
\be
\det\dw[\tau] = \det \ad(\tau)|_{\Omega^{0}(\Sg,\lk)}\;\;.
\ee
This determinant of course requires some regularization. We will say
more about it once we have combined this with another determinant that
will arise below.

We now look at the reduced action $S[\tau,A^{\lt},A^{\lk}]$.
Because of the
orthogonality of $\lt$ and $\lk$ with respect to the trace, the only
terms that survive in the action are
\be
S[\tau,A^{\lt},A^{\lk}] =
\int_{\Sg} \tau dA^{\lt} + \trac{1}{2}[\tau,A^{\lk}]A^{\lk} +
              \trac{\ep}{2}\tau*\tau\;\;.
\ee
Integrating out $A^{\lk}$,
one obtains the inverse square root of the determinant
of $\ad(\tau)$, this time acting on the space $\Omega^{1}(\Sg,\lk)$ of
$\lk$-valued one-forms,
\be
\int[dA^{\lk}] \;\;\;\;\;\;\Ra\;\;\;\;\;\;
\det{}^{-1/2}\ad(\tau)|_{\Omega^{1}(\Sg,\lk)}\;\;.
\ee
It is clear that this almost cancels against the Weyl determinant as,
modulo zero-modes, a one-form in 2d has as many degrees of fredom as
two scalars. The zero mode surplus is one constant scalar mode minus
$(2g/2)=g$ harmonic one-form modes so that the combined determinant
would simply reduce to a finite dimensional determinant,
\be
\frac{\det\ad(\tau)|_{\Omega^{1}(\Sg,\lk)}}%
{\det{}^{1/2}\ad(\tau)|_{\Omega^{1}(\Sg,\lk)}} =
(\det\ad(\tau)|_{\lk})^{\c(\Sg)/2}\;\;,
\label{detrat}
\ee
if $\tau$ were constant. As we will see shortly that only the constant modes
of $\tau$ contribute to the path integral, we will just leave it at
(\ref{detrat}) and refer to \cite{btlec} for a more careful discussion
using a heat kernel regularization of the determinants involved.

Putting all the above together we see that we are left with an
Abelian functional integral over $\tau$ and $A^{\lt}$ and  a sum over
the topological sectors,
\be
Z_{\Sg}(\ep) = \sum_{\l\in[P_{T};P_{G}]}\int_{{\cal A}[l]}[dA^{\lt}]
\int\![d\tau]
(\det\ad(\tau)|_{\lk})^{\c(\Sg)/2}
\ex{\int_{\Sg}(i\tau dA^{\lt}+\trac{\ep}{2}\tau*\tau)}\;\;.
\label{abpi1}
\ee
To evaluate this, we will use one more trick,
namely to change variables from $A^{\lt}$ to $F^{\lt}=dA^{\lt}$.
To see how to
do that let us first note that
the theory we have arrived at has still got local Abelian gauge invariance
(as well as $W$-invariance as a remnant of the non-Abelian gauge
invariance of the original action). Let us therefore choose some
gauge fixing condition $E(A^{\lt})=0$ and change variables from
$A^{\lt}$ to $(F^{\lt},E(A^{\lt}))$. This change of
variables is fine away from the gauge equivalence classes of flat
connections (i.e.\ the moduli space $T^{2g}$ of flat $T$-connections
on $\Sg$). As these flat connections do not appear in the action, they just
give a finite volume factor (depending on the
normalization of the metric on $T$) which we will not keep track of.
On the remaining modes of the gauge field this change of variables
is then well defined. What makes it so useful is the fact, demonstrated in
\cite{btbf1}, that the corresponding Jacobian
cancels precisely against the Faddeev-Popov determinant arising from the
gauge fixing.

As a consequence, after this change of variables and integration over the
ghosts, the multiplier field and $E(A^{\lt})$
all derivatives disappear from (\ref{abpi}) and the remaining path integral
over $\tau$ and $F^{\lt}$ is completely elementary.
The only thing to keep
track of is that we are not integrating over arbitrary two-forms $F^{\lt}$
but only over those which arise as the curvature two-form on some $T$-bundle
over $\Sigma$ which arises as the restriction of the trivial $G$-bundle
$P_{G}$.
As there are no non-trivial $G$-bundles on $\Sg$, the sum over topological
sectors $l\in [P_{T};P_{G}]$
will extend over all isomorphism classes of $T$-bundles on $\Sg$.
These are classified by their first Chern class in
$H^{2}(\Sg,\ZZ^{r})\sim\ZZ^{r}$, and hence the summation over $l$ in
(\ref{weyl5}) can be thought of as a summation over $r$-tuples of integers.
In particular, if we expand $F^{\lt}$ in a basis of simple roots
$\{\alpha_{a}\}$ of $\lg$ as $F^{\lt}=i\alpha_{a} F^{a}$,
then in the topological sector characterized by $l=(n^{1},\ldots,n^{r})$
we have
\be
\int_{\Sg}F^{a} = 2\pi n^{a}\;\;.\label{chern}
\ee
As all topological sectors appear, one way to enforce (\ref{chern}) in the
path integral is to
impose a periodic delta-function constraint on the curvature $F^{\lt}$
of the torus gauge field, i.e. to insert
\be
\d^{P}(\int F^{\lt}) =
\prod_{a=1}^{r}\sum_{m_{a}\in\ZZ}
\ex{im_{a}\int_{\Sg}F^{a}}    \label{period}
\ee
into the path integral. It will be convenient to write this as a sum
over the weight lattice $\Lambda$ dual to the lattice spanned by the simple
roots (or Chern classes),
\be
\d^{P}(\int F^{\lt}) =
\sum_{\lambda\in\Lambda}\ex{\int_{\Sg}\lambda F^{\lt}}\;\;.
\ee
Then the path integral to be evaluated reads
\be
Z_{\Sg}(\ep) = \sum_{\lambda\in\Lambda}\int\![F^{\lt}]\int\![d\tau]
(\det\ad(\tau)|_{\lk})^{\c(\Sg)/2}
\ex{\int_{\Sg}(i\tau+\lambda)F^{\lt}+\trac{\ep}{2}\tau*\tau}\;\;.
\label{abpi}
\ee
Let us do the $F^{\lt}$-integral first. This imposes the
delta-function constraint $\tau=i\lambda$ on $\tau$, so that in particular
only the constant modes of $\tau$ ever contribute to the partition function.
We are then left with just the sum over the (regular) elements of $\Lambda$.
Eliminating the Weyl group invariance is then the same as summing only over
$\lambda$'s in the interior of the positive Weyl chamber $C^{+}$.
Writing $\lambda$ as $\lambda=\mu+\rho$ one sees that the sum is now precisely
over all the highest weights $\mu\in\Lambda^{+}=\Lambda\cap\bar{C}^{+}$
of the unitary irreducible representations of $G$,
\be
Z_{\Sg}(\ep) = \sum_{\mu\in\Lambda^{+}}
\prod_{\alpha>0}(\alpha,\mu+\rho)^{\c(\Sg)}
\ex{-\frac{\ep}{2}(\mu+\rho,\mu+\rho)}
\;\;.
\ee
Using (\ref{weyldim}) and (\ref{cas}), we can rewrite this as
\be
Z_{\Sg}(\ep)=\ex{-\frac{\ep}{2} (\rho,\rho)}
\prod_{\alpha >0}(\alpha,\rho)^{\c(\Sg)}
\sum_{\mu\in\Lambda^{+}}d(\mu)^{\c(\Sg)}\ex{-\frac{\ep}{2}c(\mu)}\;\;.
\ee
This differs from the expression obtained in e.g.\
\cite{rusakov,fine,ewym,btym}
only by the first two terms which are precisely of the form of a standard
area and topological renormalisation respectively. What is interesting,
though, is that the form of the partition function obtained here (with
$(\mu + \rho,\mu + \rho)$ in the exponential instead of $c_{\mu}$) agrees
with what one obtains from non-Abelian localization \cite{ew2d} and is the
correct one to use for determining the intersection numbers of the moduli
space of flat $G$-connections on $\Sg$.

The uses of the Weyl integral formula are of course not restricted
to Yang-Mills theory (and its non-linear cousin, the $G/G$-model). For
instance, in principle it can also be applied to gauge theories on manifolds
of the form $M\times S^{1}$, where the `temporal gauge' $\del_{0}A_{0}=0$
can profitably be augmented by using time-independent gauge transformations
to achieve the gauge condition $A_{0}^{\lk}=0$ (see e.g.\ \cite{btver} for
an application to Chern-Simons theory).
In other words, the temporal gauge reduces $A_{0}$ to a map from $M$ to $\lg$
to which the considerations of this section regarding diagonalizability
and the ensuing form of the path integral can be applied.

In practice,
however, this method has its limits, not only because in higher dimensions
the issue of non-regular maps raises its ugly head but also because
simplifications brought about by a suitable choice of gauge alone will
not be sufficient to make the theory solvable without recourse to
more traditional techniques of quantum field theory as well. On a more
optimistic note one may hope that the technique developed here can shed
some light on the global aspects of the Abelian projection technique
introduced in \cite{tho}, currently popular in lattice gauge theories.

Finally, we want to mention that arguments as in section 4.1
permit one to also deduce a path integral analogue of the Weyl integral
formula  (\ref{w2}) valid for functions which are not conjugation invariant.
This formula is then applicable to theories with less or no gauge symmetry,
and can potentially help to disentangle the degrees of freedom of the
theory. This appears to be the case e.g.\ in the ordinary (ungauged)
Wess-Zumino-Witten model.

%\subsubsection*{Acknowledgements}

%REFERENCES
\rnc{\Large}{\normalsize}

\end{document}